\documentclass[12pt,a4paper]{article}
\pdfoutput=1
\usepackage{indentfirst}
\usepackage{graphicx}
\usepackage{subfigure}
\usepackage{threeparttable}
\usepackage{array}
\usepackage{multirow}
\usepackage{amsmath}
\usepackage{hyperref}
\hypersetup{pdfauthor={Ruifeng Yuan, Chengwen Zhong},pdftitle={A conservative implicit scheme for steady state solutions of diatomic gas flow in all flow regimes}}

\begin{document}


\title{A conservative implicit scheme for steady state solutions of diatomic gas flow in all flow regimes}
\renewcommand{\thefootnote}{\fnsymbol{footnote}}
\author{Ruifeng Yuan\footnotemark[1], Chengwen Zhong\footnotemark[1]}
\footnotetext[1]{National Key Laboratory of Science and Technology on Aerodynamic Design and Research, Northwestern Polytechnical University, Xi'an, Shaanxi 710072, China}
\footnotetext{\emph{Email addresses:} xyrfx@mail.nwpu.edu.cn (Ruifeng Yuan), zhongcw@nwpu.edu.cn (Chengwen Zhong)}
\date{Oct. 23, 2018}
\maketitle


\rule[-5pt]{\textwidth}{0.5pt}
\begin{abstract}
An implicit scheme for steady state solutions of diatomic gas flow is presented. The method solves the Rykov model equation in the finite volume discrete velocity method (DVM) framework, in which the translational and rotational degrees of freedom are taken into account. At the cell interface, a difference scheme of the model equation is used to construct a multiscale flux (similar to discrete unified gas-kinetic scheme (DUGKS)), so that the cell size is not constrained by the cell Knudsen (Kn) number. The physical local time step is implemented to preserve the multiscale property in the nonuniform-mesh case. The implicit macroscopic prediction technique is adopted to find a predicted equilibrium state at each time level and the implicit macroscopic governing equation is solved along with the implicit microscopic system. Furthermore, an efficient integral error compensation technique is applied, which makes the scheme conservative and allows more flexible discretization for particle velocity space. In the test cases, the unstructured velocity-space mesh is used, the present method is proved to be efficient and accurate.
~\\

\end{abstract}
\rule[5pt]{\textwidth}{0.5pt}


\section{Introduction}

Due to the development of hypersonic vehicle, spacecraft and micro-electromechanical system (MEMS), much effort has been devoted to study the nonequilibrium flow. Many numerical methods have been proposed for the nonequilibrium flow simulation, among which the deterministic method is always an attractive group of methods in the history. Unlike the famous direct simulation Monte Carlo (DSMC) method \cite{Bird1994Molecular}, which tries to describe the dynamic of gas particles through Monte Carlo technique, the deterministic method uses the particle velocity distribution function to describe the gas and solves the Boltzmann equation or its model equations by a regular numerical scheme to do the simulation.

The conventional discrete velocity method (DVM), or also known as the discrete ordinate method (DOM), is a classical deterministic method \cite{Goldstein1989Investigations,Yang1995Rarefied,Mieussens2000Discretev,li2004Study,Titarev2007Conservative} for the nonequilibrium flow simulation. In this category of methods, the transportation term and the collision term of the equation are totally decoupled, which makes the scheme very concise and efficient in the high Knudsen (Kn) number flow simulation. However, in order to get an accurate result, the cell size and the time step are very restricted in the transition and continuum flow regimes due to the intensive particle collision, making the method prohibitively expensive when approaching the continuum limit. Furthermore, in DVM, the governing equation is discretized in time, physical space and particle velocity space, leading to high computational cost and memory cost in 3D case. Hence, much effort has been devoted to accelerate the method. In Yang and Huang's work \cite{Yang1995Rarefied}, an implicit DVM has been presented to lift the restriction on time step, but in the scheme the equilibrium state $g^{n+1}$ is simply approximated by $g^n$, which makes the scheme hard to converge in the continuum regime \cite{Mieussens2000DISCRETE}. Mieussens proposed another version of implicit DVM scheme \cite{Mieussens2000Discretev,Mieussens2000DISCRETE}, in which $g^{n+1}$ is calculated through a linear mapping between the equilibrium state $g$ and the distribution function $f$. The linear mapping involves a large Jacobian matrix and makes the method a little complex. Moreover, Chen et al.~\cite{Chen2017A} proposed a memory saving technique and the memory consumption for the steady state calculation can be reduced to the order of the conventional macroscopic method.

It is worth noting that although the implicit algorithm can release conventional DVM from the restriction of the time step, the method still needs to use a very small cell size to get an accurate result in low cell Kn number case. This shortcoming is first overcome by Xu and Huang's unified gas-kinetic scheme (UGKS) \cite{Xu2010A}. In the scheme, the basic framework of DVM is adopted but the analytical solution of the model equation is used to construct a multiscale numerical flux at the cell interface, then the cell size is not restricted by the cell Kn number anymore and the time step for the explicit scheme is only constrained by the CFL condition. Guo et al.~\cite{guo2013discrete,guo2015discrete} proposed the discrete unified gas-kinetic scheme (DUGKS) based on the similar idea, in which the multiscale flux at the cell interface is constructed through a difference scheme of the model equation. There are also implicit versions of these multiscale methods. Mao et al.~\cite{Mao2015STUDY} presented an implicit UGKS with a similar treatment for $g^{n+1}$ to Ref.~\cite{Yang1995Rarefied}. Zhu et al.~\cite{Zhu2016Implicit} used an implicit macroscopic prediction technique to calculate a predicted $\tilde g^{n+1}$ in their implicit UGKS, which is concise and guarantees high convergence rate in all flow regimes. The method is also coupled with the multigrid algorithm \cite{Zhu2017Unified}. Yang et al.~\cite{Yang2018An} applied the memory saving technique into the implicit multiscale scheme with the macroscopic prediction, where the multiscale flux has a form similar to DUGKS.

An important thing for the implicit scheme solving the model equation in the DVM framework is the conservativeness. In the scheme, numerical integrals in the velocity space will be done per time level to calculate the equilibrium state $g$. If without any special treatment, the integral error will appear as an additional source term, making the scheme nonconservative and hard to converge. One solution for this problem is to use high-accuracy quadrature method in the velocity space but this will make the discretization of the velocity space so constrained.
Mieussens \cite{Mieussens2000Discretev,Mieussens2000DISCRETE} has used the conservation condition to find a discrete equilibrium state satisfying the conservative property at the discrete level. Although this method has to solve a nonlinear system with Newton iteration algorithm, less velocity points can be used and the overall efficiency of the scheme will be improved. This technique has been widely applied in many works \cite{Titarev2007Conservative,Huang2011Aconservative,Huang2015Aconservative,Jiang2015STUDY}.

The original UGKS and DUGKS \cite{Xu2010A,guo2013discrete,guo2015discrete} are constructed for monatomic gas. For the air, it mainly consists of oxygen and nitrogen, both are diatomic gas. In the previous works, Liu et al.~\cite{liu2014unified} have developed UGKS for diatomic gas based on the Rykov model equation \cite{Rykov1975A}. Zhang \cite{zhang2015Aunified} and Wang et al.~\cite{Wang2017Unified} presented UGKS for diatomic gas considering the vibrational degree of freedom. All of these diatomic schemes are explicit. In this paper, a multiscale implicit scheme for steady state solutions of diatomic gas flow is presented. The scheme is based on the Rykov model equation \cite{Rykov1975A} and the rotational degrees of freedom are taken into account. The multiscale flux at the cell interface is constructed through a difference scheme (the same idea of DUGKS). The macroscopic prediction technique of Zhu et al.~\cite{Zhu2016Implicit} is adopted to handle $g^{n+1}$. The physical local time step is applied to preserve the multiscale property of the scheme in the case of nonuniform mesh. Furthermore, an integral error compensation technique is applied to make the scheme conservative. This compensation procedure is very efficient and the maximum additional computation cost is less than 3.6\%. The unstructured discretization of the velocity space is applied in the test cases and it is verified that the present method is accurate and efficient, with high flexibility in the discretization of the velocity space.

The remainder of the paper is organized as follows. In Section \ref{sec:method}, the basic model of the diatomic gas is briefly introduced and then the general framework of the numerical method, the construction of the multiscale flux and the integral error compensation technique are described in order. In Section \ref{sec:numericaltest}, the efficiency and accuracy of the method are testified by three test cases. The physical local time step and the integral error compensation are validated. Section \ref{sec:conclusions} is a summary about the work of this paper.

\section{Numerical method}\label{sec:method}

\begin{figure}
\centering
\includegraphics[width=0.6\textwidth]{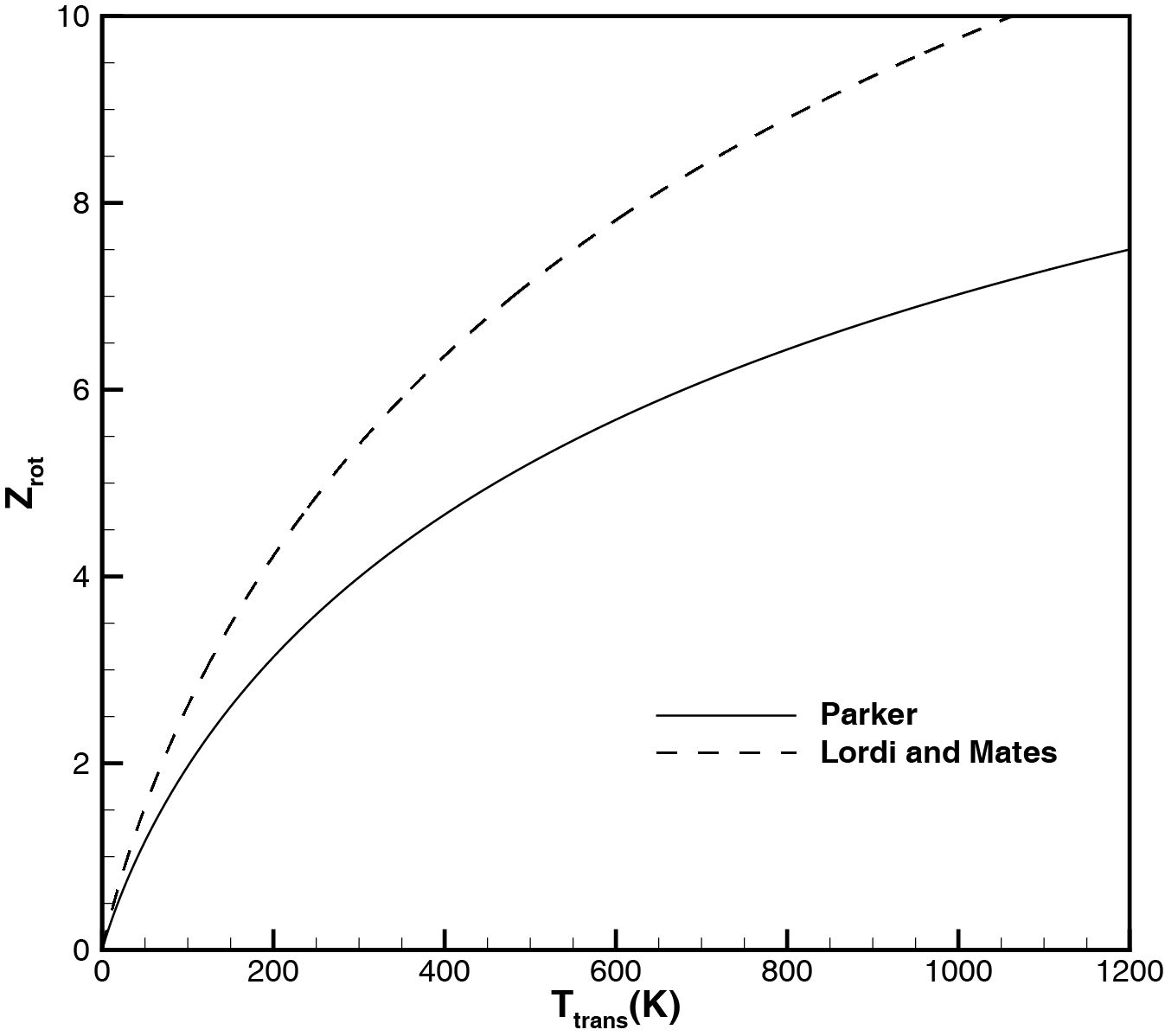}
\caption{\label{fig:method_zrot}The rotational relaxation collision number $Z_{\rm{rot}}$.}
\end{figure}

\begin{figure}
\centering
\includegraphics[width=0.75\textwidth]{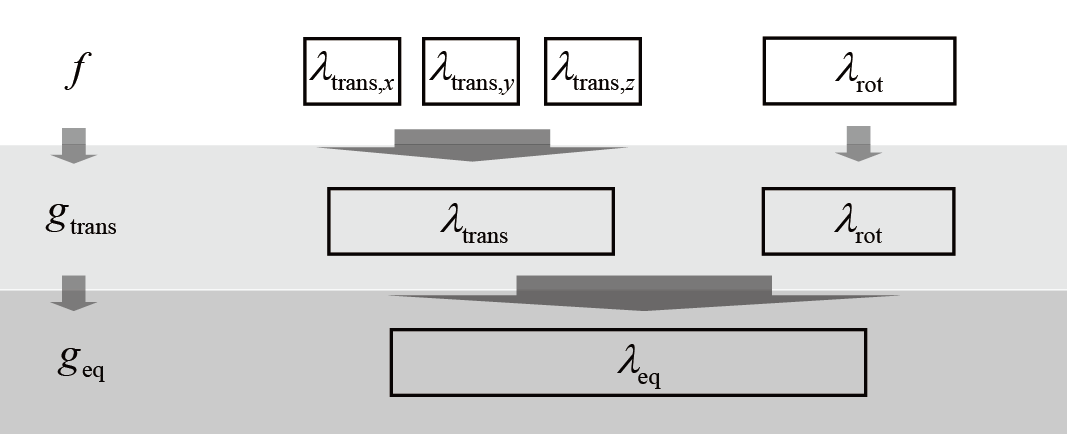}
\caption{\label{fig:method_rykovrelaxation}The relaxation process of the Rykov model.}
\end{figure}

For the diatomic molecule, there are internal degrees of freedom besides the three translational degrees of freedom. At room temperature, there are two rotational degrees of freedom. At temperature higher than 1000K, the vibrational degrees of freedom start to be excited. In this paper, only the translational and rotational degrees of freedom are considered. The basic physical model is based on the gas kinetic framework. The state of the gas is described by the particle velocity distribution function $f$, which is related to the macroscopic variables through
\begin{equation}\label{eq:intf}
\vec W = \int {\vec \psi fd\Xi},
\end{equation}
where $\vec W=(\rho,\rho\vec U,\rho E,\rho E_{\rm{rot}})^T$ is the vector of the macroscopic variables, $\rho E_{\rm{rot}}$ is the rotational energy density, $\vec \psi$ is the vector of moments $\vec \psi  = {\left( {1,\vec u,\frac{1}{2}({{\vec u}^2} + {{\vec \xi }^2}),\frac{1}{2}{{\vec \xi }^2}} \right)^T}$, $\vec u$ and $\vec \xi$ are the translational and rotational velocities of the gas particle, $d\Xi  = dudvdwd{\xi _1}d{\xi _2}$ is the velocity space element. The stress tensor $\pmb{P}$ and the heat flux $\vec q$ can also be calculated by $f$ as
\begin{equation}\label{eq:stress}
    \pmb{P} = \int {\vec c\vec cfd\Xi },
\end{equation}
\begin{equation}\label{eq:qflux}
\vec q = \int {\frac{1}{2}\vec c({{\vec c}^2} + {{\vec \xi }^2})fd\Xi } ,
\end{equation}
where $\vec c$ is the peculiar velocity $\vec c = \vec u - \vec U$. In particular, the translational heat flux ${\vec q_{{\rm{trans}}}}$ and the rotational heat flux ${\vec q_{{\rm{rot}}}}$ can be calculated respectively as
\begin{equation}\label{eq:qfluxtrans}
{\vec q_{{\rm{trans}}}} = \int {\frac{1}{2}\vec c{{\vec c}^2}fd\Xi } ,
\end{equation}
\begin{equation}\label{eq:qfluxrot}
{\vec q_{{\rm{rot}}}} = \int {\frac{1}{2}\vec c{{\vec \xi}^2}fd\Xi } .
\end{equation}
The dynamics of the distribution function $f$ is described by the Rykov model equation \cite{liu2014unified,Rykov1975A},
\begin{equation}\label{eq:rykov}
\frac{{\partial f}}{{\partial t}}{\rm{ + }}\vec u \cdot \frac{{\partial f}}{{\partial \vec x}} = \frac{{{g_{{\rm{trans}}}} - f}}{\tau } + \frac{{{g_{{\rm{eq}}}} - {g_{{\rm{trans}}}}}}{{{Z_{{\rm{rot}}}}\tau }}.
\end{equation}
In the model equation, ${{g_{{\rm{trans}}}}}$ and ${{g_{{\rm{eq}}}}}$ are equilibrium states expressed as
\begin{equation}\label{eq:eq_trans}
{g_{{\rm{trans}}}} = \rho {\left( {\frac{{{\lambda _{{\rm{trans}}}}}}{\pi }} \right)^{\frac{3}{2}}}{e^{ - {\lambda _{{\rm{trans}}}}{{\vec c}^2}}}\frac{{{\lambda _{{\rm{rot}}}}}}{\pi }{e^{ - {\lambda _{{\rm{rot}}}}{{\vec \xi }^2}}}(1 + {H_{{\rm{trans}}}}),
\end{equation}
\begin{equation}\label{eq:eq_eq}
{g_{{\rm{eq}}}} = \rho {\left( {\frac{{{\lambda _{{\rm{eq}}}}}}{\pi }} \right)^{\frac{3}{2}}}{e^{ - {\lambda _{{\rm{eq}}}}{{\vec c}^2}}}\frac{{{\lambda _{{\rm{eq}}}}}}{\pi }{e^{ - {\lambda _{{\rm{eq}}}}{{\vec \xi }^2}}}(1 + {H_{{\rm{eq}}}}),
\end{equation}
where $\lambda $ is a variable related to the temperature $T$ by $\lambda  = 1/(2RT)$ and the subscripts \emph{trans}, \emph{rot}, \emph{eq} denote the translational, rotational, thermo-equilibrium variables. The terms ${H_{{\rm{trans}}}}$ and ${H_{{\rm{eq}}}}$ are the correction terms derived from the Hermite polynomial for heat flux relaxation rate,
\begin{equation}
\begin{aligned}
{H_{{\rm{trans}}}} =  & \frac{{4(1 - \Pr )\lambda _{{\rm{trans}}}^2{{\vec q}_{{\rm{trans}}}} \cdot \vec c}}{{5\rho }}(2{\lambda _{{\rm{trans}}}}{{\vec c}^2} - 5)\\
 & + \frac{{4(1 - \sigma )\lambda _{{\rm{trans}}}^{}\lambda _{{\rm{rot}}}^{}{{\vec q}_{{\rm{rot}}}} \cdot \vec c}}{\rho }(\lambda _{{\rm{rot}}}^{}{{\vec \xi }^2} - 1),
\end{aligned}
\end{equation}
\begin{equation}
\begin{aligned}
{H_{{\rm{eq}}}} =  & {\omega _0}\frac{{4(1 - \Pr )\lambda _{{\rm{eq}}}^2{{\vec q}_{{\rm{trans}}}} \cdot \vec c}}{{5\rho }}(2{\lambda _{{\rm{eq}}}}{{\vec c}^2} - 5)\\&
 + {\omega _1}\frac{{4(1 - \sigma )\lambda _{{\rm{eq}}}^2{{\vec q}_{{\rm{rot}}}} \cdot \vec c}}{\rho }(\lambda _{{\rm{eq}}}^{}{{\vec \xi }^2} - 1),
\end{aligned}
\end{equation}
where the coefficients adopt the values \cite{Xu2015Direct} $\Pr  = 2/3$, $\sigma  = 1/1.55$, ${\omega _0} = 0.2354$ and ${\omega _1} = 0.3049$ for nitrogen in the present work. In the Rykov model Eq.~\ref{eq:rykov}, $\tau$ is the relaxation time for the translational degree of freedom and can be calculated as $\tau  = \mu /p$, where $\mu$ and $p$ are the viscosity and pressure determined by the translational temperature $T_{\rm{trans}}$. ${Z_{{\rm{rot}}}}$ is the rotational relaxation collision number accounting for the ratio of the slower inelastic translation-rotation energy relaxation relative to the elastic translational relaxation. There are many researches and models for the energy relaxation \cite{Bird1994Molecular,Parker1959Rotational,Lordi1970Rotational,Koura1998Statistical,Ivanov1998Computational}.  Parker \cite{Parker1959Rotational} employed the equation
\begin{equation}
{Z_{{\rm{rot}}}} = \frac{{Z_{{\rm{rot}}}^\infty }}{{1 + ({\pi ^{3/2}}/2)\sqrt {{T^*}/{T_{{\rm{trans}}}}} {\rm{ + (}}\pi  + {\pi ^2}{\rm{/4)(}}{T^*}/{T_{{\rm{trans}}}}{\rm{)}}}}
\end{equation}
with $Z_{{\rm{rot}}}^\infty  = 15.7$ and ${T^*} = 80.0\rm{K}$. When the data of Lordi and Mates \cite{Lordi1970Rotational} is fitted to this equation the values are $Z_{{\rm{rot}}}^\infty  = 23.0$ and ${T^*} = 91.5\rm{K}$. The variation of $Z_{{\rm{rot}}}$ with $T_{\rm{trans}}$ is illustrated in Fig.~\ref{fig:method_zrot}. In this paper, for simplicity, $Z_{{\rm{rot}}}$ is assumed as a constant depending on different test cases. As illustrated in Fig.~\ref{fig:method_rykovrelaxation}, the Rykov model describes the relaxation process that the particle velocity distribution $f$ first relaxes to the bi-temperature state ${{g_{{\rm{trans}}}}}$ through the elastic particle collision, and then relaxes to the thermo-equilibrium state ${{g_{{\rm{eq}}}}}$ through the inelastic collision with translation-rotation energy exchange. So far, the basic model of the diatomic gas has been clarified, the implicit numerical method will be constructed in the sections below.

\subsection{General framework}\label{sec:frame}

The Rykov model equation Eq.~\ref{eq:rykov} can be transformed into
\begin{equation}\label{eq:rekov2}
\frac{{\partial f}}{{\partial t}}{\rm{ + }}\vec u \cdot \frac{{\partial f}}{{\partial \vec x}} = \frac{{{g^ * } - f}}{\tau },
\end{equation}
which has a form the same with the BGK model equation \cite{bhatnagar1954model}. Here $g^ *$ is
\begin{equation}
{g^ * } = \frac{{{Z_{{\rm{rot}}}} - 1}}{{{Z_{{\rm{rot}}}}}}{g_{{\rm{trans}}}} + \frac{1}{{{Z_{{\rm{rot}}}}}}{g_{{\rm{eq}}}}.
\end{equation}
The finite volume method is used in physical space, the implicit backward Euler method is used in time, the velocity space is discretized into discrete velocity points, and then the implicit discrete governing equation can be written as
\begin{equation}\label{eq:discmic}
\frac{{{V_i}}}{{\Delta t}}\left( {f_{i,k}^{n + 1} - f_{i,k}^n} \right) + \sum\limits_{j \in N\left( i \right)} {{A_{ij}}{{\vec u}_k} \cdot {{\vec n}_{ij}}f_{ij,k}^{n + 1}}  = {V_i}\frac{{g_{i,k}^{ * ,n + 1} - f_{i,k}^{n + 1}}}{{\tau _i^{n + 1}}},
\end{equation}
where the signs $i,n,k$ correspond to the discretizations in physical space, time and velocity space respectively. $j$ denotes the neighboring cell of cell $i$ and $N\left( i \right)$ is the set of all of the neighbors of $i$. $ij$ denotes the variable at the interface between cell $i$ and $j$. $A_{ij}$ is the interface area, ${\vec n_{ij}}$ is the outward normal unit vector of interface $ij$ relative to cell $i$, and $V_i$ is the volume of cell $i$.

It's not easy to directly solve the implicit discrete equation Eq.~\ref{eq:discmic} because the term $g_{i,k}^{ * ,n + 1}$ is hard to handle. From Eq.~\ref{eq:eq_trans} and Eq.~\ref{eq:eq_eq} we know that the determination of $g_{i,k}^{ * ,n + 1}$ requires the determination of the macroscopic variable vector $\vec W_i^{n + 1}$ which is further related to $f_{i,k}^{n + 1}$ by Eq.~\ref{eq:intf}. In some of the previous implicit methods, such as Yang and Huang's scheme \cite{Yang1995Rarefied}, Mao et al.'s scheme \cite{Mao2015STUDY}, $g^{n+1}$ is approximated by $g^n$, which will slow down the convergence in continuum flow regime \cite{Mieussens2000DISCRETE}. In Mieussens's scheme \cite{Mieussens2000DISCRETE}, $g^{n+1}$ is calculated through a linear mapping between $f$ and $g$, which involves a huge matrix with large dimensions and increases the complexity of the method. Given the above, the present method adopts the idea of macroscopic variable prediction proposed by Zhu et al.~\cite{Zhu2016Implicit}, which is also applied in the method of Yang et al.~\cite{Yang2018An}. A predicted macroscopic variable $\tilde {\vec W}_i^{n + 1}$ is used to calculate $\tilde g_{i,k}^{*,n + 1}$. This predicted $\tilde {\vec W}_i^{n + 1}$ is obtained from the macroscopic governing equation to ensure the fast convergence of the scheme in continuum flow regime.

Take the moment of Eq.~\ref{eq:rekov2} for $\vec \psi $ and we can derive the macroscopic governing equation for a control volume $\Omega$
\begin{equation}\label{eq:macgovern}
\frac{\partial }{{\partial t}}\int\limits_\Omega  {\vec WdV} {\rm{ + }}\oint\limits_{\partial \Omega } {\vec FdA}  = \int\limits_\Omega  {\vec SdV}.
\end{equation}
The source term $\vec S$ is expressed as
\begin{equation}\label{eq:source}
\vec S = \int {\vec \psi \frac{{{g^ * } - f}}{\tau }d\Xi }  = {(0,\vec 0,0,\frac{{\rho {E_{{\rm{rot,eq}}}} - \rho {E_{{\rm{rot}}}}}}{{{Z_{{\rm{rot}}}}\tau }})^T},
\end{equation}
where $\rho {E_{{\rm{rot}},{\rm{eq}}}} = \int {\frac{1}{2}{\xi ^2}{g_{{\rm{eq}}}}d\Xi } $ is the rotational energy density at the thermo-equilibrium state ${g_{{\rm{eq}}}}$. The macroscopic governing equation Eq.~\ref{eq:macgovern} can be implicitly discretized as
\begin{equation}\label{eq:discmac}
\frac{{{V_i}}}{{\Delta t}}\left( {\vec W_i^{n + 1} - \vec W_i^n} \right) + \sum\limits_{j \in N(i)} {{A_{ij}}\vec F_{ij}^{n + 1}}  = {V_i}\vec S_i^{n + 1}.
\end{equation}
Replace $\vec W_i^{n + 1}$ with the predicted $\tilde {\vec W}_i^{n + 1}$, and rearrange Eq.~\ref{eq:discmac} into the incremental form
\begin{equation}\label{eq:discmac_inc}
\frac{{{V_i}}}{{\Delta t}}\Delta \tilde {\vec W}_i^{n + 1} + \sum\limits_{j \in N(i)} {{A_{ij}}\Delta \tilde {\vec F}_{ij}^{n + 1}}  = {V_i}\tilde {\vec S}_i^{n + 1} - \sum\limits_{j \in N(i)} {{A_{ij}}\vec F_{ij}^n},
\end{equation}
where the symbol $\sim$ denotes the predicted variables for the next time level. The flux $\vec F_{ij}^n$ is calculated from the distribution function $f_{ij,k}^n$ at the interface by numerical integrals in the velocity space
\begin{equation}\label{eq:discinc_flux}
\vec F_{ij}^n = \sum {{{\vec \psi }_k}{{\vec u}_k} \cdot {{\vec n}_{ij}}f_{ij,k}^n\Delta \Xi_k },
\end{equation}
where the construction of $f_{ij,k}^n$ will be detailed in Section \ref{sec:multiscaleflux}. The variation of the flux ${\Delta \tilde {\vec F}_{ij}^{n + 1}}$ is approximated by
\begin{equation}\label{eq:deltamacflux}
\Delta \tilde {\vec F}_{ij}^{n + 1} = \tilde {\vec R}_{ij}^{n + 1} - {\vec R}_{ij}^n,
\end{equation}
where $\vec R_{ij}$ has the form of the well-known Roe's flux function
\begin{equation}\label{eq:roeflux}
{\vec R_{ij}} = \frac{1}{2}\left( {{{\vec G}_{ij}}({{\vec W}_i}) + {{\vec G}_{ij}}({{\vec W}_j}) + {r_{ij}}{{\vec W}_i} - {r_{ij}}{{\vec W}_j}} \right).
\end{equation}
Here ${\vec G_{ij}}(\vec W)$ is the Euler flux
\begin{equation}
{\vec G_{ij}}(\vec W) = \left( \begin{array}{c}
\rho \vec U \cdot {{\vec n}_{ij}}\\
\rho {U_x}\vec U \cdot {{\vec n}_{ij}} + {n_{ij,x}}p\\
\rho {U_y}\vec U \cdot {{\vec n}_{ij}} + {n_{ij,y}}p\\
\rho {U_z}\vec U \cdot {{\vec n}_{ij}} + {n_{ij,z}}p\\
(\rho E + p)\vec U \cdot {{\vec n}_{ij}}\\
\rho {E_{{\rm{rot}}}}\vec U \cdot {{\vec n}_{ij}}
\end{array} \right),
\end{equation}
and $r_{ij}$ is
\begin{equation}
{r_{ij}} = \left| {{{\vec U}_{ij}} \cdot {{\vec n}_{ij}}} \right| + {a_{ij}} + 2\frac{{{\mu _{ij}}}}{{{\rho _{ij}}\Delta {x_{ij}}}},
\end{equation}
where $a_{ij}$ is the acoustic speed at the interface and $\Delta {x_{ij}}$ is the distance between cell center $i$ and $j$. The source term $\tilde {\vec S}_i^{n + 1}$ is handled as
\begin{equation}\label{eq:discsource}
\tilde {\vec S}_i^{n + 1} = {(0,\vec 0,0,\frac{{\widetilde {\rho E}_{{\rm{rot,eq,}}i}^{n + 1} - \rho E_{{\rm{rot,}}i}^n - \Delta \widetilde {\rho E}_{{\rm{rot,}}i}^{n + 1}}}{{{Z_{{\rm{rot}}}}\tau _i^n}})^T}.
\end{equation}
Note that for the conserved variables $\rho $, $\rho \vec U$ and $\rho E$, the source terms are zero. Substitute Eq.~\ref{eq:deltamacflux}, Eq.~\ref{eq:roeflux} and Eq.~\ref{eq:discsource} into Eq.~\ref{eq:discmac_inc}, and note that $\sum\limits_{j \in N(i)} {{A_{ij}}{G_{ij}}({{\vec W}_i})}  = \vec 0$ holds, then we can get the expression
\begin{equation}\label{eq:refreshmac_conserve}
\begin{aligned}
\left( {\frac{{{V_i}}}{{\Delta t}} + \frac{1}{2}\sum\limits_{j \in N(i)} {{r_{ij}}{A_{ij}}} } \right)\Delta \tilde Q_i^{n + 1} =  &  - \sum\limits_{j \in N(i)} {{A_{ij}}F_{ij,Q}^n}  + \frac{1}{2}\sum\limits_{j \in N(i)} {{r_{ij}}{A_{ij}}\Delta \tilde Q_j^{n + 1}} \\&
 - \frac{1}{2}\sum\limits_{j \in N(i)} {{A_{ij}}\left( {G_{ij,Q}^{}(\tilde{\vec W}_j^{n + 1}) - G_{ij,Q}^{}(\vec W_j^n)} \right)}
\end{aligned},
\end{equation}
where $Q$ denotes a certain conserved variable $\rho $, $\rho \vec U$ or $\rho E$. For $\rho E_{\rm{rot}}$, it has
\begin{equation}\label{eq:refreshmac_rot}
\begin{aligned}
&\left( {\frac{{{V_i}}}{{\Delta t}} + \frac{1}{2}\sum\limits_{j \in N(i)} {{r_{ij}}{A_{ij}}}  + \frac{{{V_i}}}{{{Z_{{\rm{rot}}}}\tau _i^n}}} \right)\Delta \tilde Q_i^{n + 1}\\
 =  &  - \sum\limits_{j \in N(i)} {{A_{ij}}F_{ij,Q}^n}  + \frac{1}{2}\sum\limits_{j \in N(i)} {{r_{ij}}{A_{ij}}\Delta \tilde Q_j^{n + 1}} \\&
 - \frac{1}{2}\sum\limits_{j \in N(i)} {{A_{ij}}\left( {G_{ij,Q}^{}(\tilde{\vec W}_j^{n + 1}) - G_{ij,Q}^{}(\vec W_j^n)} \right)}  + \frac{{{V_i}}}{{{Z_{{\rm{rot}}}}\tau _i^n}}\left( {\widetilde {\rho E}_{{\rm{rot,eq,}}i}^{n + 1} - Q_i^n} \right)
\end{aligned},
\end{equation}
where $Q$ denotes the rotational energy density $\rho E_{\rm{rot}}$. Eq.~\ref{eq:refreshmac_conserve} and Eq.~\ref{eq:refreshmac_rot} are solved by the SGS ( Symmetric Gauss-Seidel) method, or also known as the PRSGS (Point Relaxation Symmetric Gauss-Seidel) method \cite{Rogers1995Comparison,Yuan2002Comparison}. In each time of SGS iteration, a forward sweep from the first to the last cell and a backward sweep from the last to the first cell are implemented, during which the data of a cell is always updated by the latest data of its adjacent cells through Eq.~\ref{eq:refreshmac_conserve} and Eq.~\ref{eq:refreshmac_rot}. Such a SGS iteration procedure is totally matrix-free and easy to implement. In our work, 60 times' SGS iterations are done during one time level to get the predicted $\tilde {\vec W}_i^{n + 1}$.

Since we have get the predicted macroscopic variable vector $\tilde {\vec W}_i^{n + 1}$, it's time to deal with the microscopic implicit discrete equation Eq.~\ref{eq:discmic} for ${f_{i,k}^{n + 1}}$. Similarly, rearrange Eq.~\ref{eq:discmic} into the incremental form
\begin{equation}\label{eq:discmic_inc}
\begin{aligned}
&\left( {\frac{{{V_i}}}{{\Delta t}} + \frac{{{V_i}}}{{\tilde \tau _i^{n + 1}}}} \right)\Delta f_{i,k}^{n + 1} + \sum\limits_{j \in N(i)} {{A_{ij}}{{\vec u}_k} \cdot {{\vec n}_{ij}}\Delta f_{ij,k}^{n + 1}} \\
 = & {V_i}\frac{{\tilde g_{i,k}^{ * ,n + 1} - f_{i,k}^n}}{{\tilde \tau _i^{n + 1}}} - \sum\limits_{j \in N(i)} {{A_{ij}}{{\vec u}_k} \cdot {{\vec n}_{ij}}f_{ij,k}^n}
\end{aligned},
\end{equation}
where $\tilde g_{i,k}^{ * ,n + 1}$ and $\tilde \tau _i^{n + 1}$ is determined by the predicted $\tilde {\vec W}_i^{n + 1}$. $f_{ij,k}^n$ will be detailed in Section \ref{sec:multiscaleflux}. ${\Delta f_{ij,k}^{n + 1}}$ is simply handled by the upwind scheme and Eq.~\ref{eq:discmic_inc} is turned into
\begin{equation}\label{eq:refreshmic}
\begin{aligned}
&\left( {\frac{{{V_i}}}{{\Delta t}} + \frac{{{V_i}}}{{\tilde \tau _i^{n + 1}}} + \sum\limits_{j \in N_k^ + (i)} {{A_{ij}}{{\vec u}_k} \cdot {{\vec n}_{ij}}} } \right)\Delta f_{i,k}^{n + 1}\\
 = &{V_i}\frac{{\tilde g_{i,k}^{ * ,n + 1} - f_{i,k}^n}}{{\tilde \tau _i^{n + 1}}} - \sum\limits_{j \in N(i)} {{A_{ij}}{{\vec u}_k} \cdot {{\vec n}_{ij}}f_{ij,k}^n}  - \sum\limits_{j \in N_k^ - (i)} {{A_{ij}}{{\vec u}_k} \cdot {{\vec n}_{ij}}\Delta f_{j,k}^{n + 1}}
\end{aligned},
\end{equation}
where $ N_k^ + (i)$ is the set of $i$'s neighboring cells satisfying ${\vec u_k} \cdot {\vec n_{ij}} \ge 0$ while for $ N_k^ - (i)$ it satisfies ${\vec u_k} \cdot {\vec n_{ij}} < 0$. Eq.~\ref{eq:refreshmic} is solved by the SGS method to obtain ${f_{i,k}^{n + 1}}$ and 2 times' SGS iterations are done per time level.

Here, suppose $f_{i,k}^{n}$ and ${\vec W}_{i}^{n}$ are known, the calculation procedure from time level $n$ to $n+1$ is listed as follows:
\begin{description}
    \item[Step 1.] Reconstruct variables in the cell and calculate $f_{ij,k}^n$ at the interface (detailed in Section \ref{sec:multiscaleflux}).
    \item[Step 2.] Do numerical integrals of $f_{ij,k}^n$ in the velocity space and calculate the terms at the $n$th time level on the right of Eq.~\ref{eq:refreshmac_conserve} and Eq.~\ref{eq:refreshmac_rot}.
    \item[Step 3.] Solve Eq.~\ref{eq:refreshmac_conserve} and Eq.~\ref{eq:refreshmac_rot} by SGS iterations to get the predicted $\tilde {\vec W}_i^{n + 1}$.
    \item[Step 4.] Calculate ${\tilde g_{i,k}^{ * ,n + 1}}$ and $\tilde \tau _i^{n + 1}$ and then solve Eq.~\ref{eq:refreshmic} by SGS iterations to get ${f_{i,k}^{n + 1}}$ at the next time level.
    \item[Step 5.] Do numerical integrals of ${f_{i,k}^{n + 1}}$ in the velocity space to get ${\vec W}_{i}^{n+1}$ at the next time level (see Section \ref{sec:compensation} for more details).
\end{description}

\subsection{Multiscale numerical flux}\label{sec:multiscaleflux}

\begin{figure}
\centering
\includegraphics[width=0.4\textwidth]{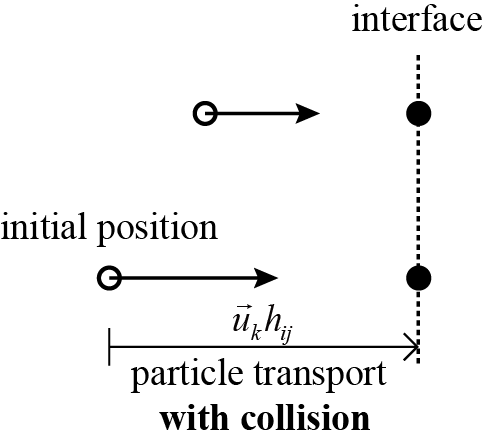}
\caption{\label{fig:method_multiscaleflux}Illustration of particles inside a cell transfer to the interface after the physical time step $h_{ij}$.}
\end{figure}

The microscopic interface flux ${{{\vec u}_k} \cdot {{\vec n}_{ij}}f_{ij,k}^n}$ and the macroscopic flux $\vec F_{ij}^n$ are both determined by the interface distribution function $f_{ij,k}^n$. The construction of $f_{ij,k}^n$ is very important and it is about whether the scheme is multiscale and applicable to all flow regimes. If the distribution function $f_{ij,k}^n$ is straightly got via the reconstruction of the initial data in the cell at the time level $n$, the scheme will meet problems in the continuum flow regime (more precisely, in the case of low cell Kn number) and yield a more dissipating result (see the conventional DVM results in Section \ref{sec:test1}). The mechanism is just  illustrated in Fig.~\ref{fig:method_multiscaleflux}. First of all, the initial distribution function data is always stored inside the cell. Suppose particles with velocity $\vec u_k$ inside the cell will transfer to the interface after a time $h_{ij}$. If $h_{ij}$ is much larger than the mean collision interval of particle, or the particle trajectory is much larger than the mean free path, particles will suffer sufficient collision before they arrive the interface. Then the distribution function at the interface will be very close to the equilibrium state and almost \emph{uncorrelated} to the initial distribution function data inside the cell. Thus, directly reconstruct $f_{ij,k}^n$ from the initial data $f_{i,k}^n$ will introduce something like \emph{information pollution} into the scheme when the cell scale is much larger than the mean free path. In UGKS, which is presented by Xu and Huang \cite{Xu2010A}, this problem is solved by using the analytical solution of the model equation to describe the evolution of the interface distribution function during the time step. In DUGKS, which is presented by Guo et al.~\cite{guo2013discrete,guo2015discrete}, a discrete temporal difference scheme of the model equation in the Lagrangian description is used at the interface to get the distribution function. Here, the idea of DUGKS is adopted. The initial distribution function is stored inside the cell and we evolve the initial data to the interface with a physical time step $h_{ij}$ through a temporal difference scheme of the Rykov equation Eq.~\ref{eq:rekov2}. Consider that this difference scheme is only used to get the proper instantaneous interface distribution function $f_{ij,k}^n$ at the corresponding cell scale, the temporal accuracy of this difference scheme is not important. Moreover, given that the collision term of the model equation will be very stiff in the continuum flow regime, the backward Euler method is used for the temporal difference. After an evolution time step $h_{ij}$, the interface distribution function $f_{ij,k}^n$ is obtained as (suppose the initial time $t^n=0$)
\begin{equation}\label{eq:interfacef0}
f_{ij,k}^n = f({\vec x_{ij}},h_{ij},{\vec u_k}) = f({\vec x_{ij}} - \vec u_kh_{ij},0,{\vec u_k}) + h_{ij}\frac{{{g^ * }({{\vec x}_{ij}},h_{ij},{{\vec u}_k}) - f({{\vec x}_{ij}},h_{ij},{{\vec u}_k})}}{{\tau _{ij}^n}},
\end{equation}
where
\begin{equation}
f({\vec x_{ij}} - {\vec u_k}h_{ij},0,{\vec u_k}) = \left\{ \begin{array}{l}
f_{i,k}^n + ({{\vec x}_{ij}} - {{\vec x}_i} - {{\vec u}_k}h_{ij})\nabla f_{i,k}^n\,\,\,\,,\;\;\;{{\vec u}_k} \cdot {{\vec n}_{ij}} \ge 0\\
f_{j,k}^n + ({{\vec x}_{ij}} - {{\vec x}_j} - {{\vec u}_k}h_{ij})\nabla f_{j,k}^n\,\,,\;\;\;{{\vec u}_k} \cdot {{\vec n}_{ij}} < 0
\end{array} \right. .
\end{equation}
This construction is similar to the method of Yang et al.~\cite{Yang2018An}. $\nabla f_{i,k}^n$ and $\nabla f_{j,k}^n$ can be obtained through the reconstruction of the initial distribution function data. As previously mentioned, the temporal accuracy is not important for Eq.~\ref{eq:interfacef0} and ${{g^ * }({{\vec x}_{ij}},h_{ij},{{\vec u}_k})}$ can be approximated by ${{g^ * }({{\vec x}_{ij}},0,{{\vec u}_k})}$, then Eq.~\ref{eq:interfacef0} can be arranged as
\begin{equation}\label{eq:interfacef1}
f_{ij,k}^n = \frac{{\tau _{ij}^n}}{{\tau _{ij}^n + h_{ij}}}f\left( {{{\vec x}_{ij}} - {{\vec u}_k}h_{ij},0,{{\vec u}_k}} \right) + \frac{h_{ij}}{{\tau _{ij}^n + h_{ij}}}g^ *\left( {{{\vec x}_{ij}},0,{{\vec u}_k}} \right).
\end{equation}
Here, ${{g^ * }({{\vec x}_{ij}},0,{{\vec u}_k})}$ and $\tau _{ij}^n$ are calculated by the same way as the method of GKS \cite{xu2001gas}. For ${{g^ * }({{\vec x}_{ij}},0,{{\vec u}_k})}$, it should be determined by the interface macroscopic variables $\vec W_{ij}^n$, which can be constructed as
\begin{equation}
\vec W_{ij}^n = \int_{\vec u \cdot {{\vec n}_{ij}} \ge 0} {\vec \psi g_{\rm{trans}}^ld\Xi  + } \int_{\vec u \cdot {{\vec n}_{ij}} < 0} {\vec \psi g_{\rm{trans}}^rd\Xi },
\end{equation}
where $g_{\rm{trans}}^l$ and $g_{\rm{trans}}^r$ are obtained through the reconstruction of the initial macroscopic variables. For $\tau _{ij}^n$, it is calculated as
\begin{equation}
\tau _{ij}^n = \frac{{\mu (\vec W_{ij}^n)}}{{p(\vec W_{ij}^n)}} + \frac{{\left| {{{{\rho ^l}} \mathord{\left/
 {\vphantom {{{\rho ^l}} {\lambda _{{\rm{trans}}}^l}}} \right.
 \kern-\nulldelimiterspace} {\lambda _{{\rm{trans}}}^l}} - {{{\rho ^r}} \mathord{\left/
 {\vphantom {{{\rho ^r}} {\lambda _{{\rm{trans}}}^r}}} \right.
 \kern-\nulldelimiterspace} {\lambda _{{\rm{trans}}}^r}}} \right|}}{{\left| {{{{\rho ^l}} \mathord{\left/
 {\vphantom {{{\rho ^l}} {\lambda _{{\rm{trans}}}^l}}} \right.
 \kern-\nulldelimiterspace} {\lambda _{{\rm{trans}}}^l}} + {{{\rho ^r}} \mathord{\left/
 {\vphantom {{{\rho ^r}} {\lambda _{{\rm{trans}}}^r}}} \right.
 \kern-\nulldelimiterspace} {\lambda _{{\rm{trans}}}^r}}} \right|}}h_{ij},
\end{equation}
where ${{\rho ^l}}, {\lambda _{{\rm{trans}}}^l}, {{\rho ^r}}, {\lambda _{{\rm{trans}}}^r}$ are all obtained from the reconstruction and the second part on the right can be regarded as the artificial viscosity. More details about the construction of ${{g^ * }({{\vec x}_{ij}},0,{{\vec u}_k})}$ and $\tau _{ij}^n$ please refer to Ref.~\cite{xu2001gas}.

Last but not least, the determination of the physical time step $h_{ij}$ is also important. As mentioned above, the physical time step $h_{ij}$ is applied to evolve the initial particle data (namely, the distribution function) inside the cell to the surface, so $h_{ij}$ should be constrained by the CFL condition. Meanwhile, $h_{ij}$ should match the cell scale. If $h_{ij}$ is too small relative to the cell scale, $f_{ij,k}^n$ will degenerate to the direct reconstruction of the initial data and lose the multiscale property. Thus, $h_{ij}$ should be determined by the local CFL condition. The physical local time step $h_i$ for the cell $i$ can be expressed as
\begin{equation}
{h_i} = \frac{{{V_i}}}{{\mathop {\max }\limits_k (\left| {{u_{k,x}}{A_{i,x}}} \right| + \left| {{u_{k,y}}{A_{i,y}}} \right| + \left| {{u_{k,z}}{A_{i,z}}} \right|)}}{\rm{CFL}},
\end{equation}
where $A_{i,x},A_{i,y},A_{i,z}$ are projection areas of cell $i$ in $x,y,z$ directions. Then the physical local time step $h_{ij}$ for the interface $ij$ is
\begin{equation}
{h_{ij}} = \min ({h_i},{h_j}).
\end{equation}
The physical local time step is very important in the application of the multiscale scheme. This is discussed further in our numerical test in Section \ref{sec:test1}.

\subsection{Integral error compensation}\label{sec:compensation}

In the calculation procedure from time level $n$ to $n+1$ (detailed in Section \ref{sec:frame}), at \textbf{Step 5}, integrals of ${f_{i,k}^{n + 1}}$ will be done in the velocity space to obtain ${\vec W}_{i}^{n+1}$. If simply do the numerical quadrature of ${f_{i,k}^{n + 1}}$ to get ${\vec W}_{i}^{n+1}$, one may get into trouble due to the integral error.

When the microscopic discrete governing equation Eq.~\ref{eq:discmic} converges, it should come to the following fixed point
\begin{equation}
{V_i}\frac{{g_{i,k}^* - {f_{i,k}}}}{{{\tau _i}}} - \sum\limits_{j \in N\left( i \right)} {{A_{ij}}{{\vec u}_k} \cdot {{\vec n}_{ij}}{f_{ij,k}}}  = 0,
\end{equation}
and the distribution function at the cell center $f_{i,k}$ will converge to
\begin{equation}\label{eq:micfixedpoint}
{f_{i,k}} = g_{i,k}^* + \frac{{{\tau _i}}}{{{V_i}}}\sum\limits_{j \in N\left( i \right)} {{A_{ij}}{{\vec u}_k} \cdot {{\vec n}_{ij}}{f_{ij,k}}}.
\end{equation}
Here, let $\left[  \ldots  \right]$ and $\left\langle  \ldots  \right\rangle $ denote the numerical integration and analytic integration in the velocity space respectively, i.e.
\begin{equation}
[...] = \sum {...\Delta {\Xi _k}} \quad \quad \quad \quad  < ... >  = \int {...d\Xi }.
\end{equation}
Then, if the macroscopic variables are calculated simply through the numerical quadrature of ${f_{i,k}}$, it will be
\begin{equation}\label{eq:discinc_mac}
{\vec W'_i} = [{\vec \psi _k}{f_{i,k}}] = [{\vec \psi _k}g_k^*({\vec W'_i})] + \frac{{{\tau _i}}}{{{V_i}}}\sum\limits_{j \in N\left( i \right)} {{A_{ij}}[{{\vec \psi }_k}{{\vec u}_k}\cdot{{\vec n}_{ij}}{f_{ij,k}}} ],
\end{equation}
where ${\vec W'_i}$ is the numerically-integrated macroscopic vector. Substitute the macroscopic flux Eq.~\ref{eq:discinc_flux} into Eq.~\ref{eq:discinc_mac}, we will get
\begin{equation}\label{eq:macfixedpoint0}
\frac{1}{{{V_i}}}\sum\limits_{j \in N\left( i \right)} {{A_{ij}}{{\vec F}_{ij}}}  = \frac{{{{\vec W'}_i} - [{{\vec \psi }_k}g_k^*({{\vec W'}_i})]}}{{{\tau _i}}}.
\end{equation}
Due to the integral error, there is
\begin{equation}
{\vec \delta _i} =  < \vec \psi g_{}^*({\vec W'_i}) >  - [{\vec \psi _k}g_k^*({\vec W'_i})],
\end{equation}
and substitute this error $\vec \delta _i$ into Eq.~\ref{eq:macfixedpoint0}, it turns into
\begin{equation}\label{eq:macfixedpoint1}
\frac{1}{{{V_i}}}\sum\limits_{j \in N\left( i \right)} {{A_{ij}}{{\vec F}_{ij}}}  = \frac{{{{\vec W}_i}^\prime  -  < \vec \psi g_{}^*({{\vec W}_i}^\prime ) > }}{{{\tau _i}}} + \frac{{{{\vec \delta }_i}}}{{{\tau _i}}}.
\end{equation}
It can be seen that, in Eq.~\ref{eq:macfixedpoint1}, ${{({{\vec W'}_i} -  < \vec \psi g_{}^*({{\vec W}_i}^\prime ) > )} \mathord{\left/
 {\vphantom {{({{\vec W}_i} -  < \vec \psi g_i^* > )} {{\tau _i}}}} \right.
 \kern-\nulldelimiterspace} {{\tau _i}}}$ is the exact source term while ${{{{\vec \delta }_i}} \mathord{\left/
 {\vphantom {{{{\vec \delta }_i}} {{\tau _i}}}} \right.
 \kern-\nulldelimiterspace} {{\tau _i}}}$ is the error term. This error term is just like an additional source term and will add mass, momentum, energy into the scheme persistently, which makes the scheme nonconservative and hard to converge, especially when $\tau_i$ is very small. The solution is to compensate ${\vec W'}_{i}$ for the integral error $\vec \delta _i$, i.e.
 \begin{equation}\label{eq:discinc_mac_compensate}
{\vec W_i} = [{\vec \psi _k}{f_{i,k}}] +  < \vec \psi g_{}^*({\vec W_i}) >  - [{\vec \psi _k}g_k^*({\vec W_i})],
 \end{equation}
 where ${\vec W_i}$ is the compensated macroscopic vector. Substitute Eq.~\ref{eq:discinc_flux} and Eq.~\ref{eq:micfixedpoint} into Eq.~\ref{eq:discinc_mac_compensate} will yield
\begin{equation}
\frac{1}{{{V_i}}}\sum\limits_{j \in N\left( i \right)} {{A_{ij}}{{\vec F}_{ij}}}  = \frac{{{{\vec W}_i} -  < \vec \psi g_{}^*({{\vec W}_i}) > }}{{{\tau _i}}},
\end{equation}
which just has an exact source term on the right side and the scheme can converge properly. Thus, at \textbf{Step 5} of the calculation procedure from time level $n$ to $n+1$ (detailed in Section \ref{sec:frame}), ${\vec W}_{i}^{n+1}$ will be calculated as
\begin{equation}
\vec W_i^{n + 1} = [{\vec \psi _k}f_{i,k}^{n + 1}] +  < \vec \psi \tilde g_i^{*,n + 1} >  - [{\vec \psi _k}\tilde g_{i,k}^{*,n + 1}].
\end{equation}

Similarly, when calculating the stress or the heat flux, such a compensation should be taken into account. Take the calculation of the translational heat flux ${\vec q_{{\rm{trans}},i}}$ as an example, if directly do the numerical quadrature of ${f_{i,k}}$ to calculate ${{\vec q}'_{{\rm{trans}},i}}$, suppose Eq.~\ref{eq:micfixedpoint} holds, then it will be
\begin{equation}\label{eq:qfluxnumericalint}
{{\vec q}'_{{\rm{trans}},i}} = [\frac{1}{2}{\vec c_k}\vec c_k^2{f_{i,k}}] = [\frac{1}{2}{\vec c_k}\vec c_k^2g_{i,k}^*] + \frac{{{\tau _i}}}{{{V_i}}}\sum\limits_{j \in N\left( i \right)} {{A_{ij}}[\frac{1}{2}{{\vec c}_k}\vec c_k^2{{\vec u}_k} \cdot {{\vec n}_{ij}}{f_{ij,k}}} ].
\end{equation}
It can be seen that in Eq.~\ref{eq:qfluxnumericalint}, if ${\tau _i}$ is very small, the integral error of the term $[\frac{1}{2}{\vec c_k}\vec c_k^2g_{i,k}^*]$ will cover up the real heat flux (this is observed in our test case in Section \ref{sec:test1}). So ${\vec q_{{\rm{trans}},i}}$ should be calculated taking into account the integral error as
\begin{equation}
{\vec q_{{\rm{trans}},i}} = [\frac{1}{2}{\vec c_k}\vec c_k^2{f_{i,k}}] +  < \frac{1}{2}\vec c{\vec c^2}g_i^* >  - [\frac{1}{2}{\vec c_k}\vec c_k^2g_{i,k}^*].
\end{equation}

The application of the above compensation reduces the accuracy requirement of the numerical quadrature method in the velocity space. Without the compensation, one may need to spend effort on the high-precision quadrature in the velocity space, which may be very laborious when handling practical engineering problems, especially problems with large temperature difference where the resolution of the discretization for velocity space in the low temperature region should be very high. With the above compensation technique, the discretization for velocity space can be more flexible and easy. In our numerical tests below, the unstructured discretization is applied in the velocity space along with the above compensation technique, which makes the numerical simulation both efficient and accurate.

\section{Numerical results and discussions}\label{sec:numericaltest}

In this section, test cases are carried out to verify the present method. First, the lid-driven cavity flow is simulated to assess the efficiency and the accuracy of the method for different flow regimes. Then the shock structures at different Mach numbers are calculated to validate the present method in the highly nonequilibrium flow. Finally the test case of hypersonic flow passing a flat plate is performed to further verify the present method for nonequilibrium flow simulation. In all test cases the working gas is nitrogen, with three translational and two rotational degrees of freedom for the molecule.

\subsection{Lid-driven cavity flow}\label{sec:test1}

The test case of lid-driven cavity flow is very suitable to test if the method can accurately simulate the viscosity effect of the flow. Here, the cavity flows at different flow regimes are simulated. In all of the simulations, the Mach number, which is defined by the upper wall velocity $U_{\rm{wall}}$ and the acoustic velocity, is 0.16. The VHS molecular model with $\omega=0.74$ is applied to approximate the nitrogen \cite{Bird1994Molecular}. Consider the wall temperature $T_{\rm{wall}}=273K$, the rotational relaxation collision number $Z_{\rm{rot}}$ is set as 3.5. The diffuse reflection boundary condition with full thermal accommodation \cite{Li2005Application} is implemented on the wall of the cavity.

\begin{figure}
\centering
\subfigure[\label{fig:test1_mesh1000}]{\includegraphics[width=0.47\textwidth]{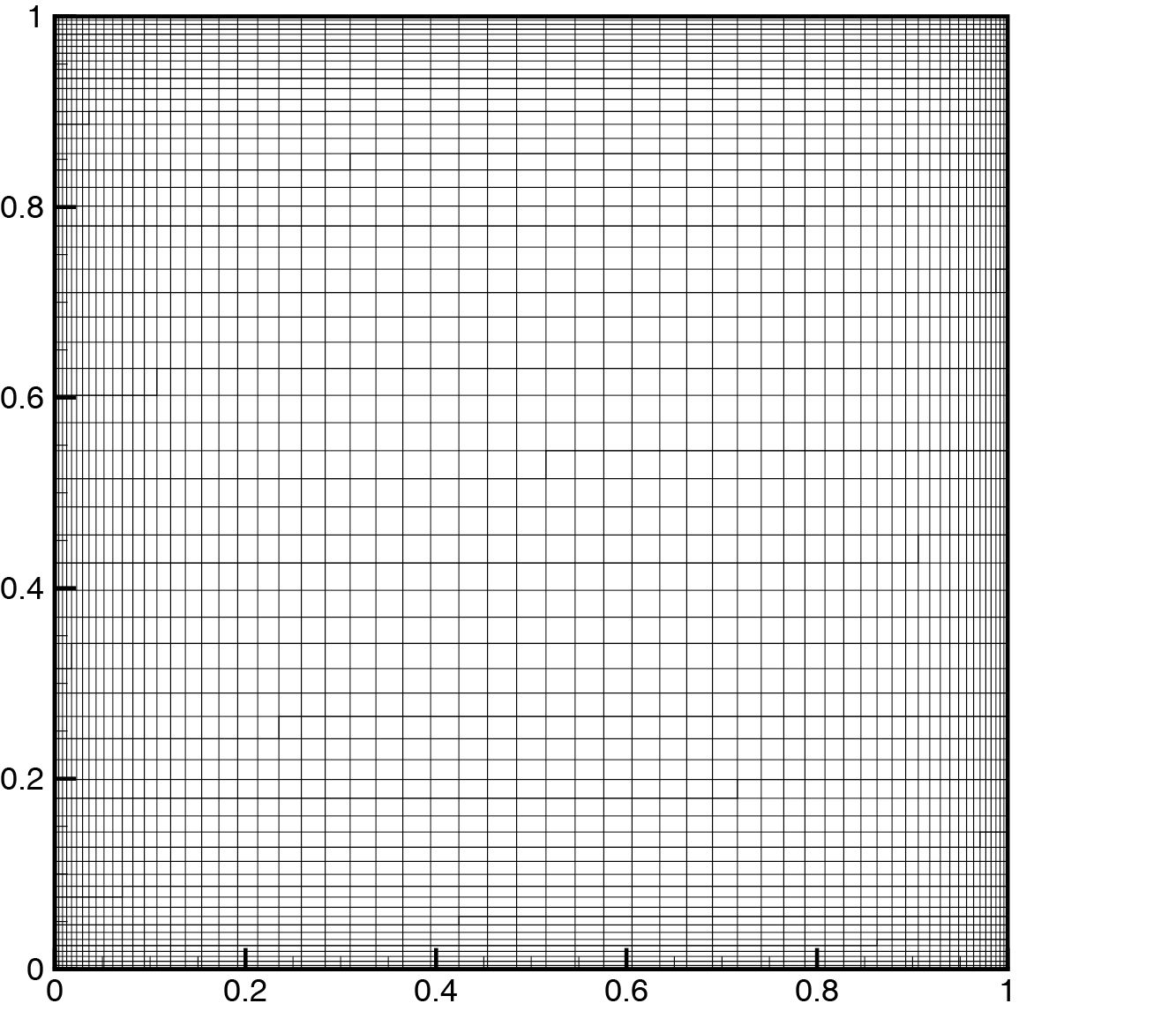}}\hspace{0.02\textwidth}%
\subfigure[]{\includegraphics[width=0.47\textwidth]{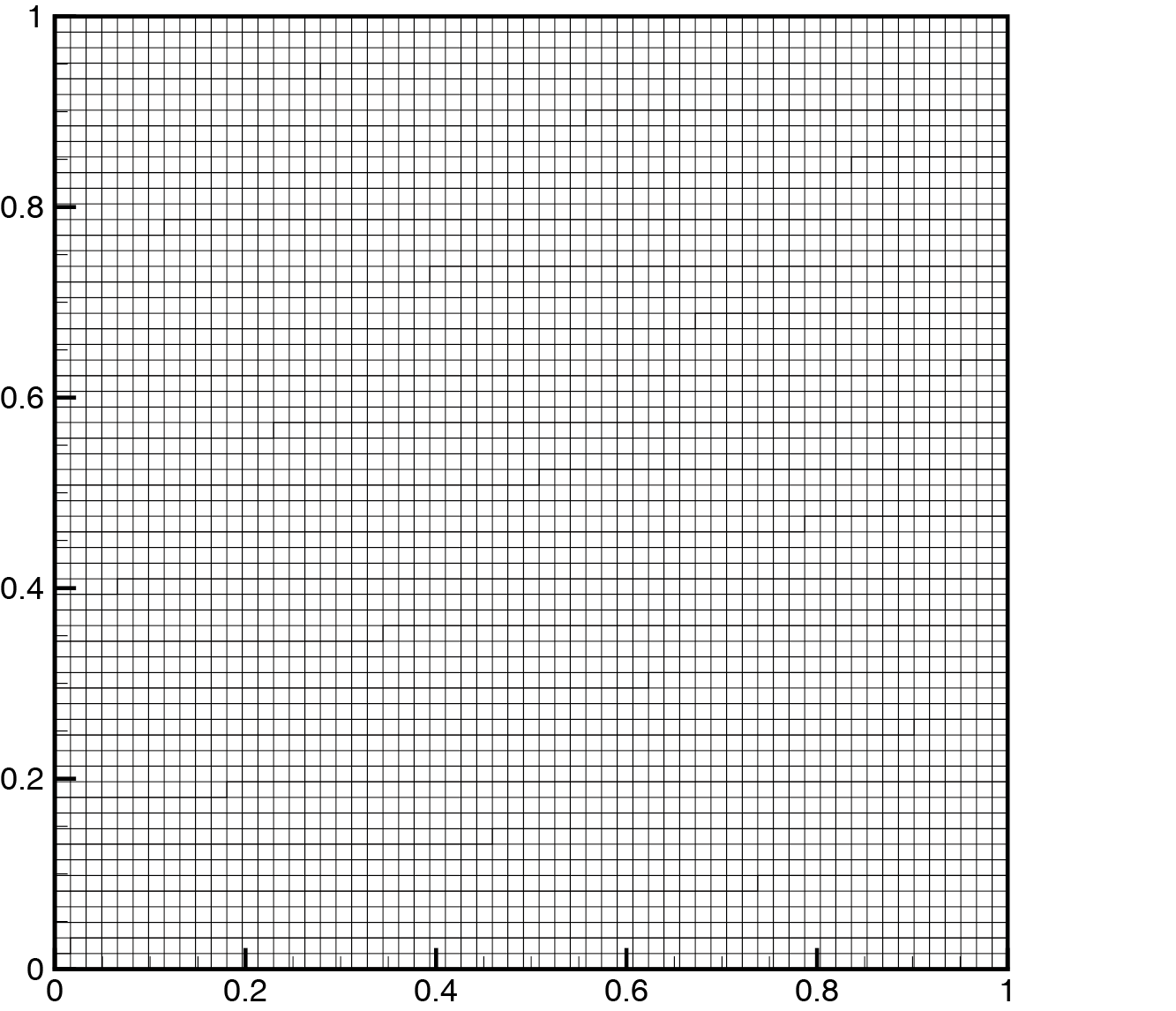}}
\caption{\label{fig:test1_mesh}(a) The nonuniform $61\times61$ mesh used for the cavity flow at Re=1000 and (b) the uniform $61\times61$ mesh used for Kn=0.075, 1, 10.}
\end{figure}

\begin{figure}
\centering
\subfigure[]{\includegraphics[width=0.47\textwidth]{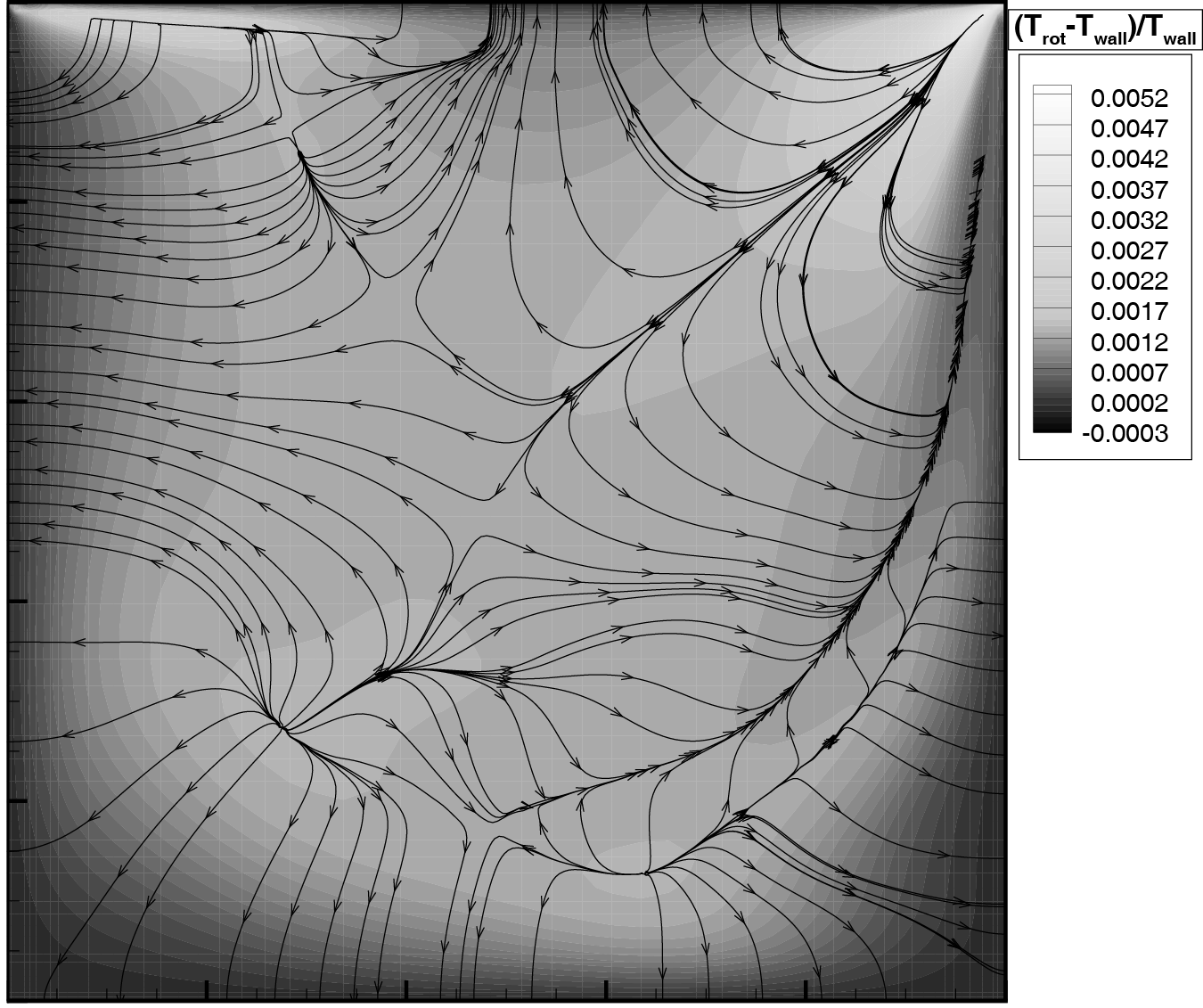}}\hspace{0.02\textwidth}%
\subfigure[]{\includegraphics[width=0.47\textwidth]{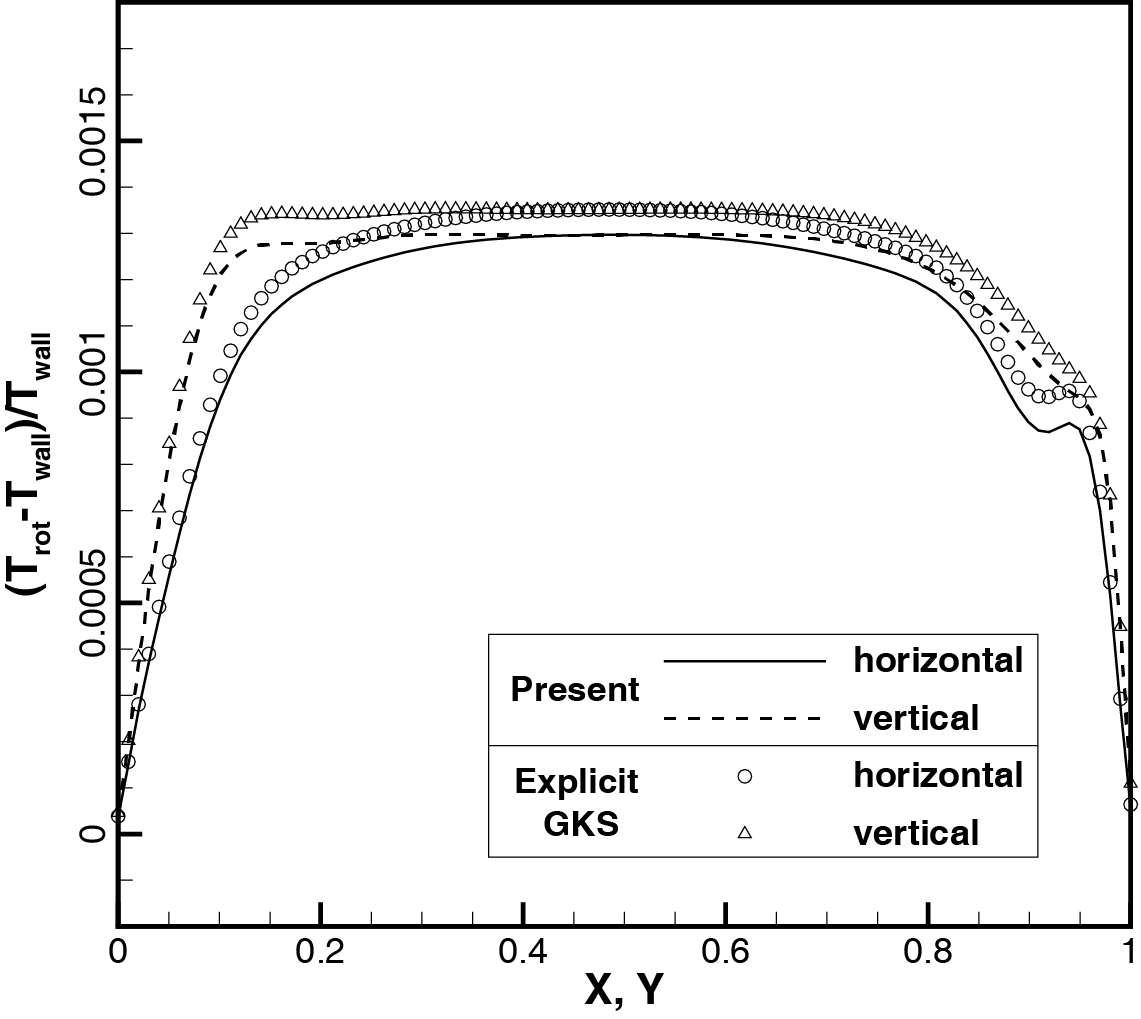}}\\
\subfigure[]{\includegraphics[width=0.47\textwidth]{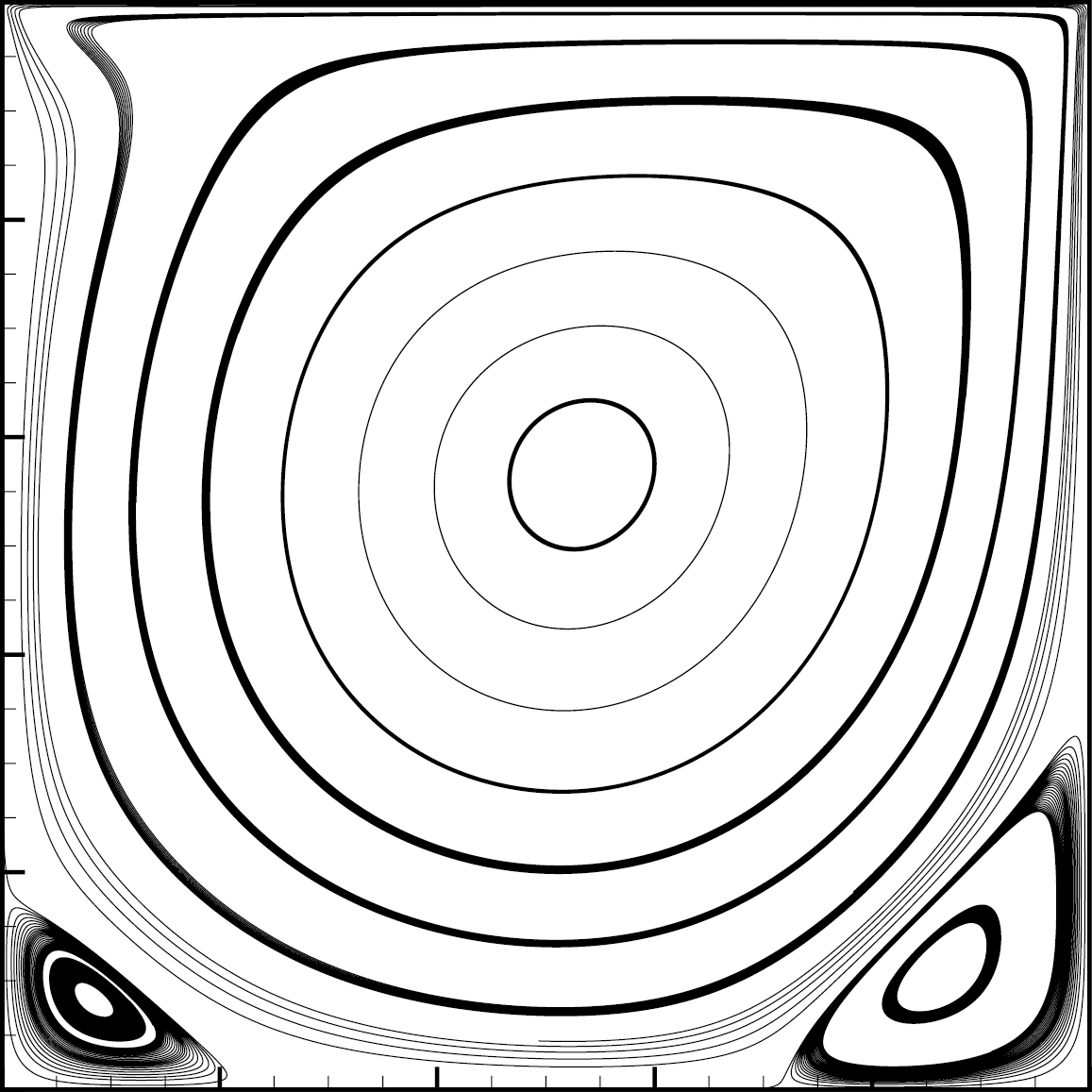}}\hspace{0.02\textwidth}%
\subfigure[]{\includegraphics[width=0.47\textwidth]{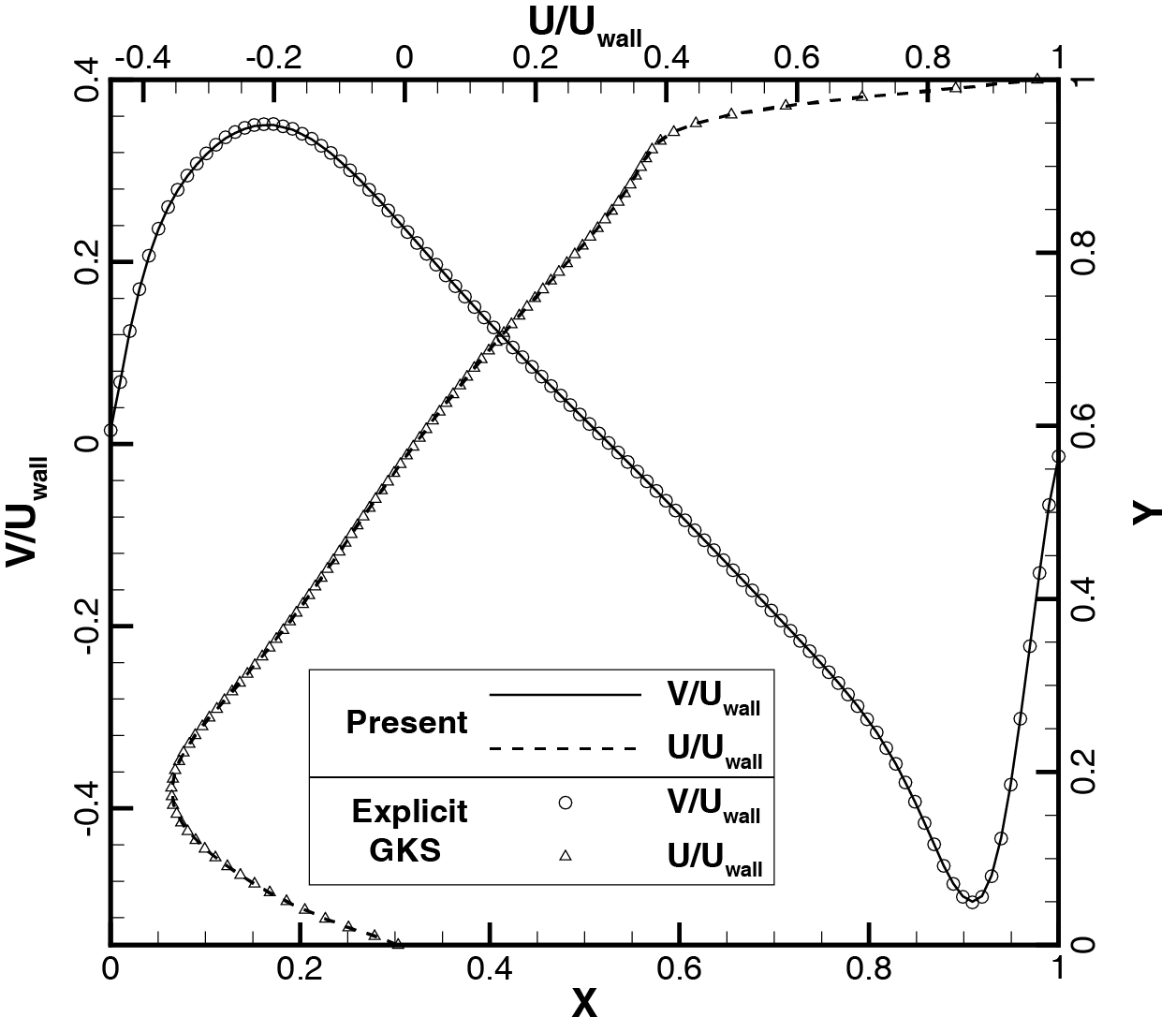}}
\caption{\label{fig:test1_1000result}Cavity flow at Re=1000. (a) The rotational temperature contours and rotational heat flux, (b) the rotational temperature distributions along the horizontal and vertical central lines, (c) the streamlines, (d) the vertical velocity $V$ along the horizontal central line and the horizontal velocity $U$ along the vertical central line. The reference result is calculated by GKS without discretization of velocity space \cite{liu2014unified} (identical to Navier-Stokes solution).}
\end{figure}

\begin{figure}
\centering
\subfigure[]{\includegraphics[width=0.47\textwidth]{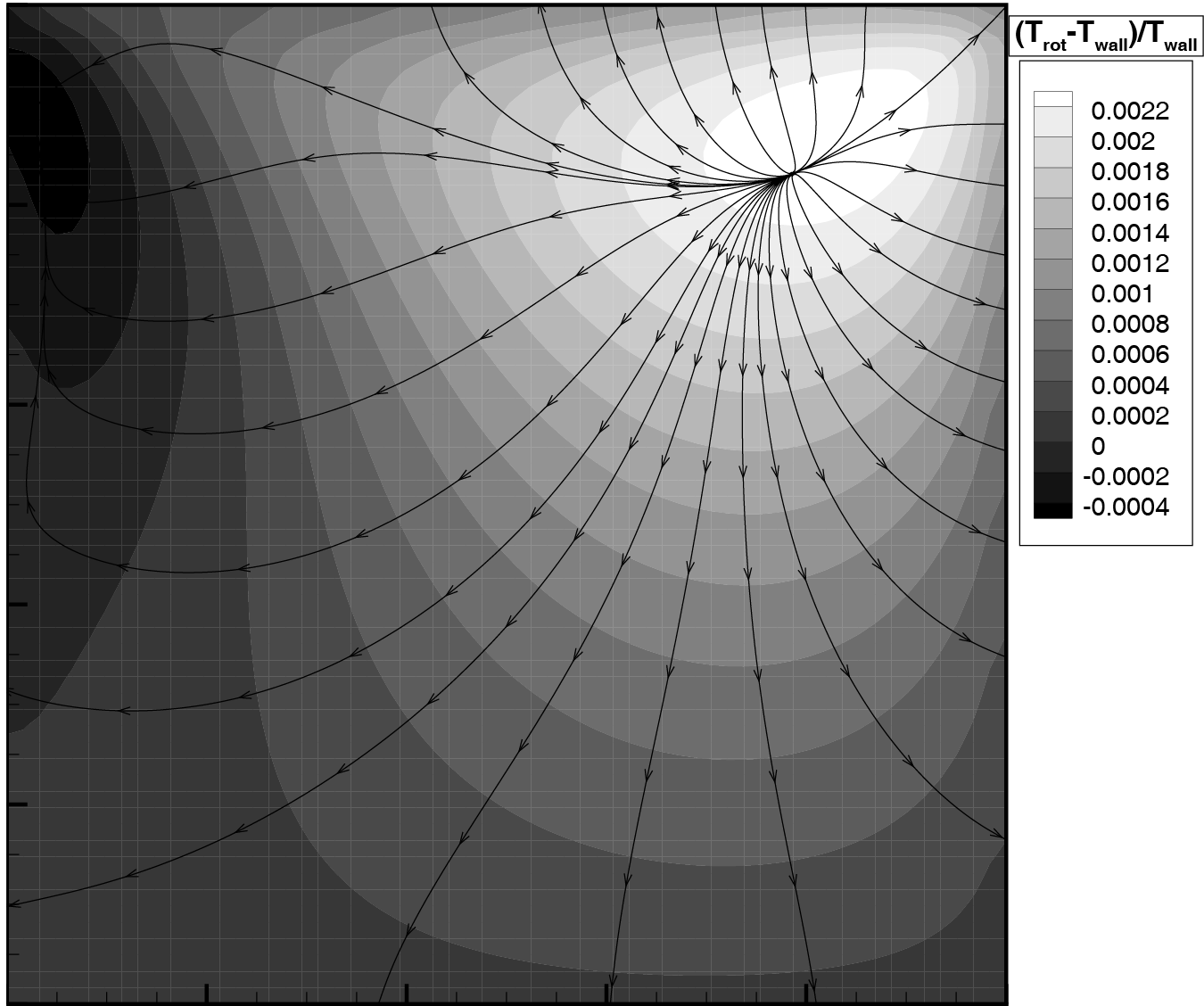}}\hspace{0.02\textwidth}%
\subfigure[]{\includegraphics[width=0.47\textwidth]{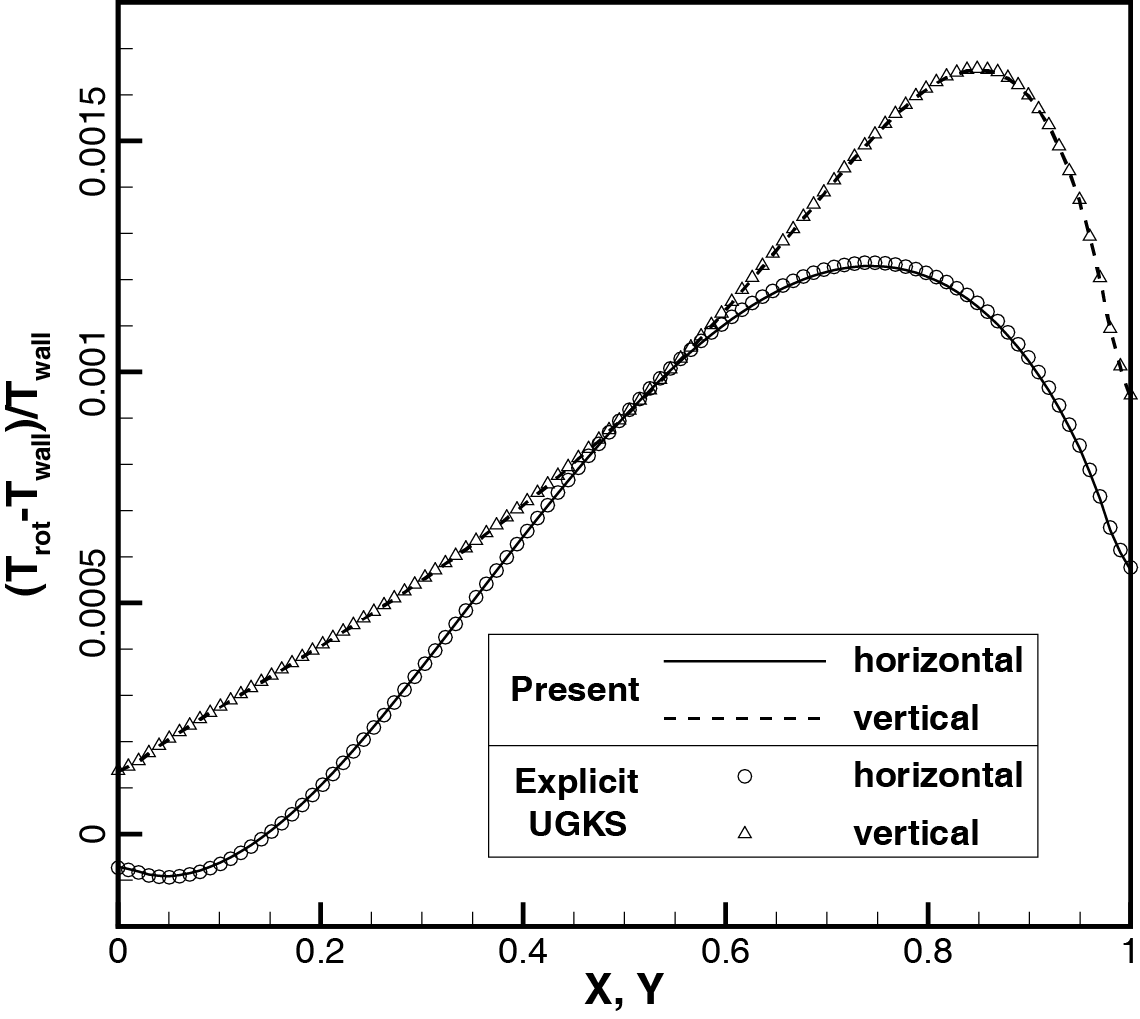}}\\
\subfigure[]{\includegraphics[width=0.47\textwidth]{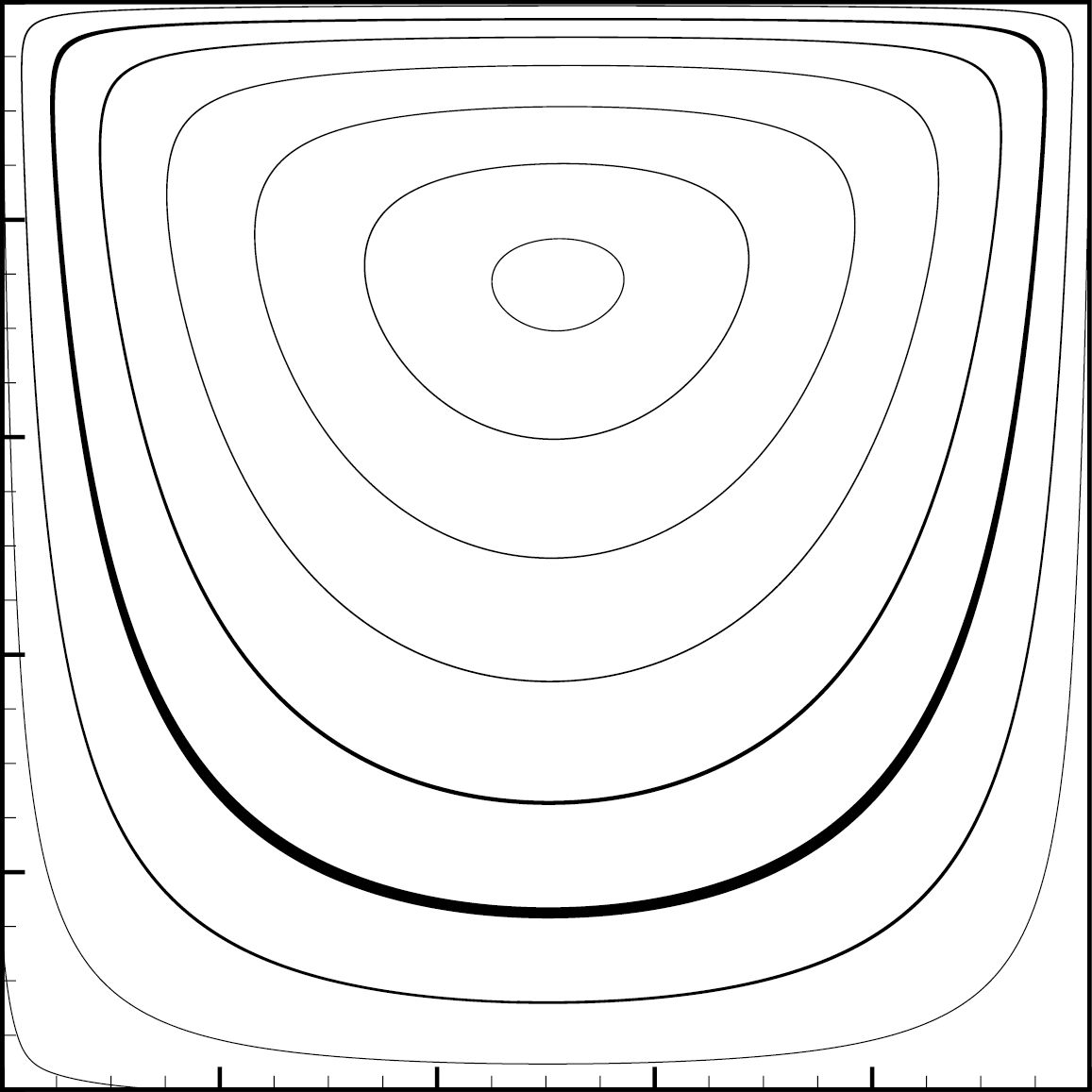}}\hspace{0.02\textwidth}%
\subfigure[]{\includegraphics[width=0.47\textwidth]{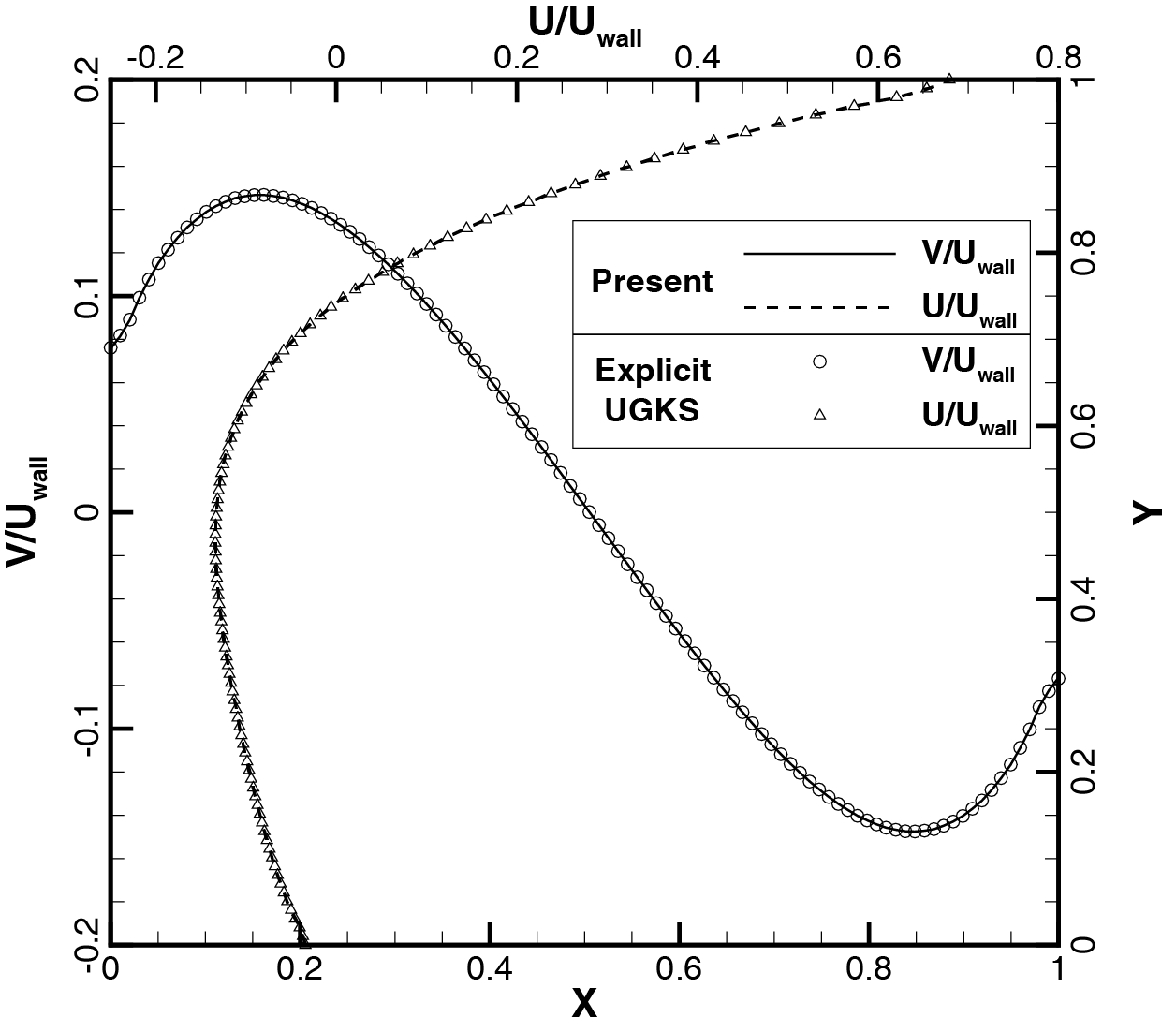}}
\caption{\label{fig:test1_75result}Cavity flow at Kn=0.075. (a) The rotational temperature contours and rotational heat flux, (b) the rotational temperature distributions along the horizontal and vertical central lines, (c) the streamlines, (d) the vertical velocity $V$ along the horizontal central line and the horizontal velocity $U$ along the vertical central line. The reference result is calculated by UGKS \cite{liu2014unified}.}
\end{figure}

\begin{figure}
\centering
\subfigure[]{\includegraphics[width=0.47\textwidth]{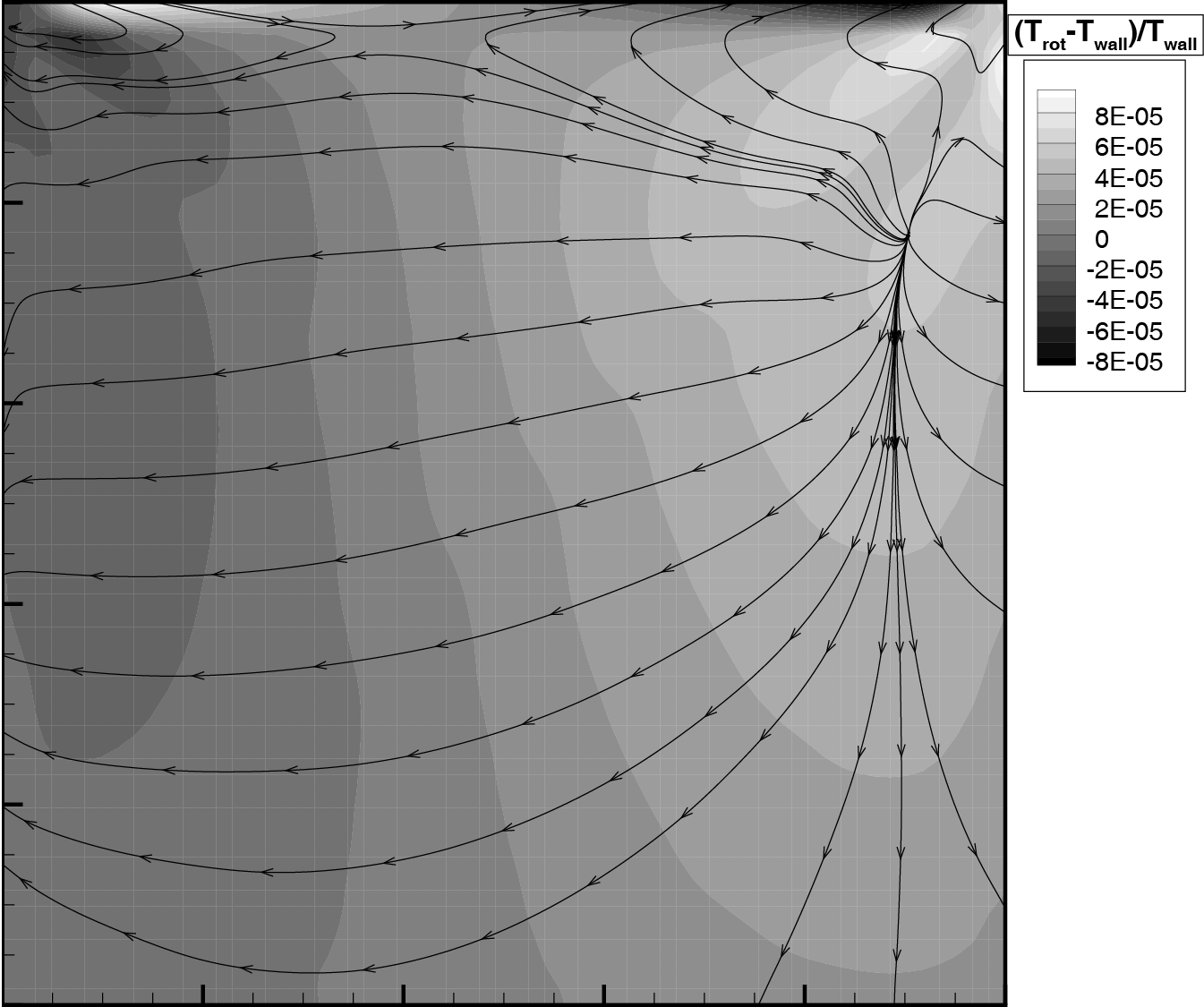}}\hspace{0.02\textwidth}%
\subfigure[]{\includegraphics[width=0.47\textwidth]{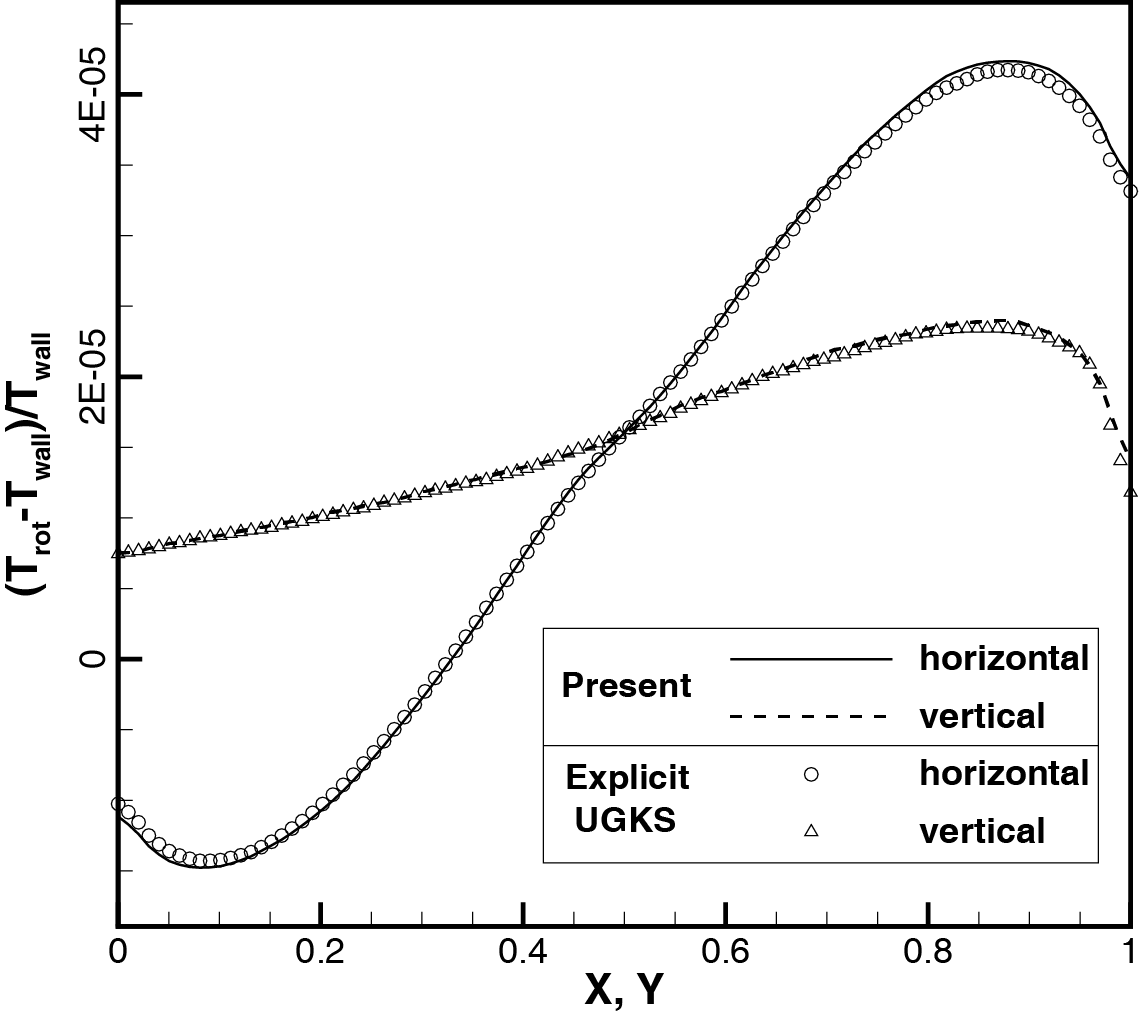}}\\
\subfigure[]{\includegraphics[width=0.47\textwidth]{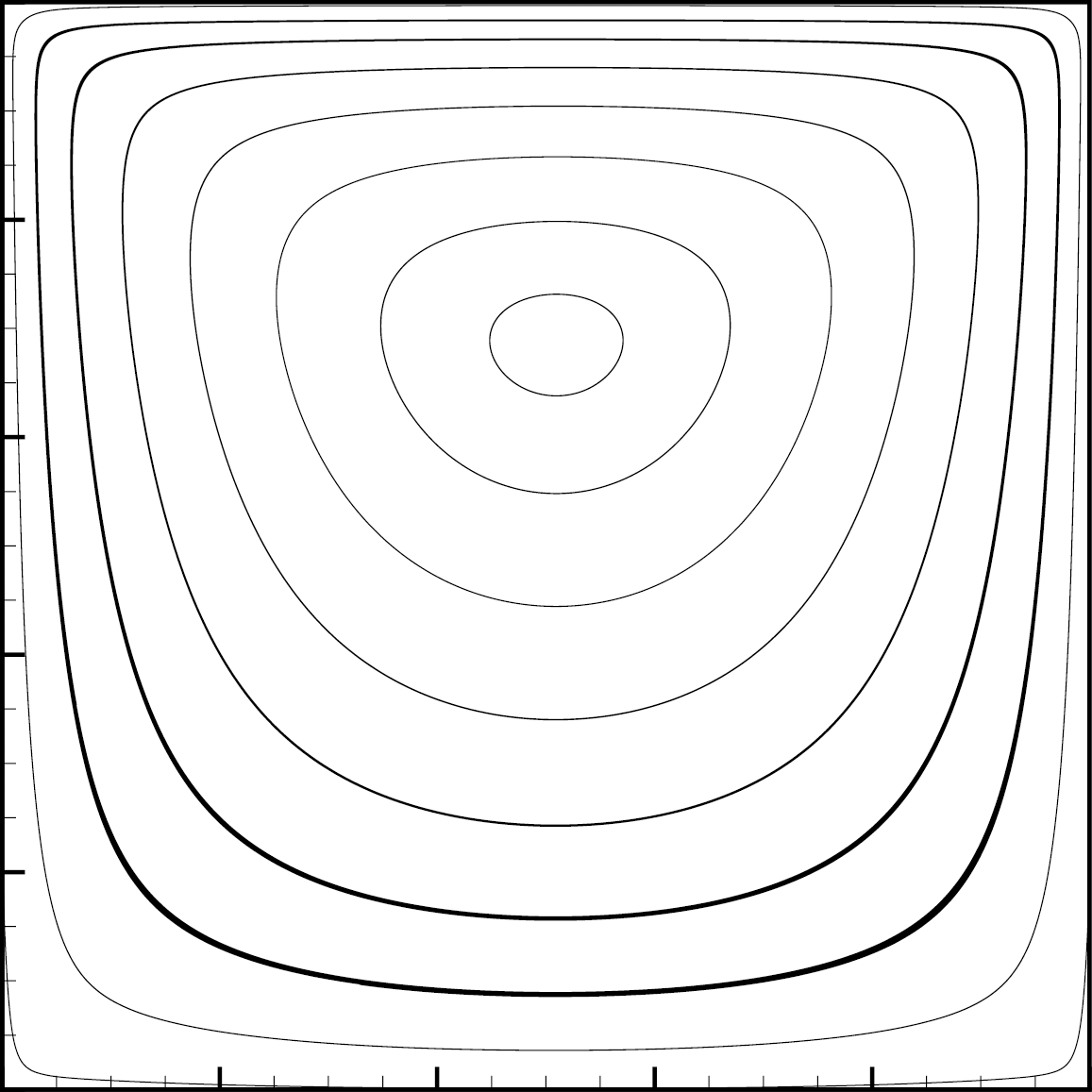}}\hspace{0.02\textwidth}%
\subfigure[]{\includegraphics[width=0.47\textwidth]{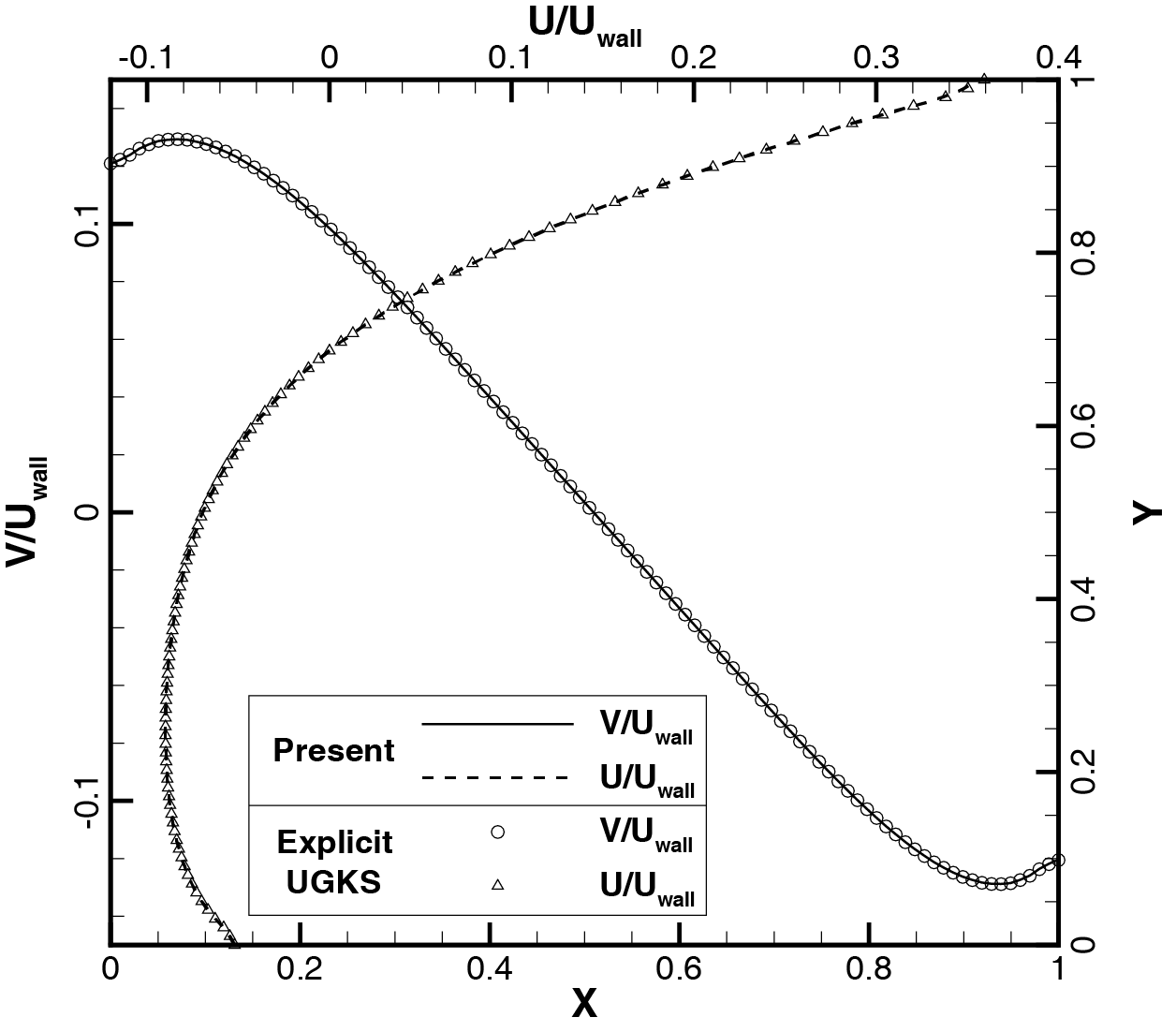}}
\caption{\label{fig:test1_10result}Cavity flow at Kn=10. (a) The rotational temperature contours and rotational heat flux, (b) the rotational temperature distributions along the horizontal and vertical central lines, (c) the streamlines, (d) the vertical velocity $V$ along the horizontal central line and the horizontal velocity $U$ along the vertical central line. The reference result is calculated by UGKS \cite{liu2014unified}.}
\end{figure}

First, the cases of Re=1000 and Kn=0.075, 1, 10 are simulated. As is shown in Fig.~\ref{fig:test1_mesh}, a nonuniform $61 \times 61$ mesh with a mesh size $0.004L$ ($L$ is the width of the cavity) near the wall is used for the case Re=1000 while a uniform $61 \times 61$ mesh is used for the cases Kn=0.075, 1, 10. For the case Re=1000, Gauss-Hermite quadrature with 12 velocity points is adopted. For the case Kn=0.075, an unstructured discretization of velocity space with 729 cells is used, and the mid-point quadrature is applied. For the case Kn=1, 10, a more refined unstructured velocity-space mesh with 6286 cells is used. The computational efficiency compared with the explicit diatomic UGKS method of Liu et al.~\cite{liu2014unified} is shown in Tab.~\ref{tab:caveff}. All of the simulations in the table are run on a single core of a computer with Intel(R) Xeon(R) CPU X5670 @ 2.93GHz. The convergence criterion is that the global root-mean-square residuals of the macroscopic variables less than $10^{-9}$, where the residual vector is defined as
\begin{equation}
\overrightarrow {\rm{Rsd}}_i^n = \vec S_i^n - \frac{1}{{{V_i}}}\sum\limits_{j \in N(i)} {{A_{ij}}\vec F_{ij}^n}.
\end{equation}
It can be seen in Tab.~\ref{tab:caveff} that the present method is 1--2 orders of magnitude faster than the explicit UGKS in all flow regimes. The results for Re=1000, Kn=0.075, 10 are shown in Fig.~\ref{fig:test1_1000result}, Fig.~\ref{fig:test1_75result}, Fig.~\ref{fig:test1_10result} respectively. For the case Re=1000, the present results are compared with the results obtained from the diatomic GKS method (a degenerate version of Liu et al.'s UGKS method \cite{liu2014unified} without discretization of velocity space) which can give a Navier-Stokes solution in the continuum regime. It is shown that for this case the present velocity distribution agrees well with the GKS result. The present rotational temperature distribution deviates a little from the GKS result but the shape of the two sets of curves are same. The maximum rotational temperature deviation between the two results is only $6 \times {10^{ - 5}}{T_{{\rm{wall}}}}$ and we found that the temperature is a very sensitive variable in the case Re=1000. This deviation may result from several factors, such as the different basic physical model between the present method and GKS, or the different treatment for the wall boundary. After all, the present method is based on the gas-kinetic theory with particle velocity space discretization while GKS is identical to a scheme based on Navier-Stokes equation in the continuum regime. For the cases Kn=0.075, 10, the reference results are calculated by the diatomic UGKS method of Liu et al.~\cite{liu2014unified}. It can be seen that the present velocity and temperature distributions match the UGKS's results perfectly.

\begin{figure}
\centering
\subfigure[]{\includegraphics[width=0.47\textwidth]{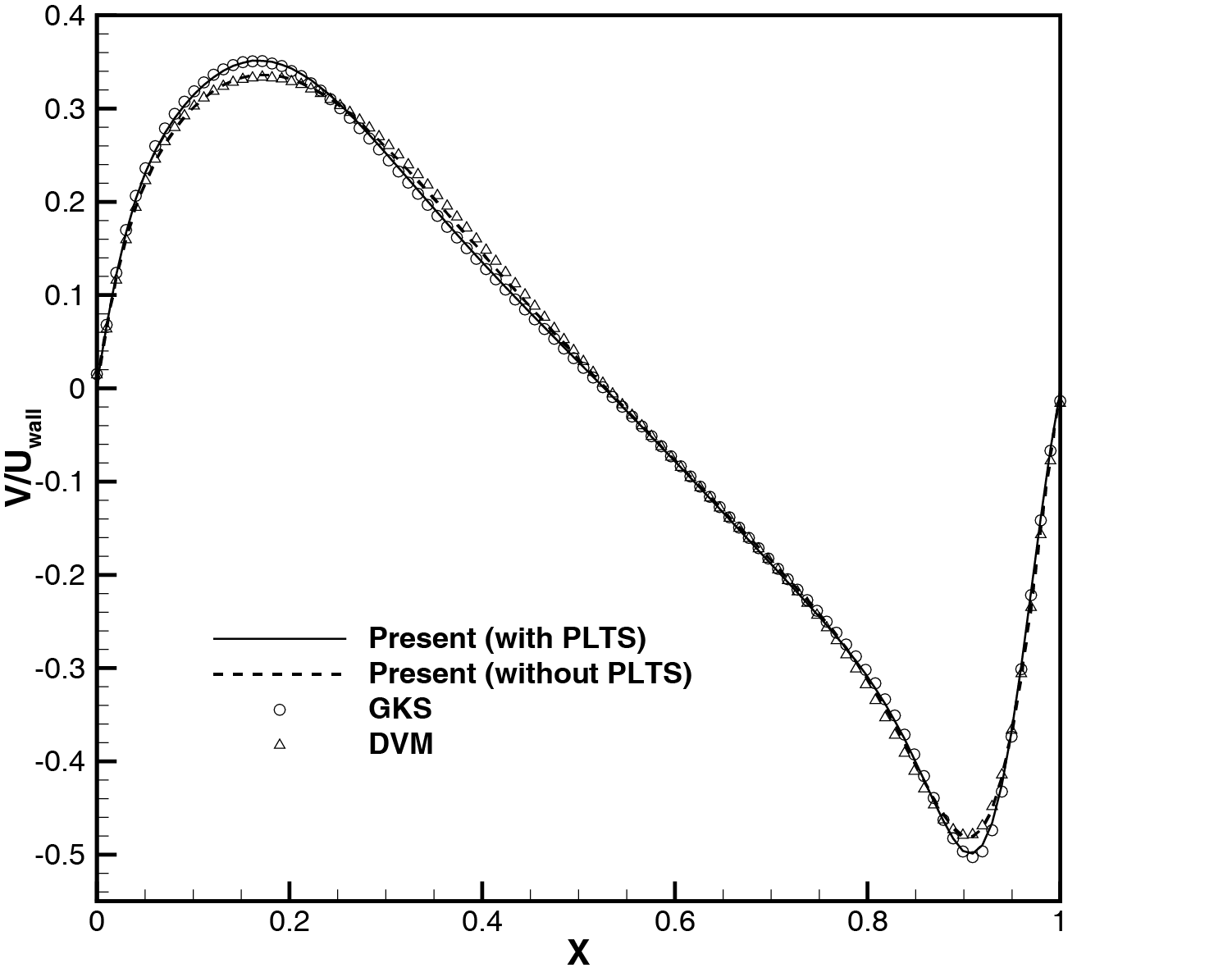}}\hspace{0.02\textwidth}%
\subfigure[]{\includegraphics[width=0.47\textwidth]{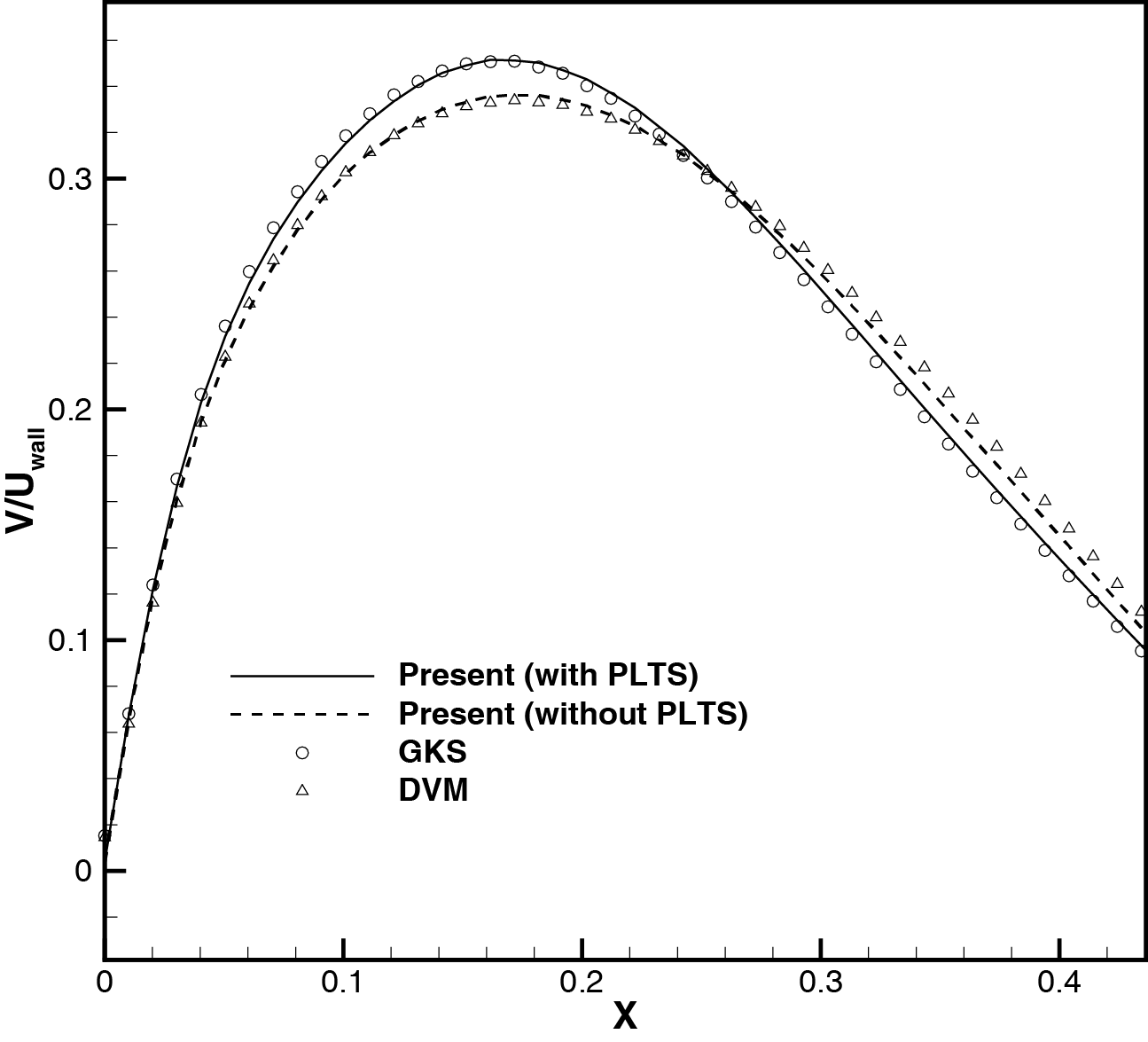}}
\caption{\label{fig:test1_lts}Cavity flow simulations with and without physical local time step (PLTS) at Re=1000, comparison of the vertical velocity $V$ along the horizontal central line.}
\end{figure}

After that, the effect of the physical local time step is validated through the case of cavity flow at Re=1000. In the validation, a nonuniform $83\times83$ mesh similar to Fig.~\ref{fig:test1_mesh1000} is used. The mesh near the cavity wall is further refined to $0.0004L$ to restrict the physical global CFL time step and the maximum mesh size is $0.03L$. The mean free path is around $0.0002L$. So the minimum mesh size is 2 times the mean free path and the maximum mesh size is 150 times the mean free path. The distributions of the vertical velocity $V$ along the horizontal central line of the cavity, calculated with the physical local and global time steps, are shown and compared with the results of GKS and conventional DVM in Fig.~\ref{fig:test1_lts}. It can be seen that without the physical local time step the result will converge to DVM's result, which means that the method loses the multiscale property and will give a more dissipating result in the continuum regime. Thus, the physical local time step is very important for the multiscale kinetic scheme.

\begin{figure}
\centering
\subfigure[]{\includegraphics[width=0.3\textwidth]{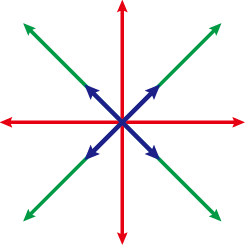}}\hspace{0.02\textwidth}%
\subfigure[\label{fig:test1_micspac792}]{\includegraphics[width=0.3\textwidth]{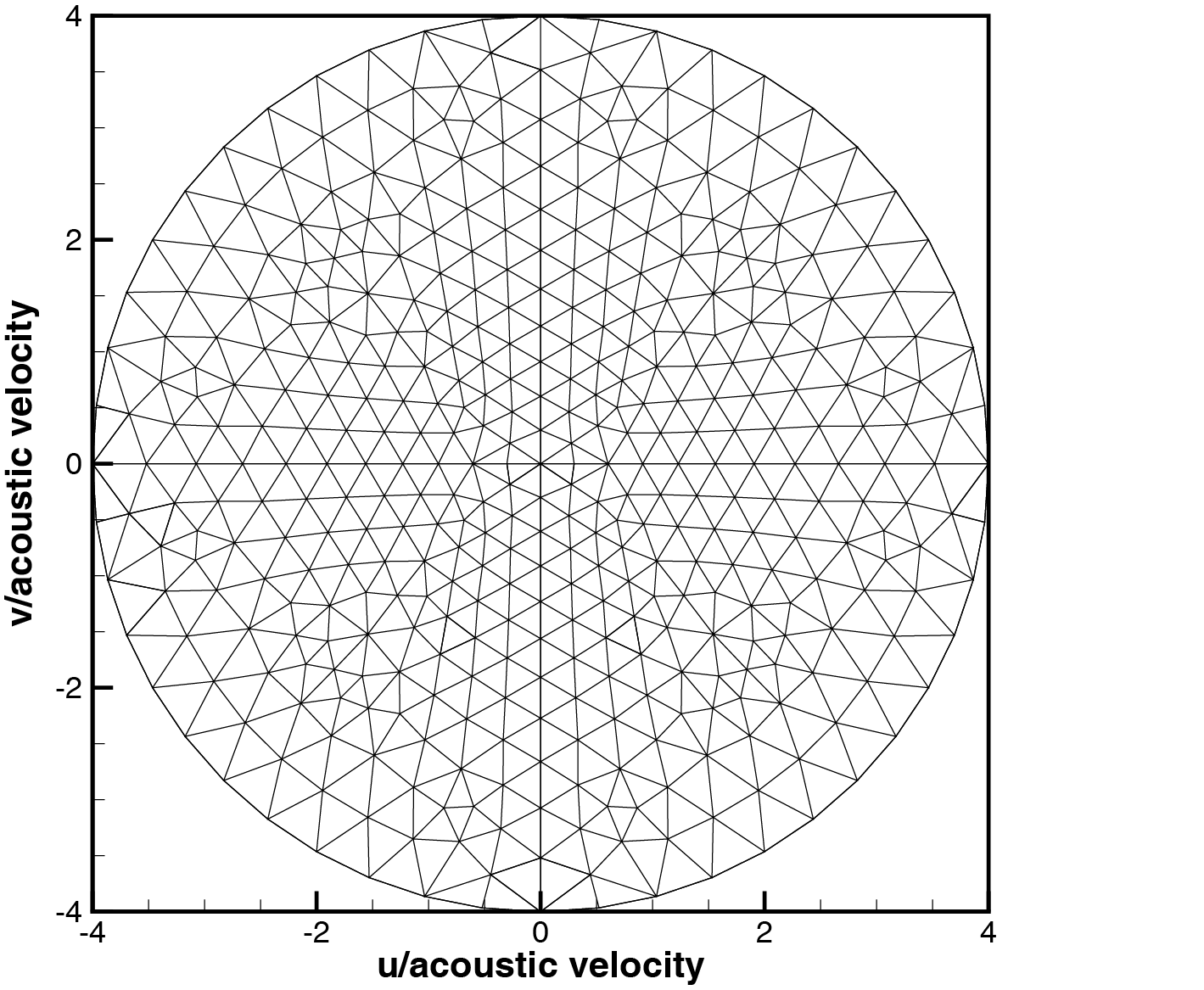}}\hspace{0.02\textwidth}%
\subfigure[]{\includegraphics[width=0.3\textwidth]{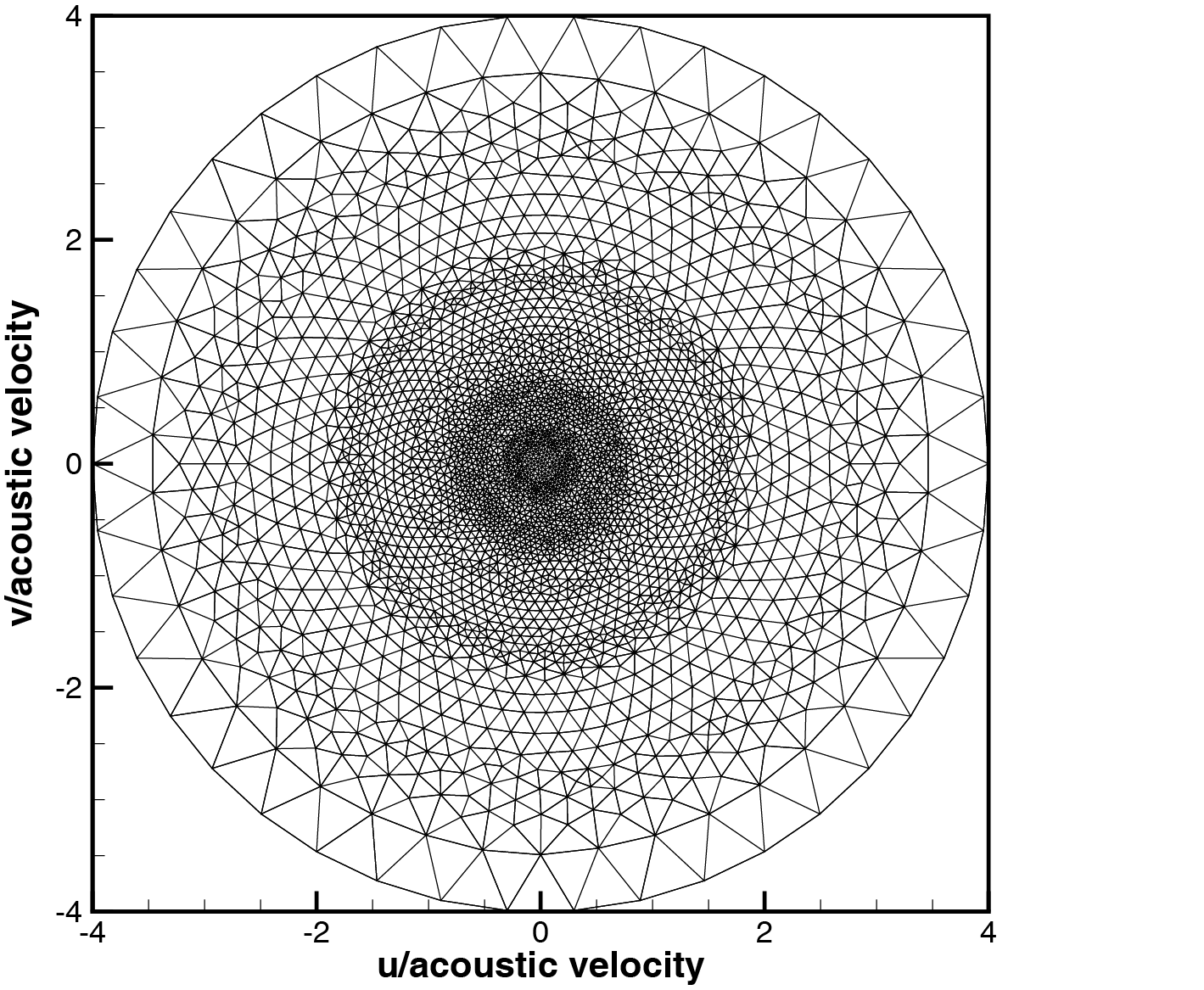}}
\caption{\label{fig:test1_micspac}Different velocity space discretizations. (a) The 12 velocity points of Gauss-Hermite quadrature formula \cite{Shan2006Kinetic}, (b) the unstructured 792 cells' and (c) the unstructured 6286 cells' velocity space discretizations with mid-point quadrature formula.}
\end{figure}

\begin{figure}
\centering
\subfigure[]{\includegraphics[width=0.47\textwidth]{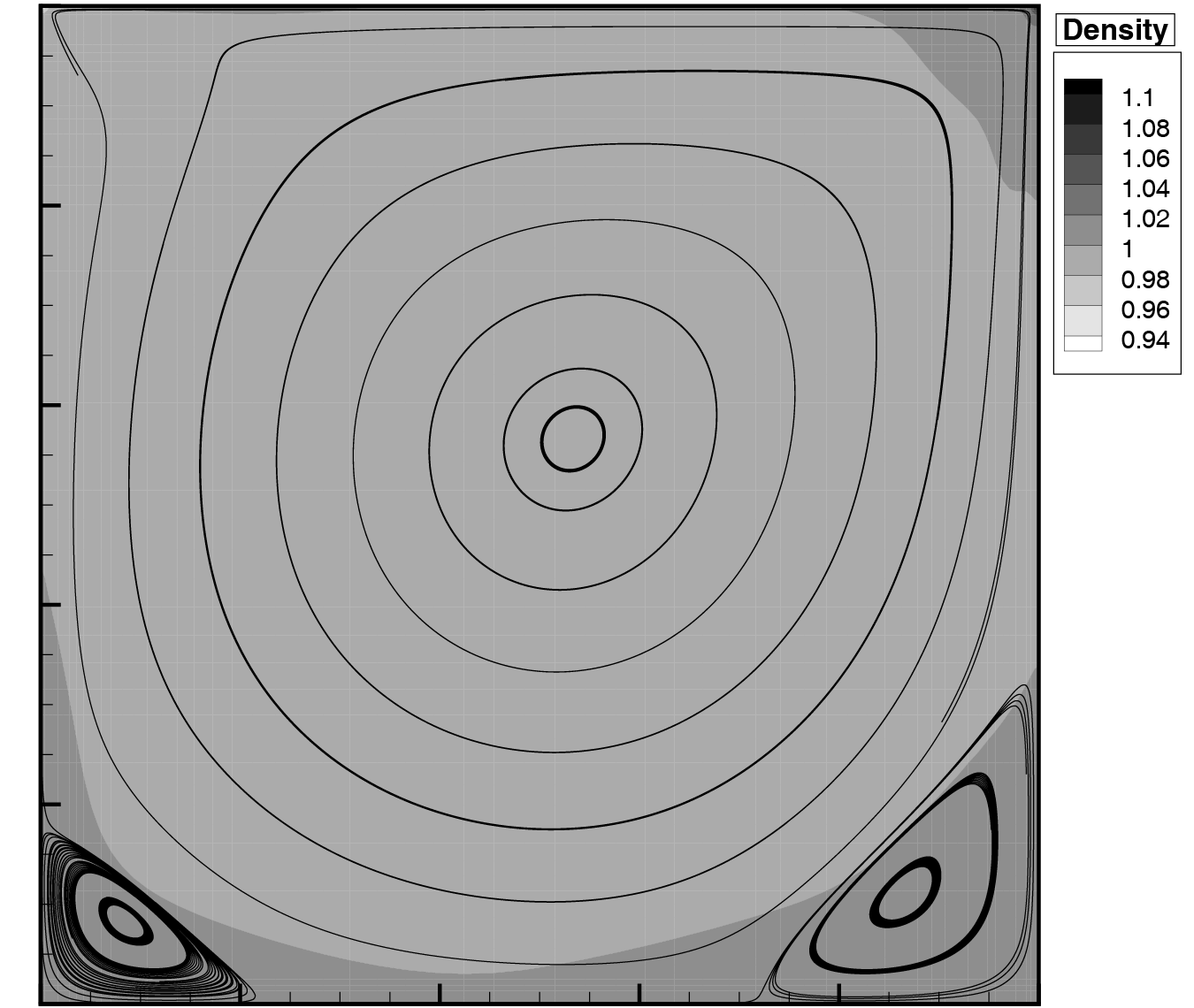}}\hspace{0.02\textwidth}%
\subfigure[]{\includegraphics[width=0.47\textwidth]{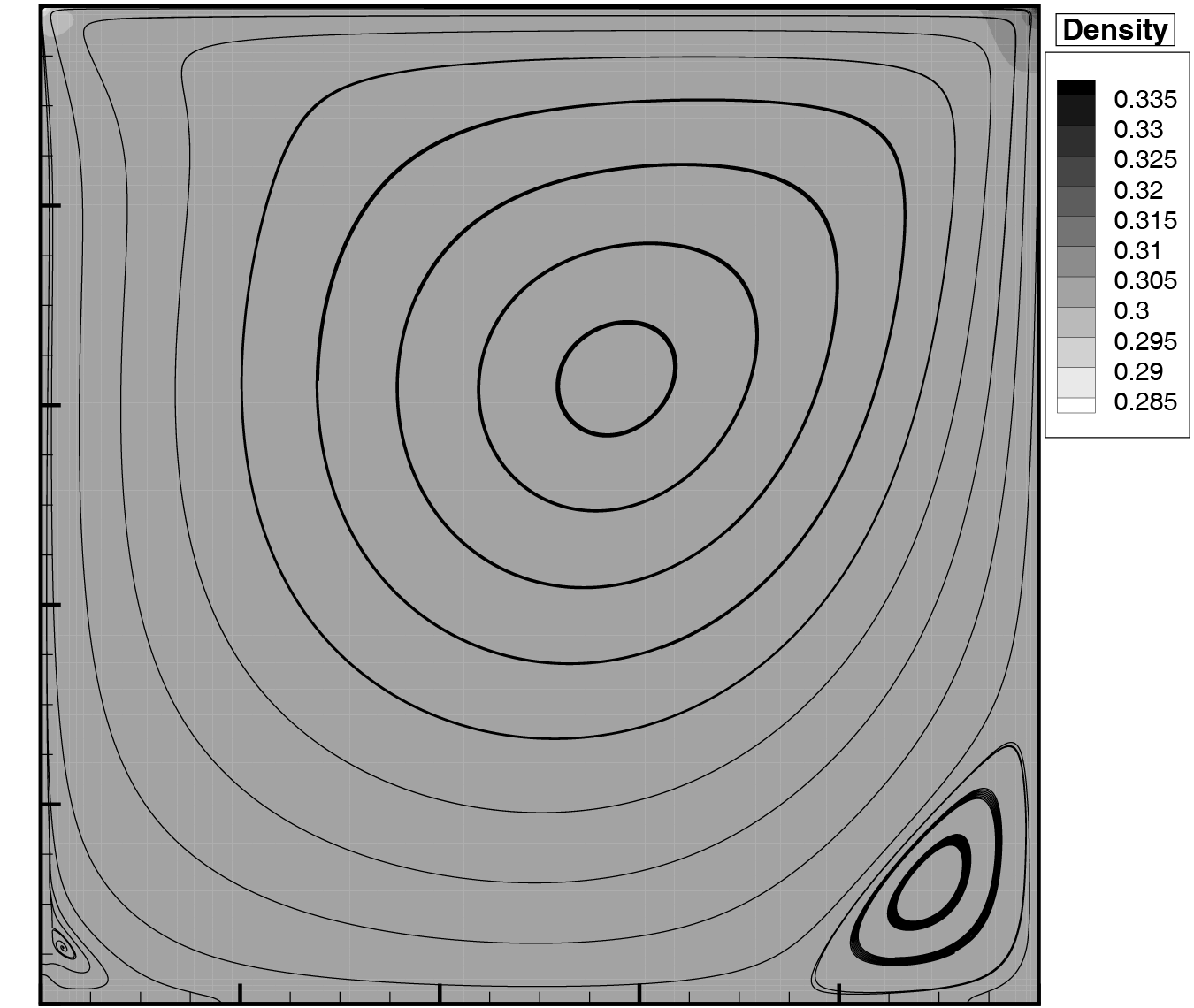}}\\
\subfigure[]{\includegraphics[width=0.47\textwidth]{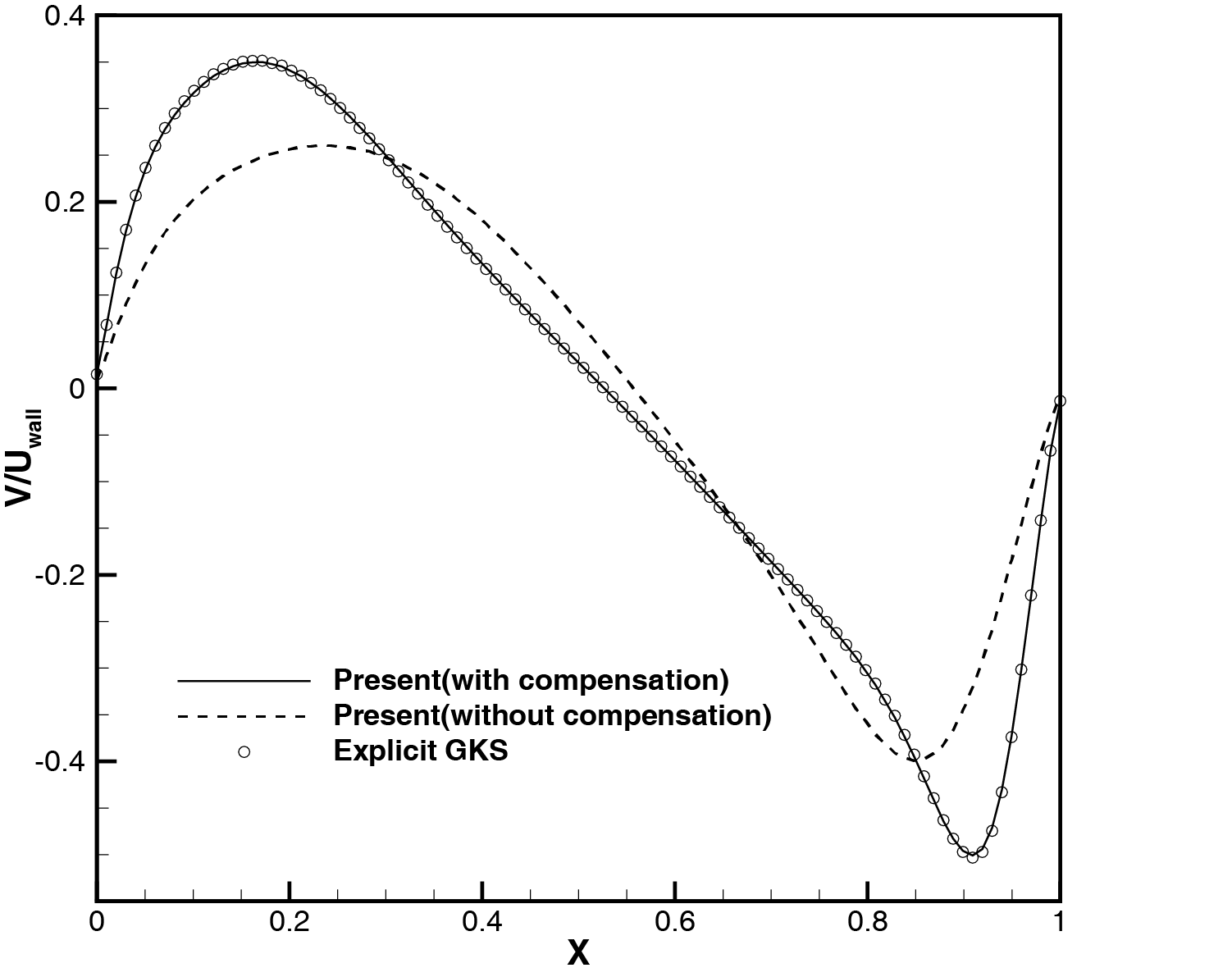}}\hspace{0.02\textwidth}%
\subfigure[]{\includegraphics[width=0.47\textwidth]{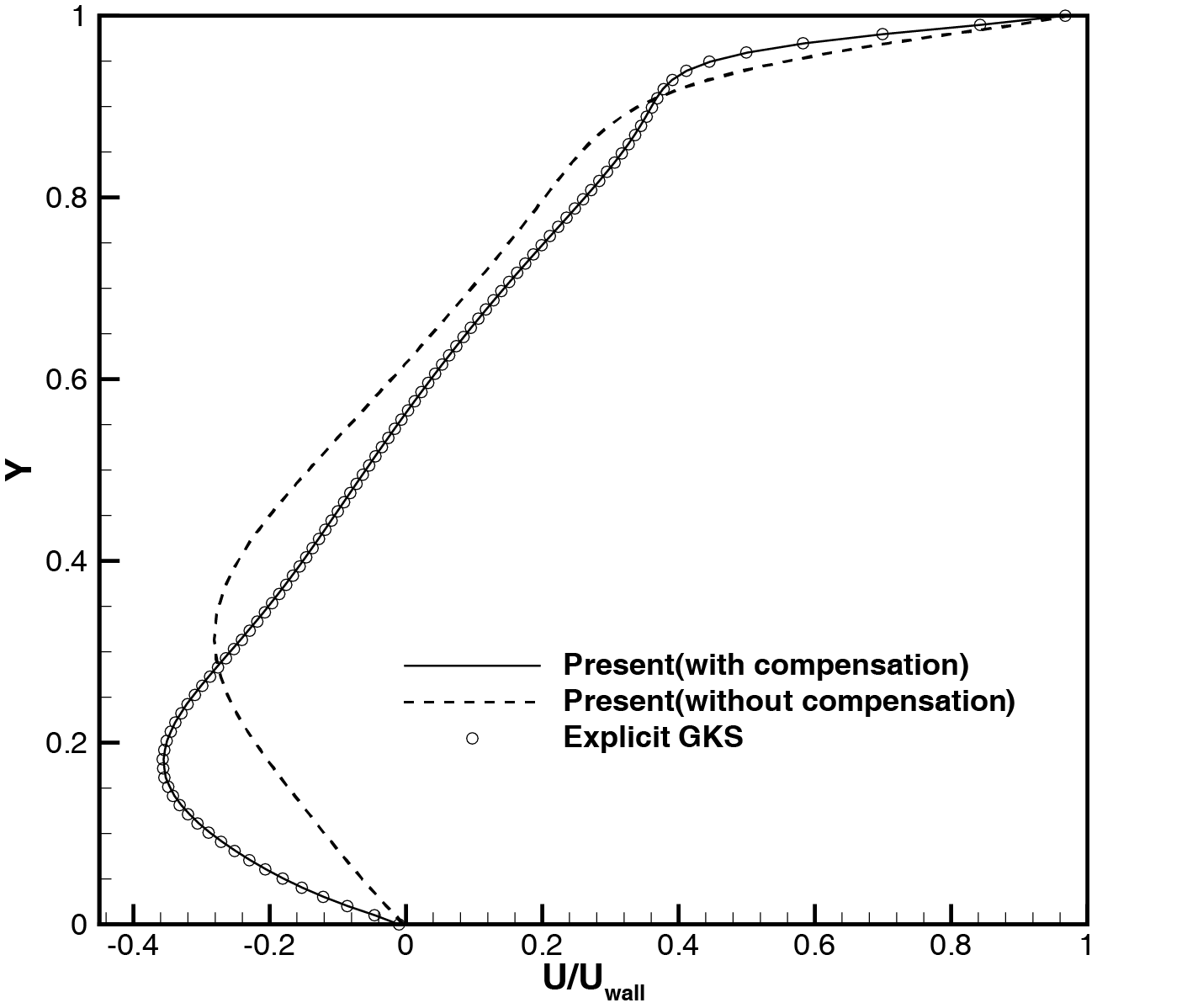}}
\caption{\label{fig:test1_fixvec}Cavity flow simulations with and without integral error compensation at Re=1000, step=1000. (a) The density contours and streamlines with compensation, (b) the density contours and streamlines without compensation, (c) the vertical velocity $V$ along the horizontal central line, (d) the horizontal velocity $U$ along the vertical central line.}
\end{figure}

\begin{figure}
\centering
\subfigure[]{\includegraphics[width=0.47\textwidth]{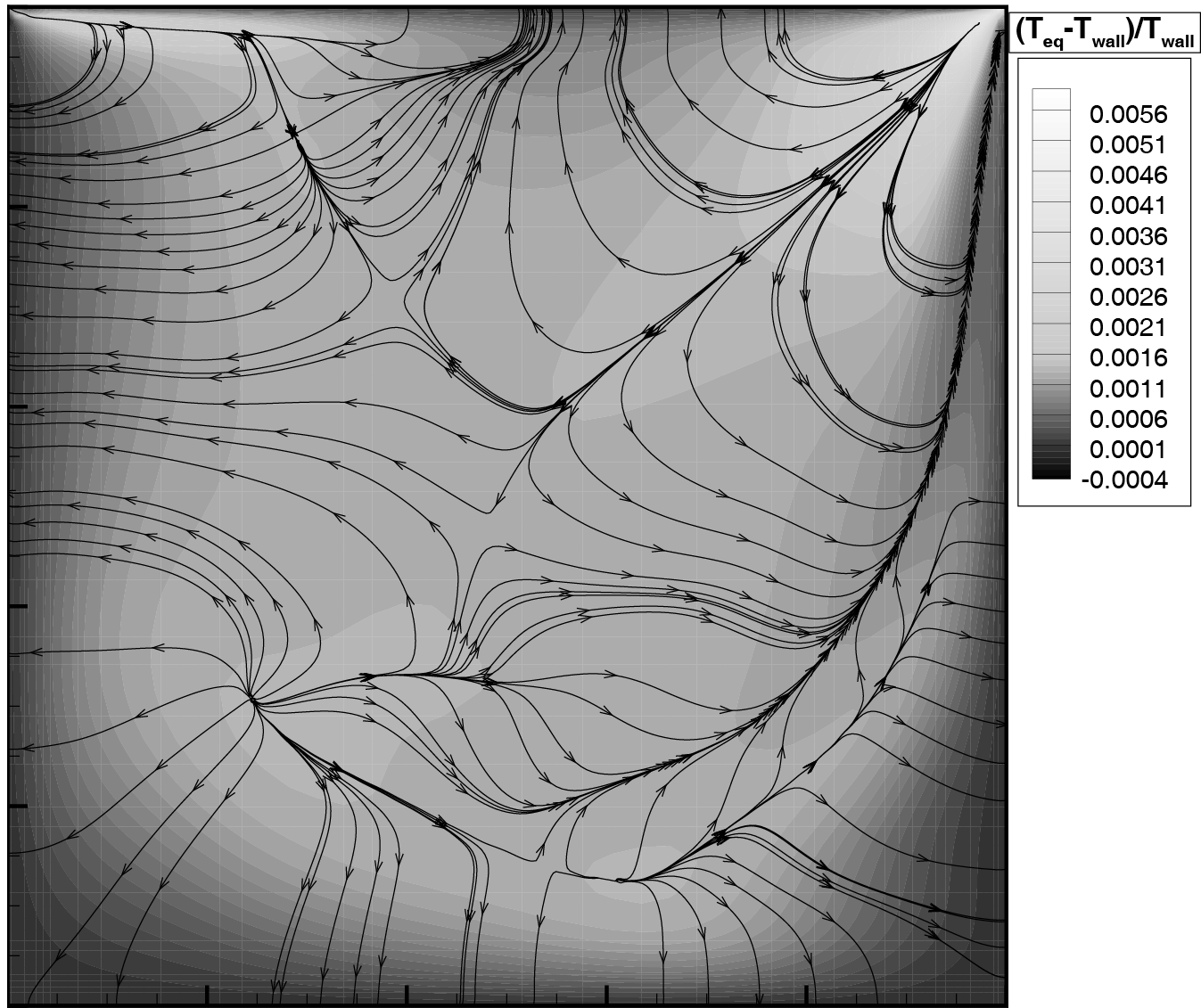}}\hspace{0.02\textwidth}%
\subfigure[]{\includegraphics[width=0.47\textwidth]{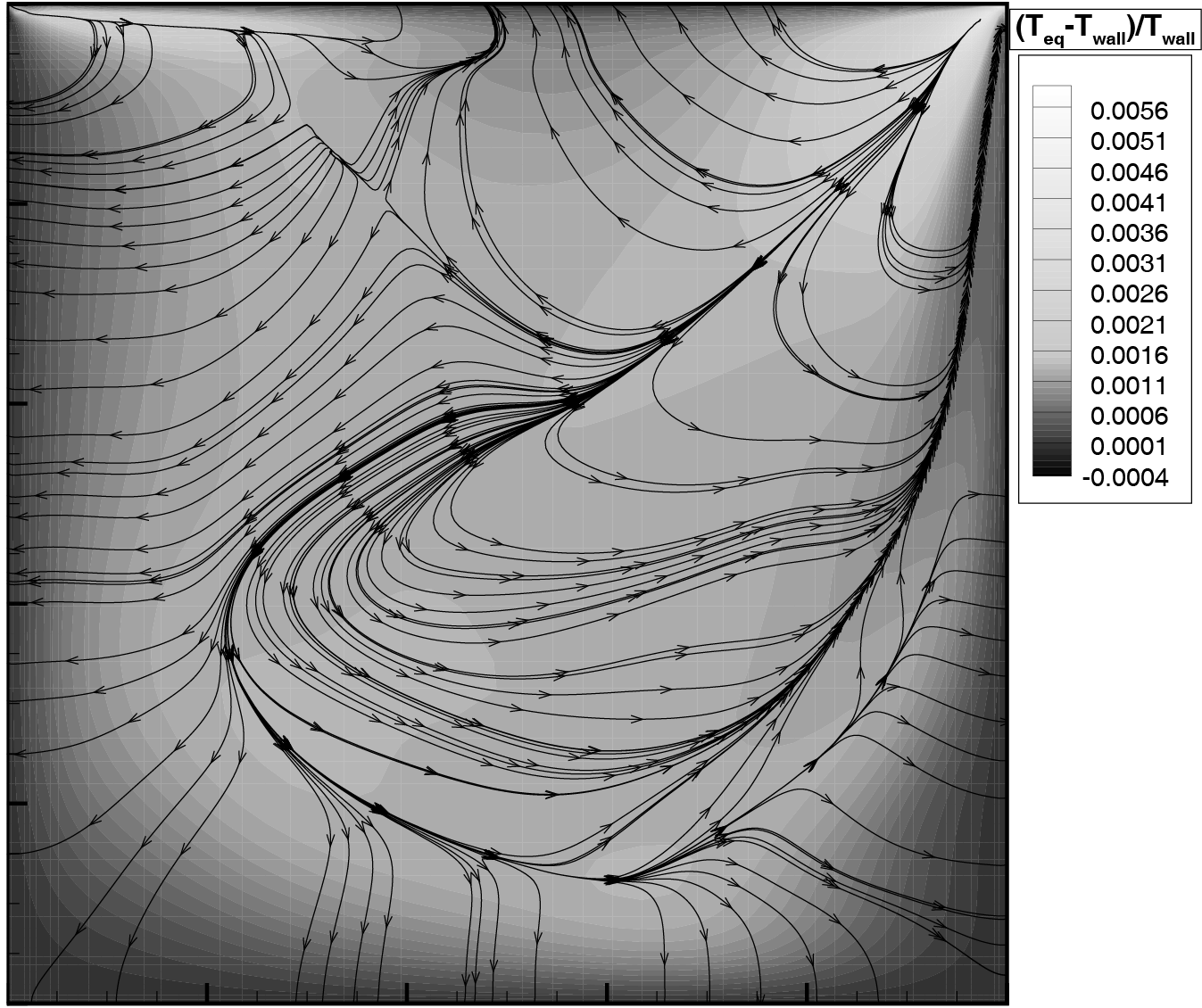}}
\caption{\label{fig:test1_fixqflux}Cavity flow simulations with and without integral error compensation for heat flux at Re=1000. (a) The equilibrium temperature contours and heat flux with compensation, (b) the equilibrium temperature contours and heat flux without compensation.}
\end{figure}

At last, the test case of the cavity flow at Re=1000 is performed again to validate the compensation technique presented in Section \ref{sec:compensation}. The calculations, with and without integral error compensation, have run 1000 implicit iterations and the results are shown in Fig.~\ref{fig:test1_fixvec}. Here the unstructured 792 cells' velocity space discretization as shown in Fig.~\ref{fig:test1_micspac792} is used, whose integral accuracy is only of the order of $10^{-3}$. As seen in Fig.~\ref{fig:test1_fixvec}, the calculation with the integral error compensation can give a result agree well with the result of GKS while the calculation without the compensation cannot. From the density contours we can see the calculation without the compensation suffers from a serious mass loss and the maximum density reduces to around 0.335 (initially 1.0 the whole flow field), while such a mass loss is not observed in the result calculated with the compensation. It is also noted that the global root-mean-square residuals of the calculations with and without the integral error compensation after 1000 implicit iterations are $1.6\times10^{-9}$ and $7.7\times10^{-3}$ respectively. Continue to do the implicit iterations, the calculation with the compensation will easily meet the convergence criterion (residuals $<10^{-9}$) at the 1015th step while the calculation without the compensation will not even after 30000 iterations. The effect of the integral error compensation for heat flux has also been tested and shown in Fig.~\ref{fig:test1_fixqflux}. In this set of tests the Gauss-Hermite quadrature with 12 velocity points is applied, which has a much higher integral accuracy, of the order of $10^{-7}$, than the 792 cells' unstructured discretization. It is shown in Fig.~\ref{fig:test1_fixqflux} that, without the compensation, the calculation fails to give a right heat flux and the heat flow is not along the negative temperature gradient direction. The last thing to notice is that the compensation procedure is very efficient. We have performed computations of different physical-space/velocity-space discretization scales and the maximum additional computation cost due to the compensation procedure is less than 3.6\%.

\subsection{Shock structure}

\begin{figure}
\centering
\subfigure[Ma=1.53]{\includegraphics[width=0.47\textwidth]{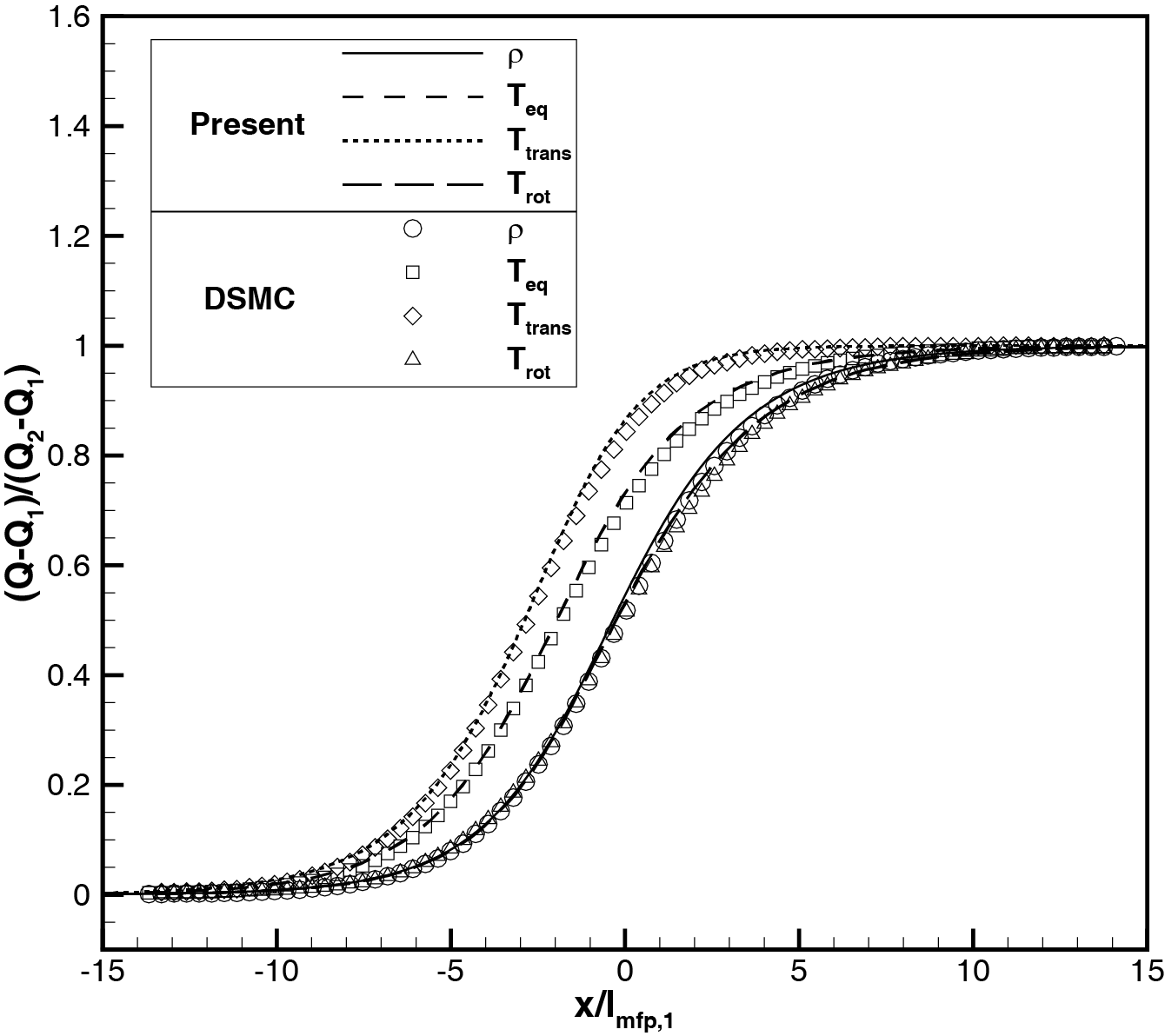}}\hspace{0.02\textwidth}%
\subfigure[Ma=4.0]{\includegraphics[width=0.47\textwidth]{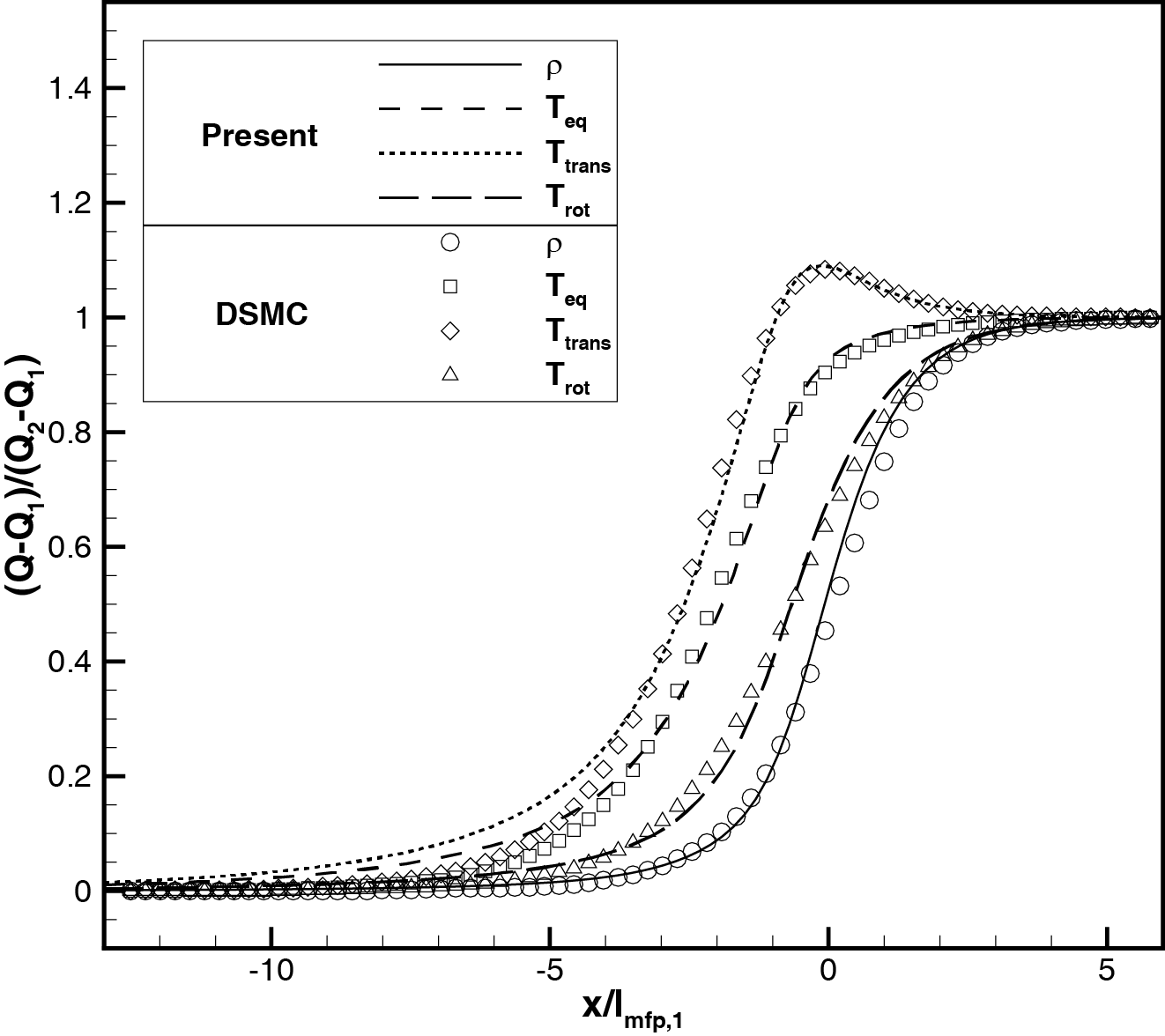}}\\
\subfigure[Ma=5.0]{\includegraphics[width=0.47\textwidth]{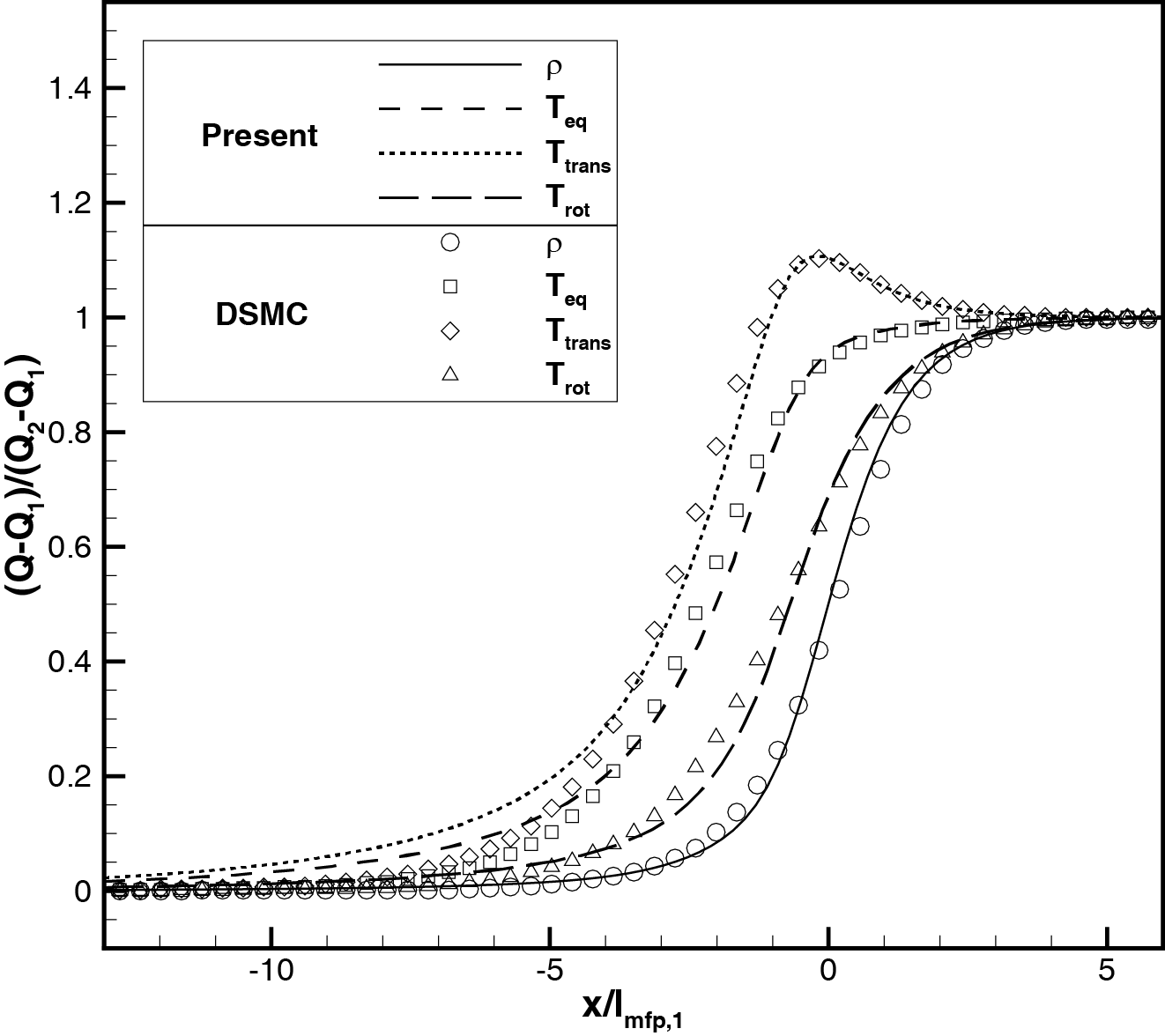}}\hspace{0.02\textwidth}%
\subfigure[Ma=7.0]{\includegraphics[width=0.47\textwidth]{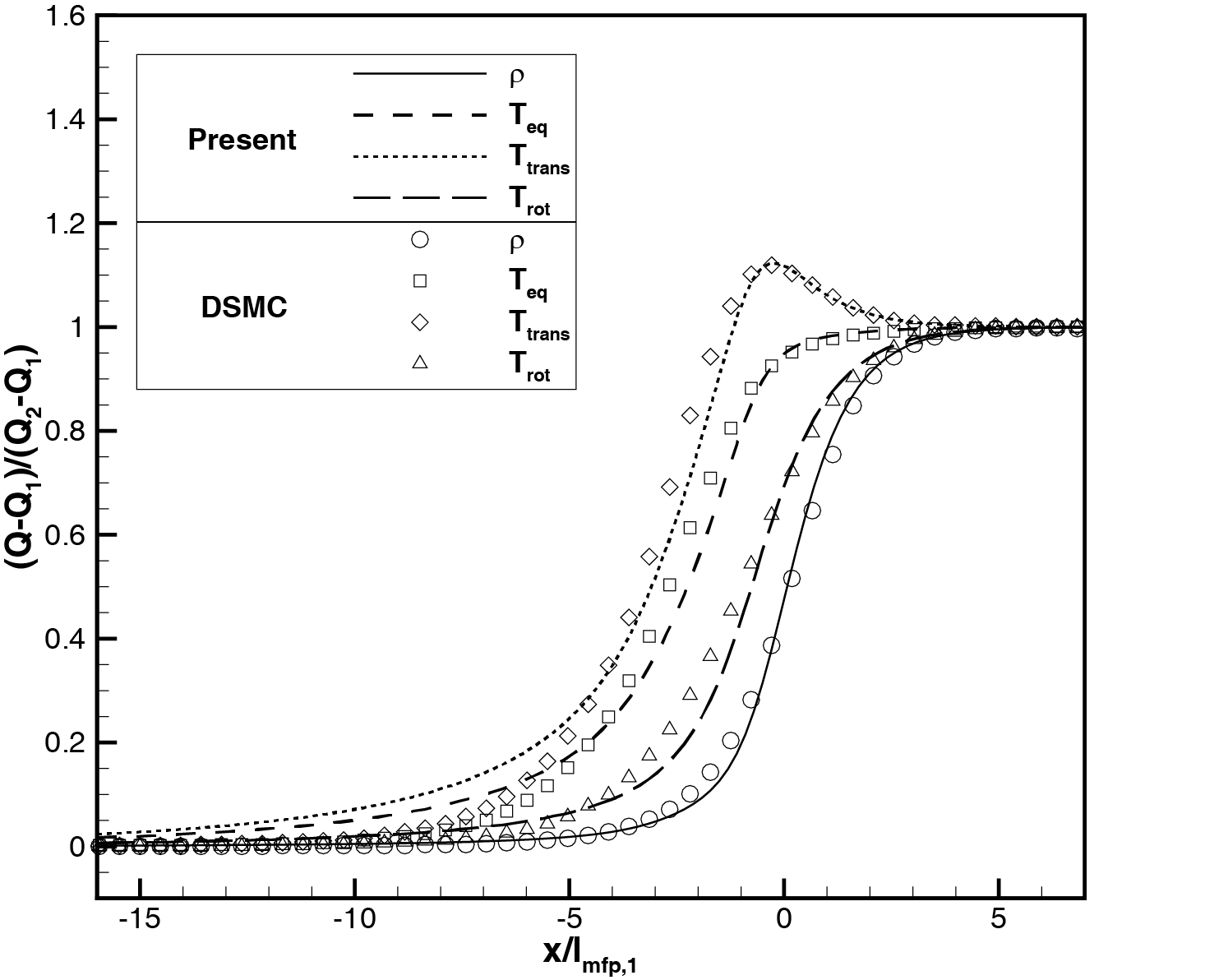}}
\caption{\label{fig:test2_cmpdsmc}Shock structures at different Mach numbers compared with DSMC's results \cite{liu2014unified}. $Q_1$ and $Q_2$ are the upstream and downstream far-field values respectively, $l_{{\rm{mfp,1}}}$ is the upstream mean free path.}
\end{figure}

\begin{figure}
\centering
\subfigure[Ma=1.7]{\includegraphics[width=0.47\textwidth]{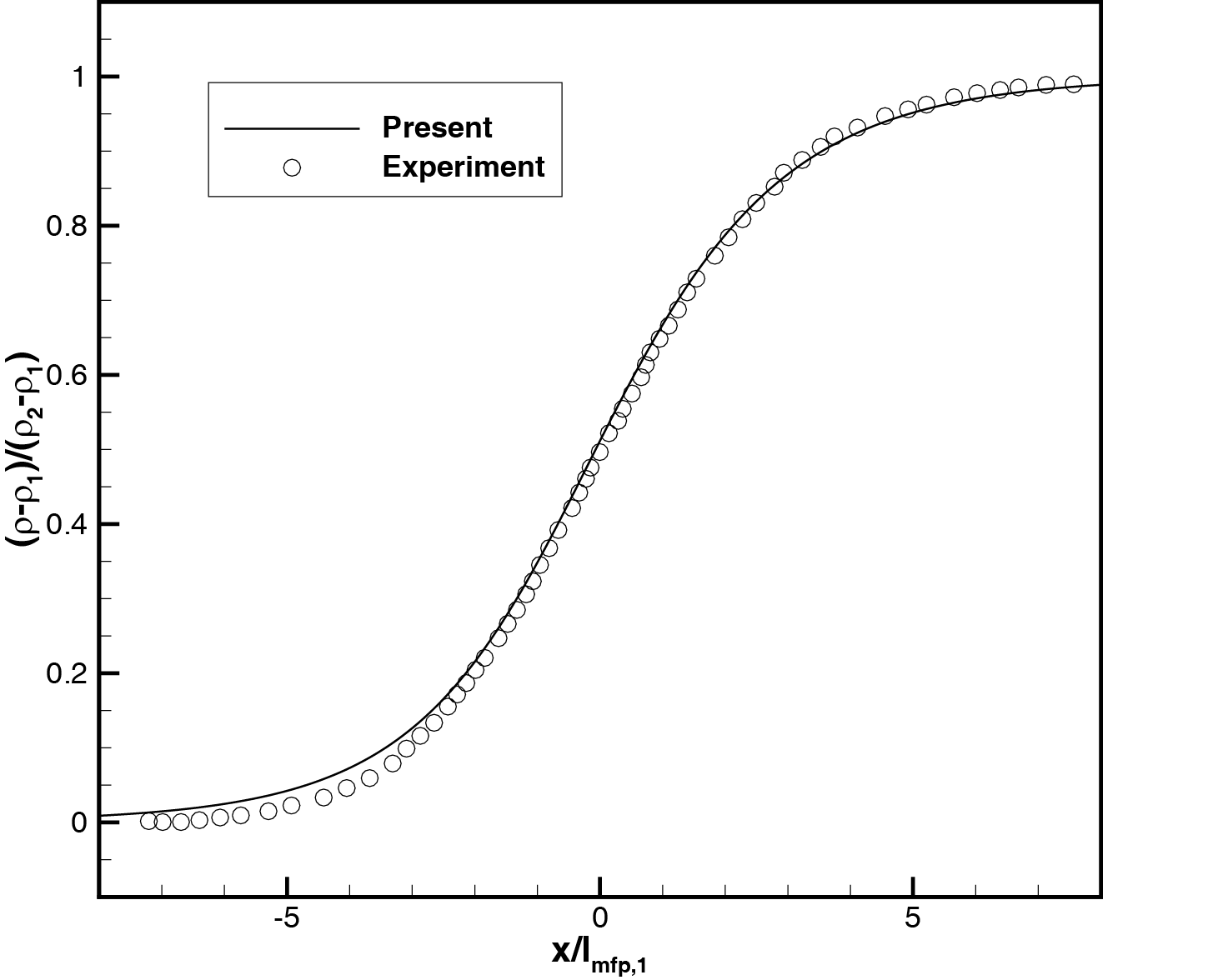}}\hspace{0.02\textwidth}%
\subfigure[Ma=3.8]{\includegraphics[width=0.47\textwidth]{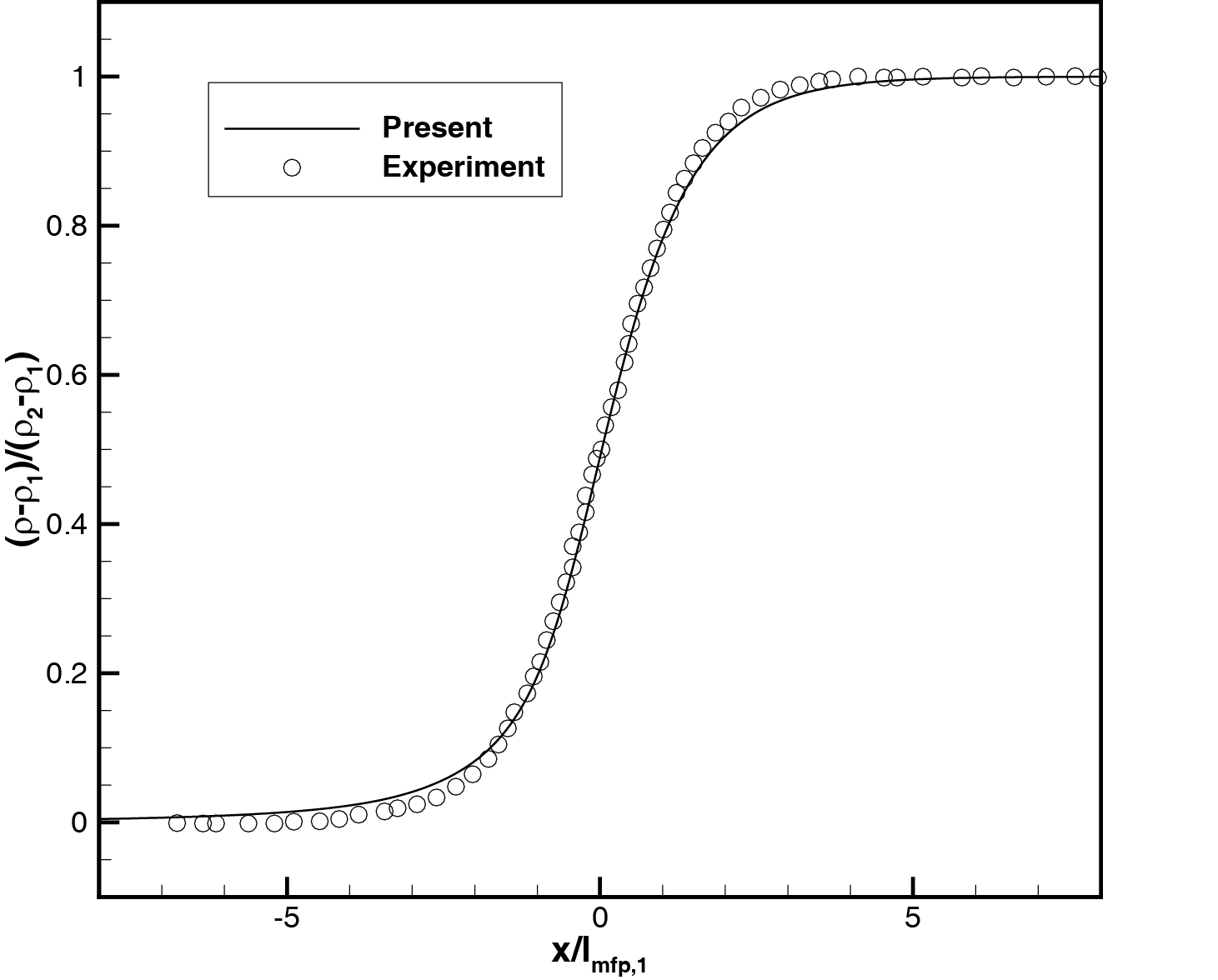}}\\
\subfigure[Ma=6.1]{\includegraphics[width=0.47\textwidth]{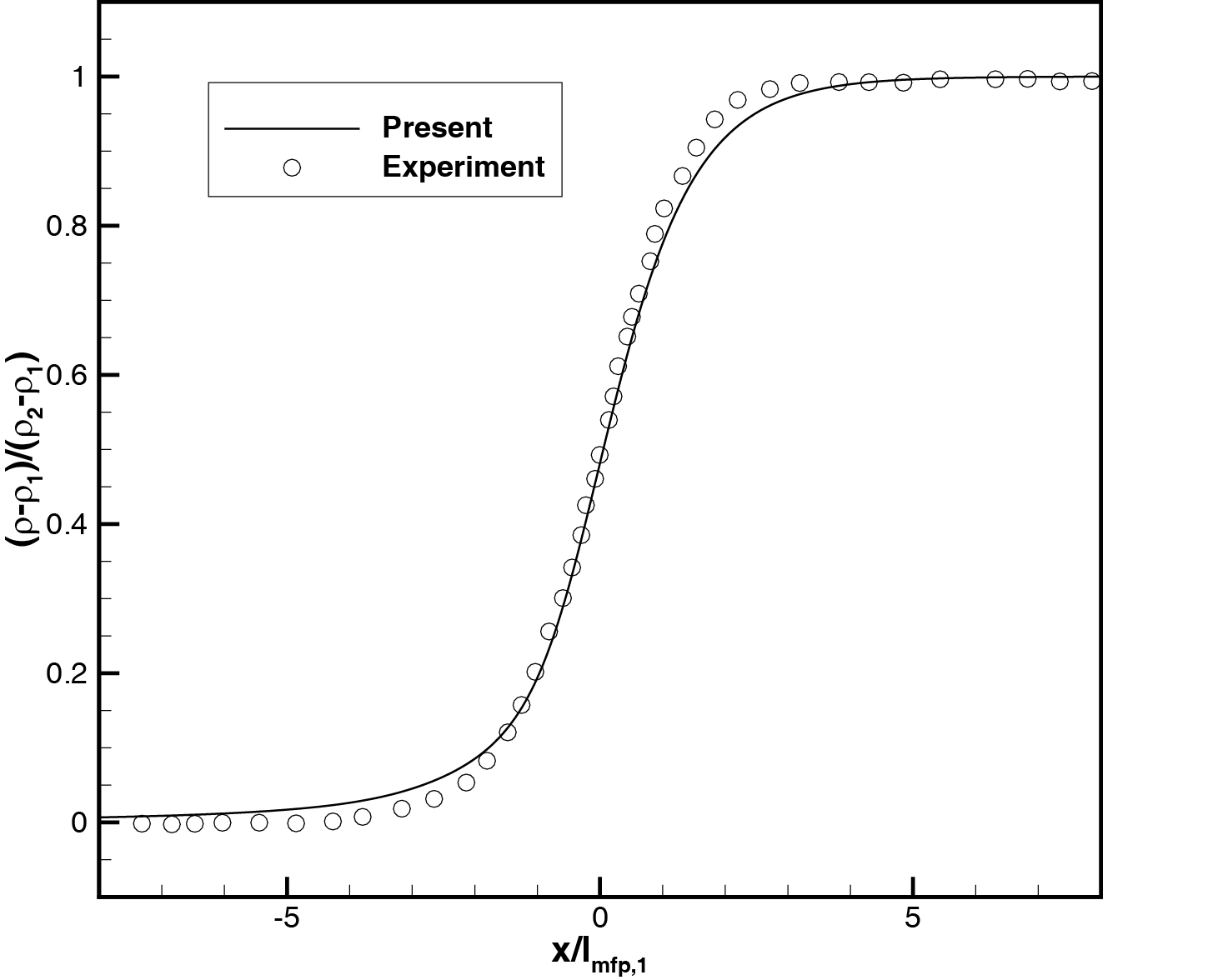}}\hspace{0.02\textwidth}%
\subfigure[Ma=10.0]{\includegraphics[width=0.47\textwidth]{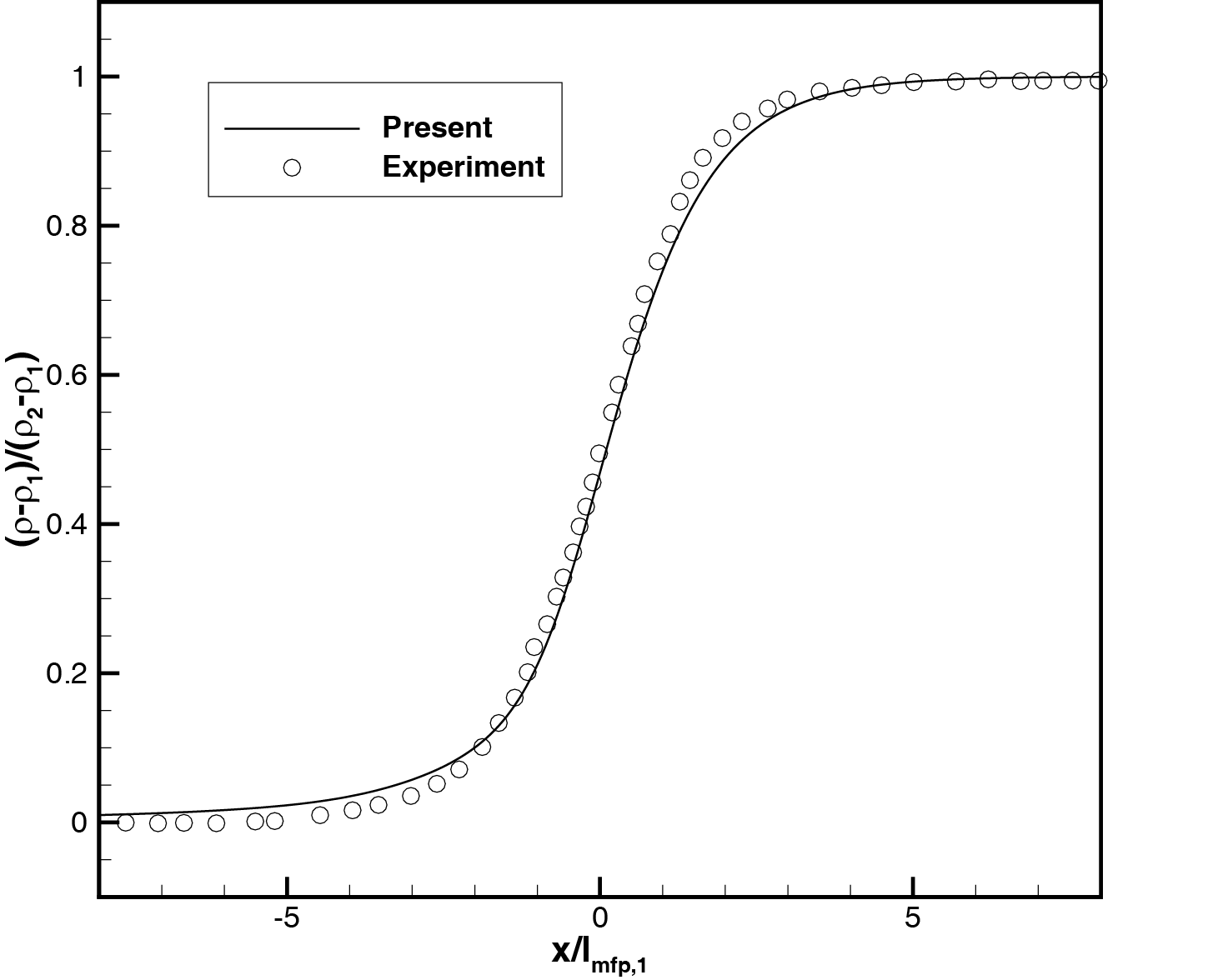}}
\caption{\label{fig:test2_cmpexp}Shock structures at different Mach numbers compared with experimental results \cite{alsmeyer1976density}. $\rho_1$ and $\rho_2$ are the upstream and downstream far-field densities respectively, $l_{{\rm{mfp,1}}}$ is the upstream mean free path.}
\end{figure}

The test case of the shock structure is conducted to verify if the present method can simulate the highly nonequilibrium flow in the shock layer. The VHS molecular model with $\omega=0.72$ is applied and the rotational relaxation collision number $Z_{\rm{rot}}$ is set as a constant $Z_{\rm{rot}}=2.4$. The computational domain is set as $[ - 400{l_{{\rm{mfp,1}}}},400{l_{{\rm{mfp,1}}}}]$ where $l_{{\rm{mfp,1}}}$ is the upstream mean free path. A large range of particle velocity space $[ - 30{a_1},30{a_1}]$ is used with a uniform 1200-cell discretization, where $a_1$ is the upstream acoustic velocity. For this test case the position of the shock at the final steady state sensitively depends on the initial value of the flow field, and a position-correction operation has been done to avoid the shock shift due to the implicit iterations: intermittently calculate the total mass in the whole computational domain and add the losing mass to the downstream field. The comparisons of the density and temperature distributions between the present and DSMC results \cite{liu2014unified} are plotted in Fig.~\ref{fig:test2_cmpdsmc}. The present density distributions agree well with the results of DSMC. For the temperature, in the downstream field the present temperature curves agree well with the DSMC's results while in the upstream field the present temperature curves are generally higher than the DSMC's results. This is due to the common defect of the relaxation-type kinetic models which have a single relaxation time for particles with different velocity. In these models the relaxation rate of the high speed particles is underestimated, and the high energy high speed particles will incorrectly transport upstream for a very long distance, leading to the overheating of the upstream flow. The recipe for this problem is beyond the scope of this paper. The comparisons of the density distributions at different Mach numbers between the present and experimental results \cite{alsmeyer1976density} are shown in Fig.~\ref{fig:test2_cmpexp}. The two sets of results are in good consistence.

\subsection{Hypersonic flow passing a flat plate}

When the hypersonic gas flow passes through a flat plate, shock wave and boundary layer interaction occur and there will be a strong thermal nonequilibrium between translational and rotational temperatures. The hypersonic rarefied nitrogen flow over a flat plate with a sharp leading edge is simulated by the present method. The condition is the same with the run34 case in Ref.~\cite{Tsuboi2005Experimental}. The freestream Mach number $\rm{Ma}$, temperature $T_{\infty}$ and pressure $p_{\infty}$ are 4.89, 116K and 2.12Pa respectively. The VHS molecular model with $\omega=0.75$ is applied and the freestream mean free path ${l_{{\rm{mfp}},\infty }}$ is around 0.78mm. The temperature of the plate surface $T_{w}$ is 290K, according to which the rotational relaxation collision number $Z_{\rm{rot}}$ is set as $3.5$. At the plate surface, the diffuse reflection boundary condition with full thermal accommodation \cite{Li2005Application} is applied.

\begin{figure}
\centering
\includegraphics[width=0.6\textwidth]{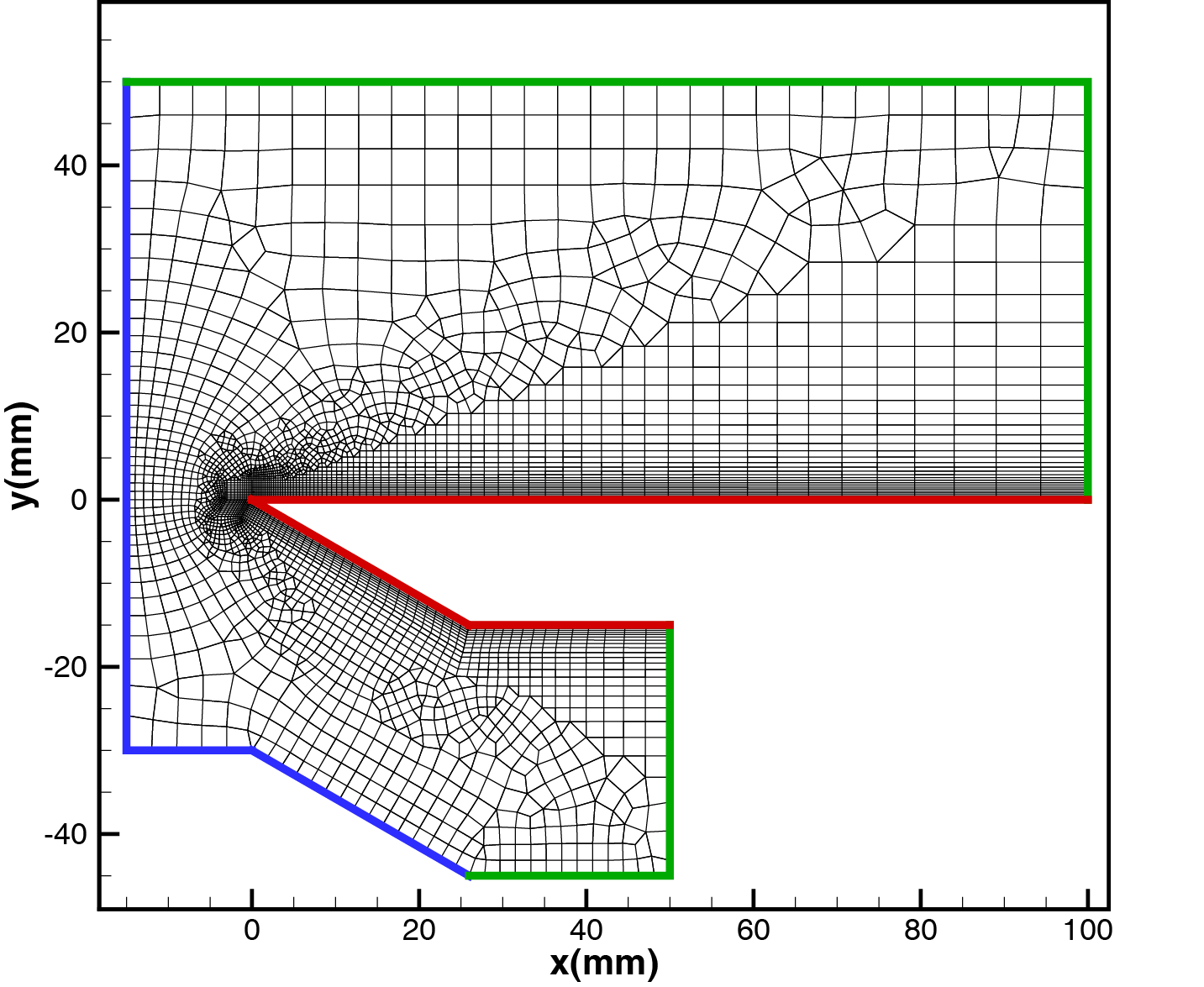}
\caption{\label{fig:test3_mesh}Mesh for the hypersonic flow passing a flat plate (3869 cells). Blue edge: inlet boundary. Green edge: outlet boundary. Red edge: solid boundary.}
\end{figure}

\begin{figure}
\centering
\includegraphics[width=0.6\textwidth]{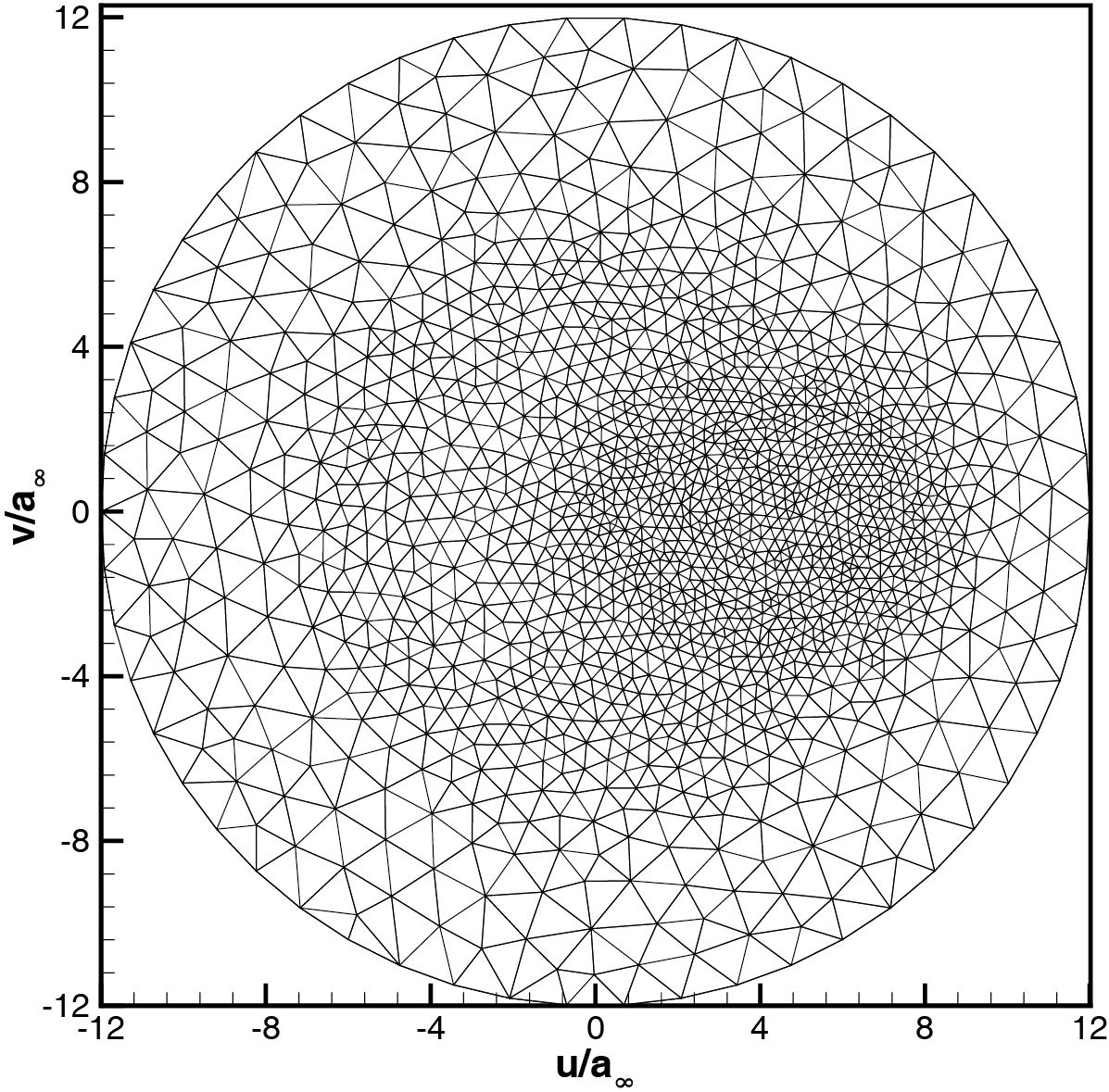}
\caption{\label{fig:test3_vecmesh}Particle velocity space mesh for the hypersonic flow passing a flat plate (2838 cells). $a_{\infty}$ is the freestream acoustic velocity.}
\end{figure}

\begin{figure}
\centering
\subfigure[Density]{\includegraphics[width=0.47\textwidth]{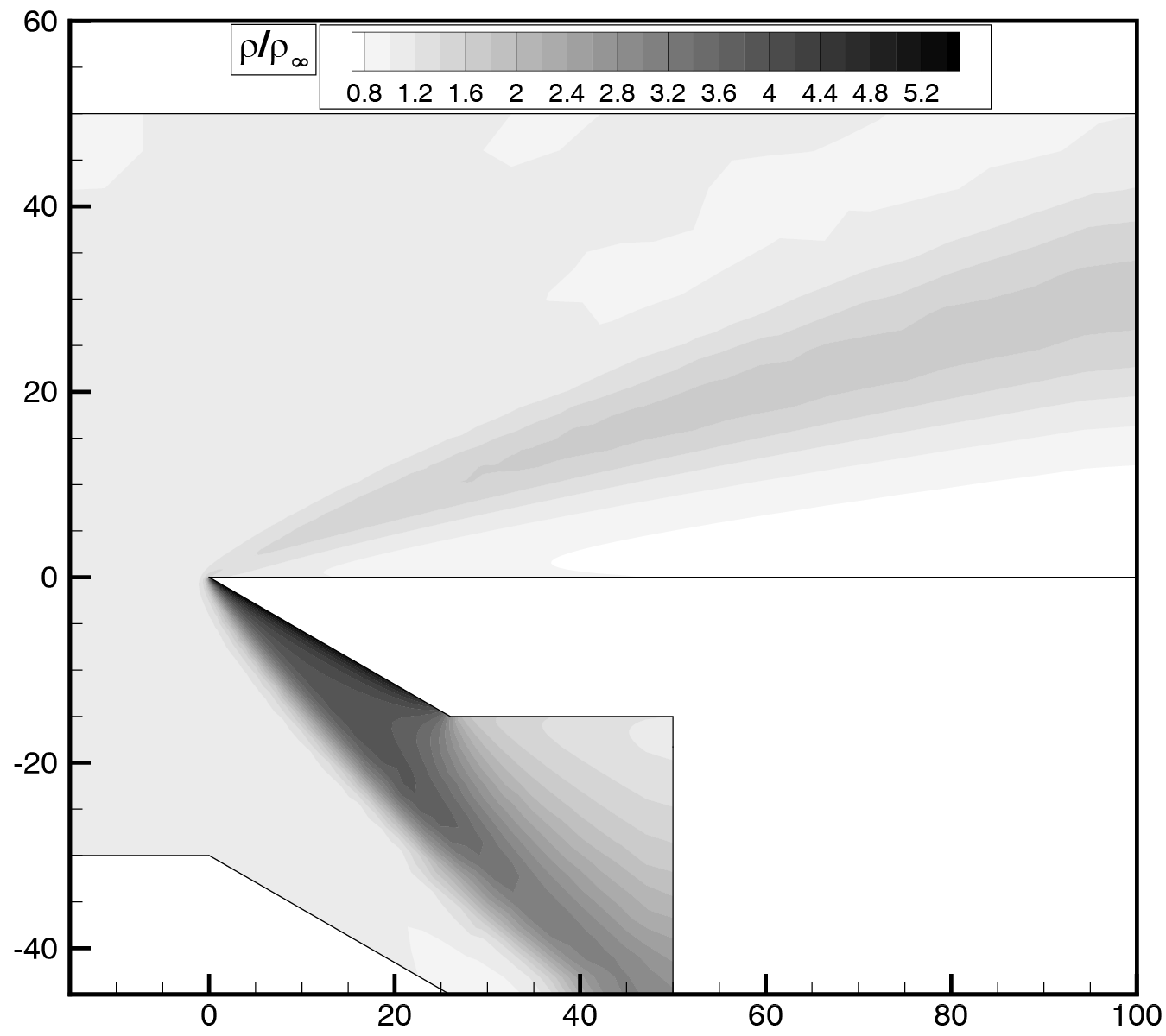}}\hspace{0.02\textwidth}%
\subfigure[Equilibrium temperature]{\includegraphics[width=0.47\textwidth]{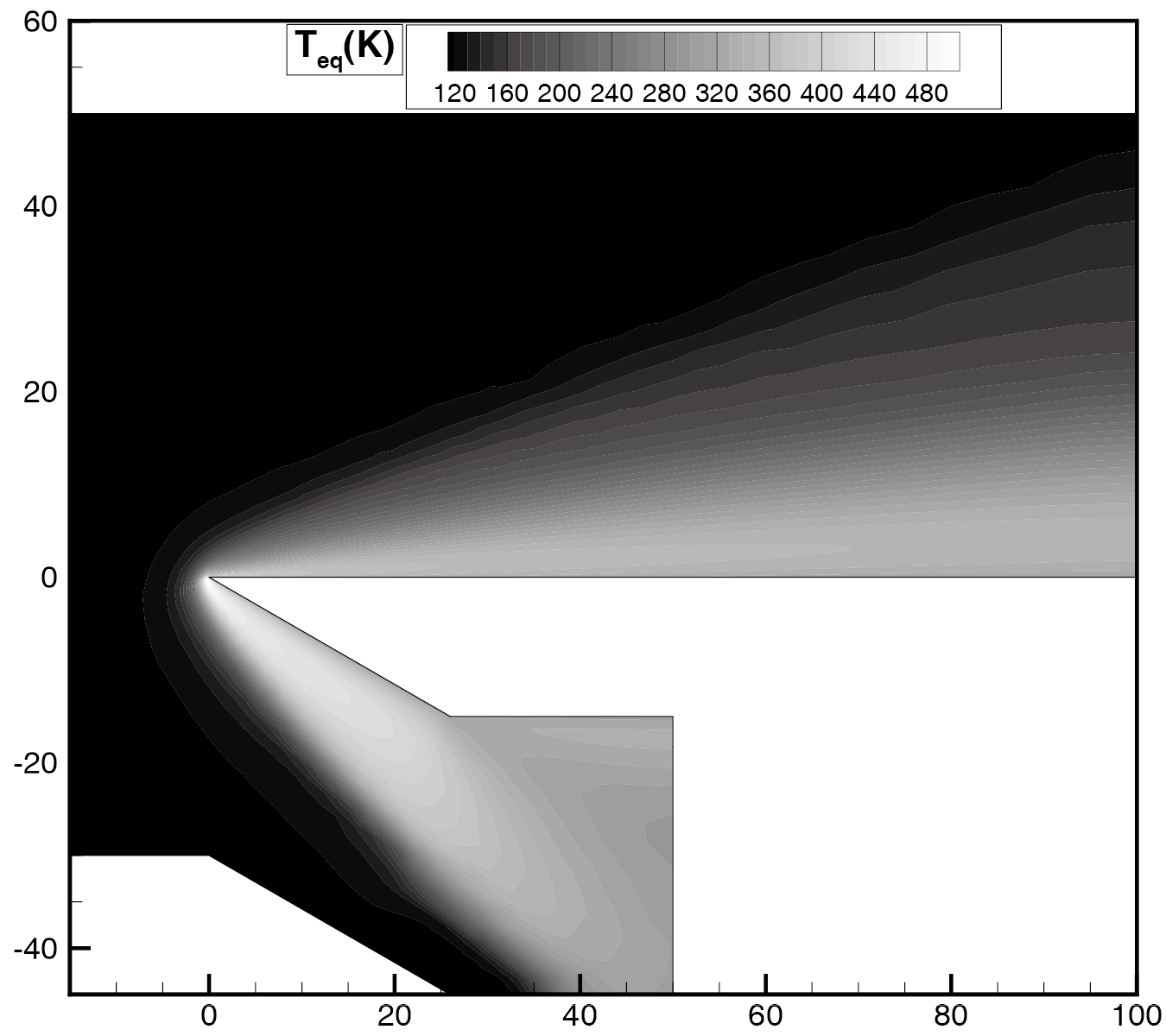}}\\
\subfigure[Translational temperature]{\includegraphics[width=0.47\textwidth]{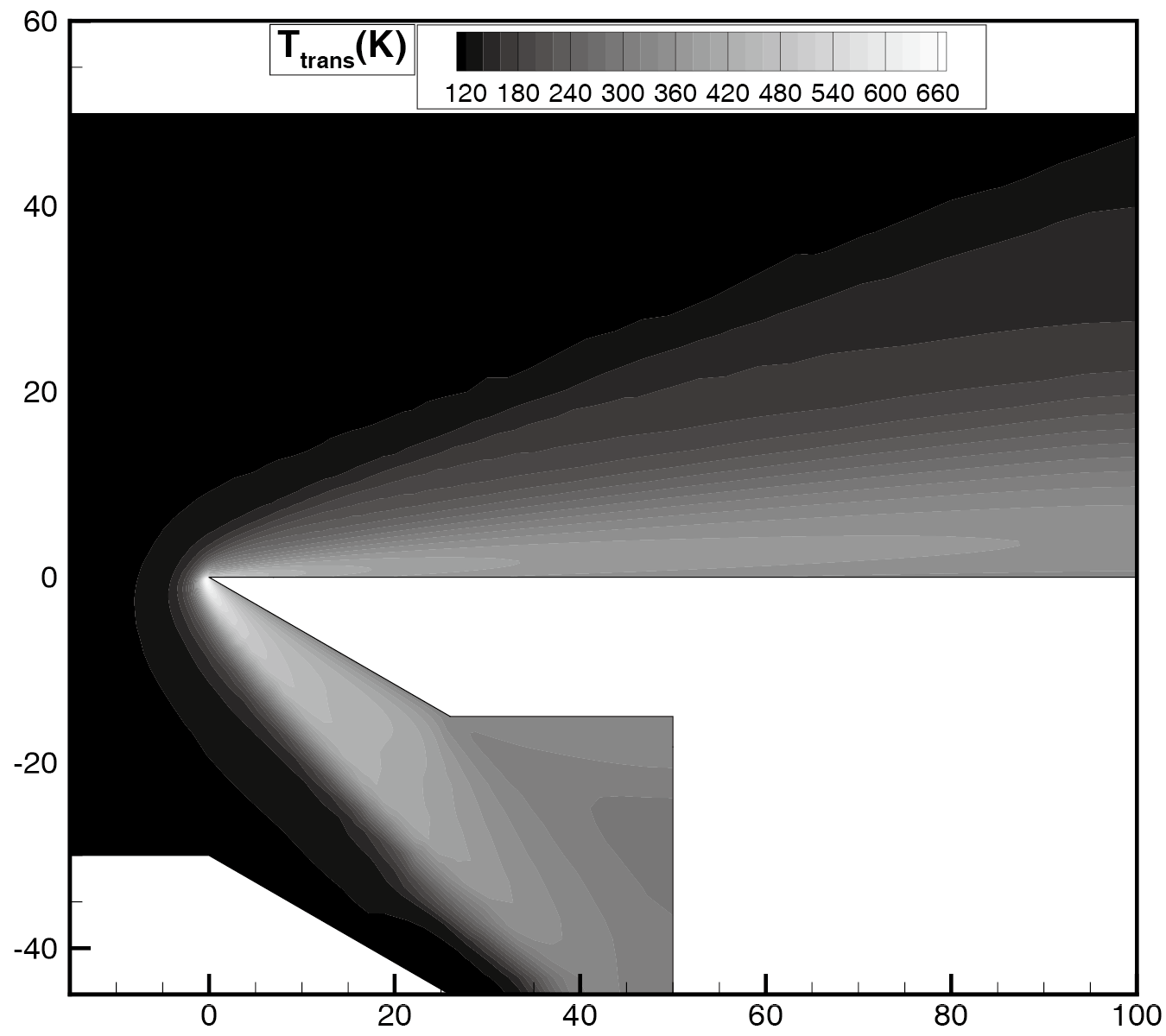}}\hspace{0.02\textwidth}%
\subfigure[Rotational temperature]{\includegraphics[width=0.47\textwidth]{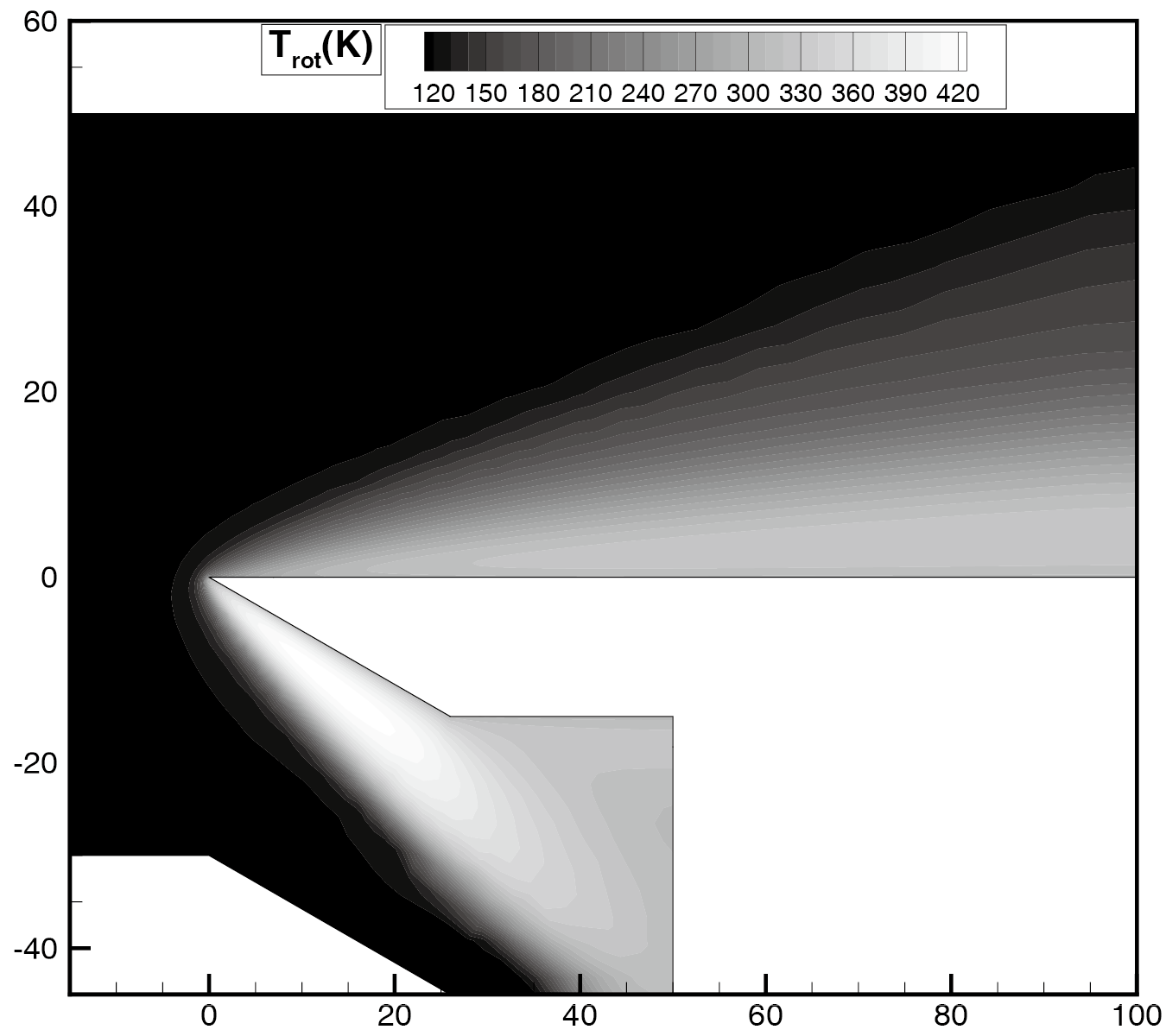}}
\caption{\label{fig:test3_contour}Hypersonic flow passing a flat plate.}
\end{figure}

\begin{figure}
\centering
\subfigure[$x=5\rm{mm}$]{\includegraphics[width=0.47\textwidth]{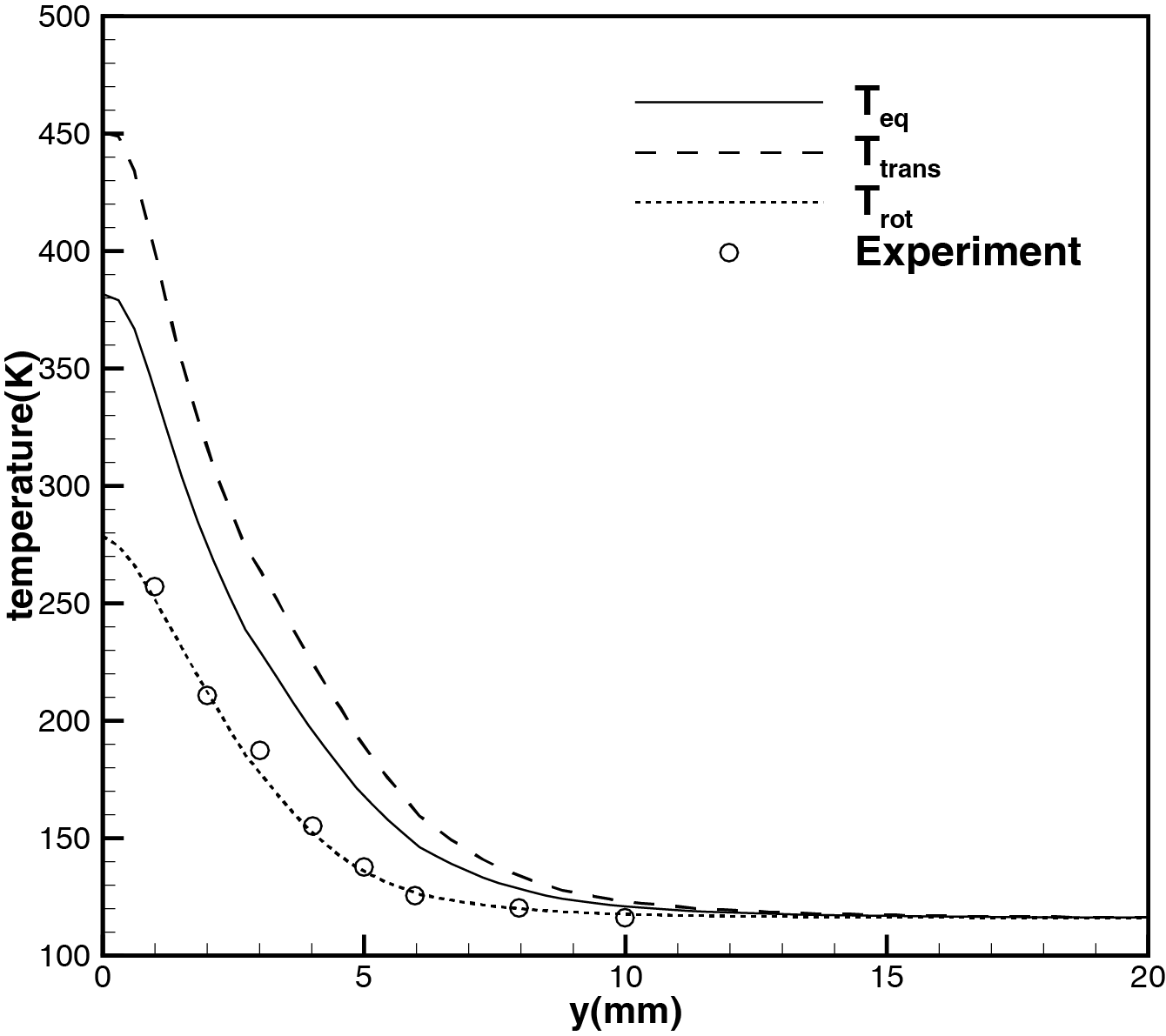}}\hspace{0.02\textwidth}%
\subfigure[$x=20\rm{mm}$]{\includegraphics[width=0.47\textwidth]{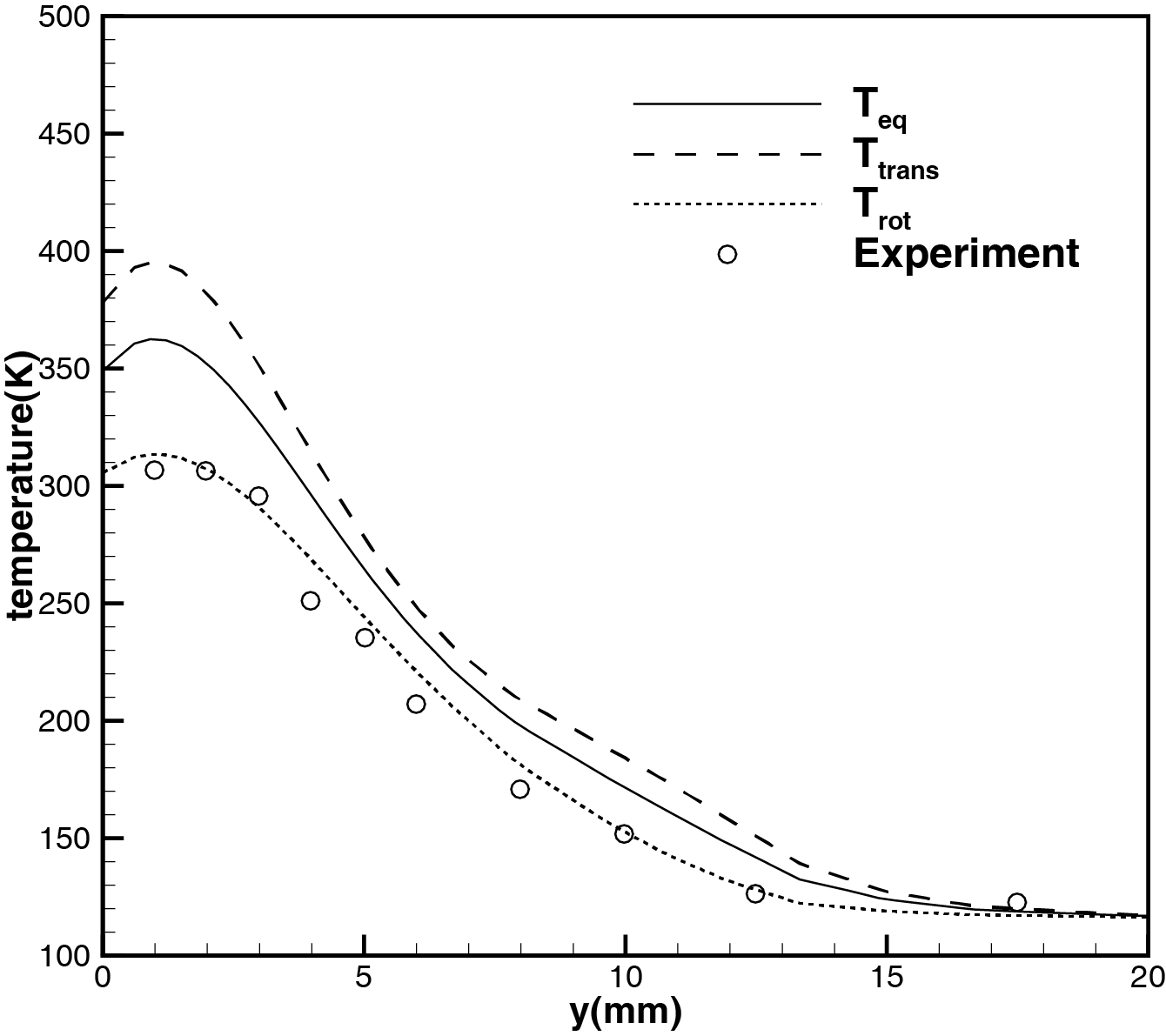}}
\caption{\label{fig:test3_Tprofile}Temperature profiles of the hypersonic flow passing a flat plate compared with the experimental rotational temperature distributions \cite{Tsuboi2005Experimental}.}
\end{figure}

\begin{figure}
\centering
\subfigure[Density distribution]{\includegraphics[width=0.47\textwidth]{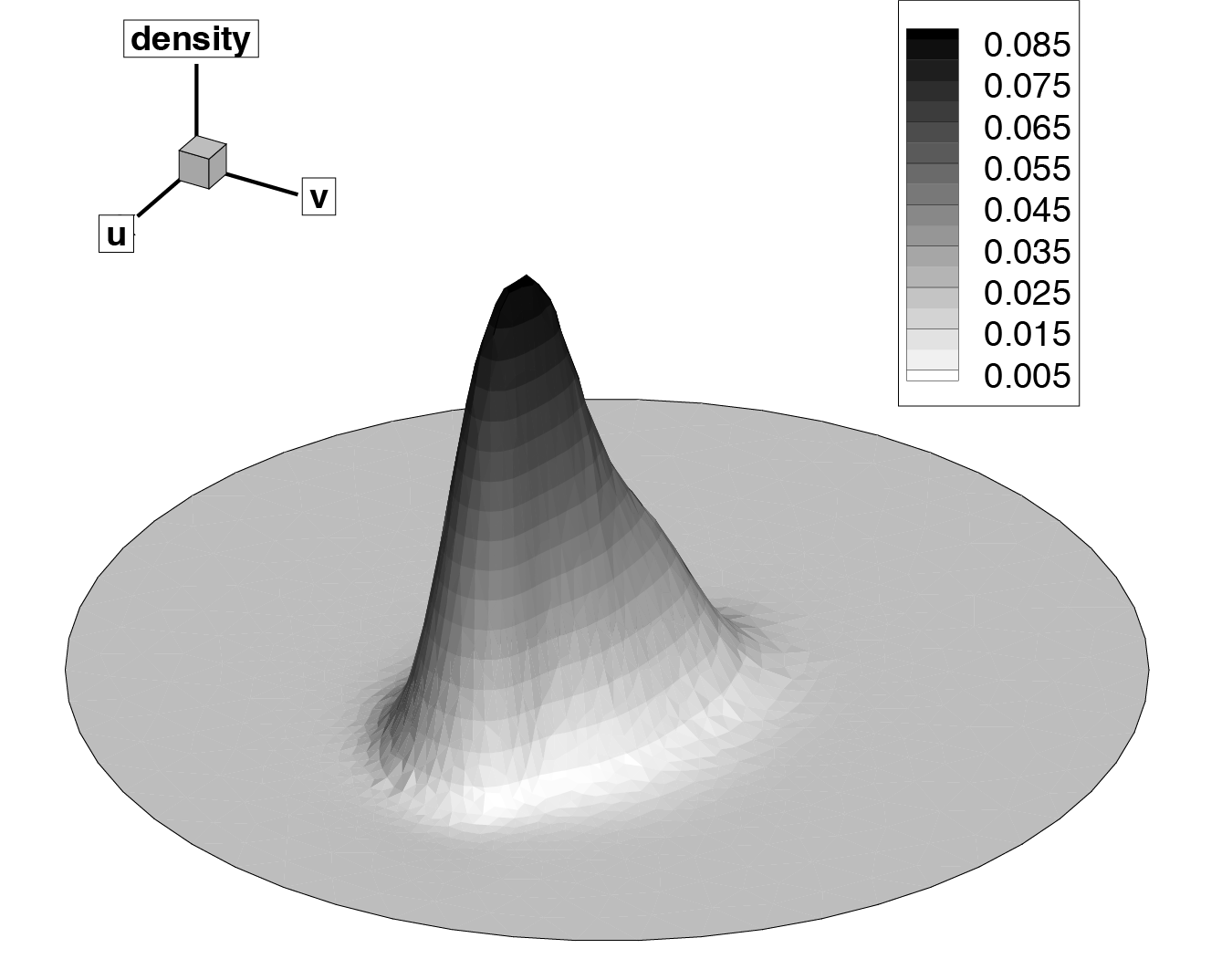}}\hspace{0.02\textwidth}%
\subfigure[Rotational energy distribution]{\includegraphics[width=0.47\textwidth]{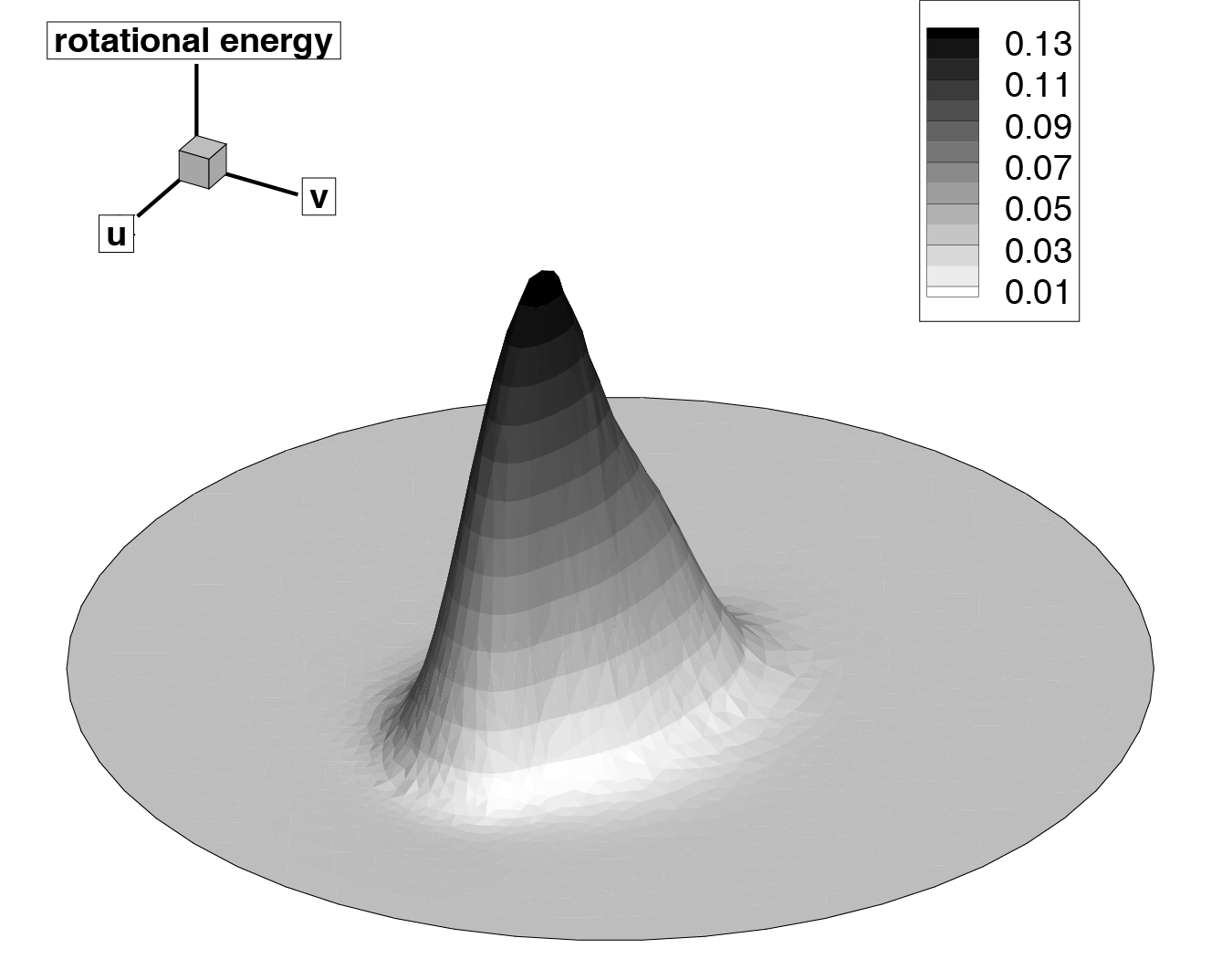}}
\caption{\label{fig:test3_x5y1}Distribution functions in the particle velocity space at $x=5.1\rm{mm}, y=0.93\rm{mm}$ for the hypersonic flow passing a flat plate.}
\end{figure}

\begin{figure}
\centering
\subfigure[Density distribution]{\includegraphics[width=0.47\textwidth]{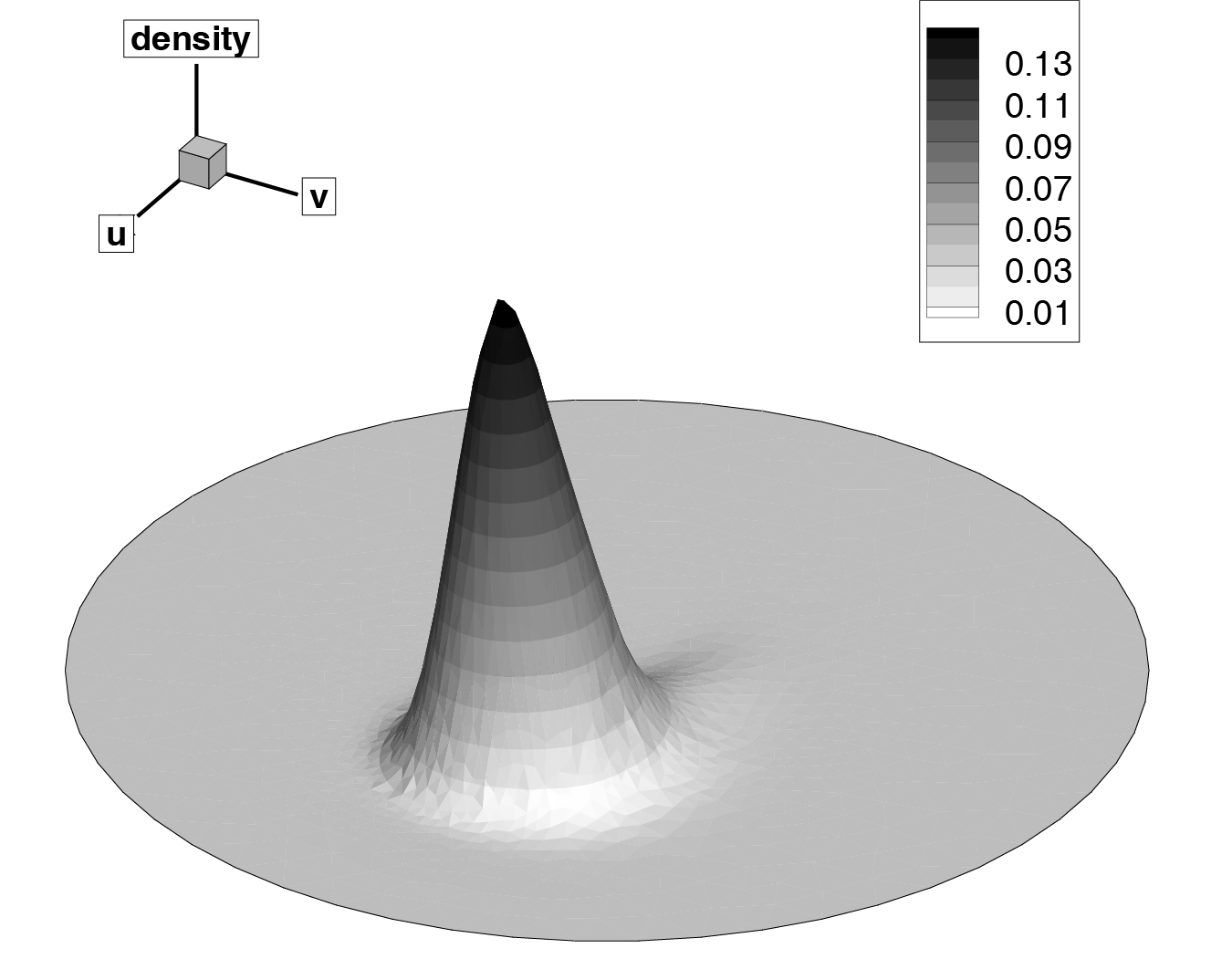}}\hspace{0.02\textwidth}%
\subfigure[Rotational energy distribution]{\includegraphics[width=0.47\textwidth]{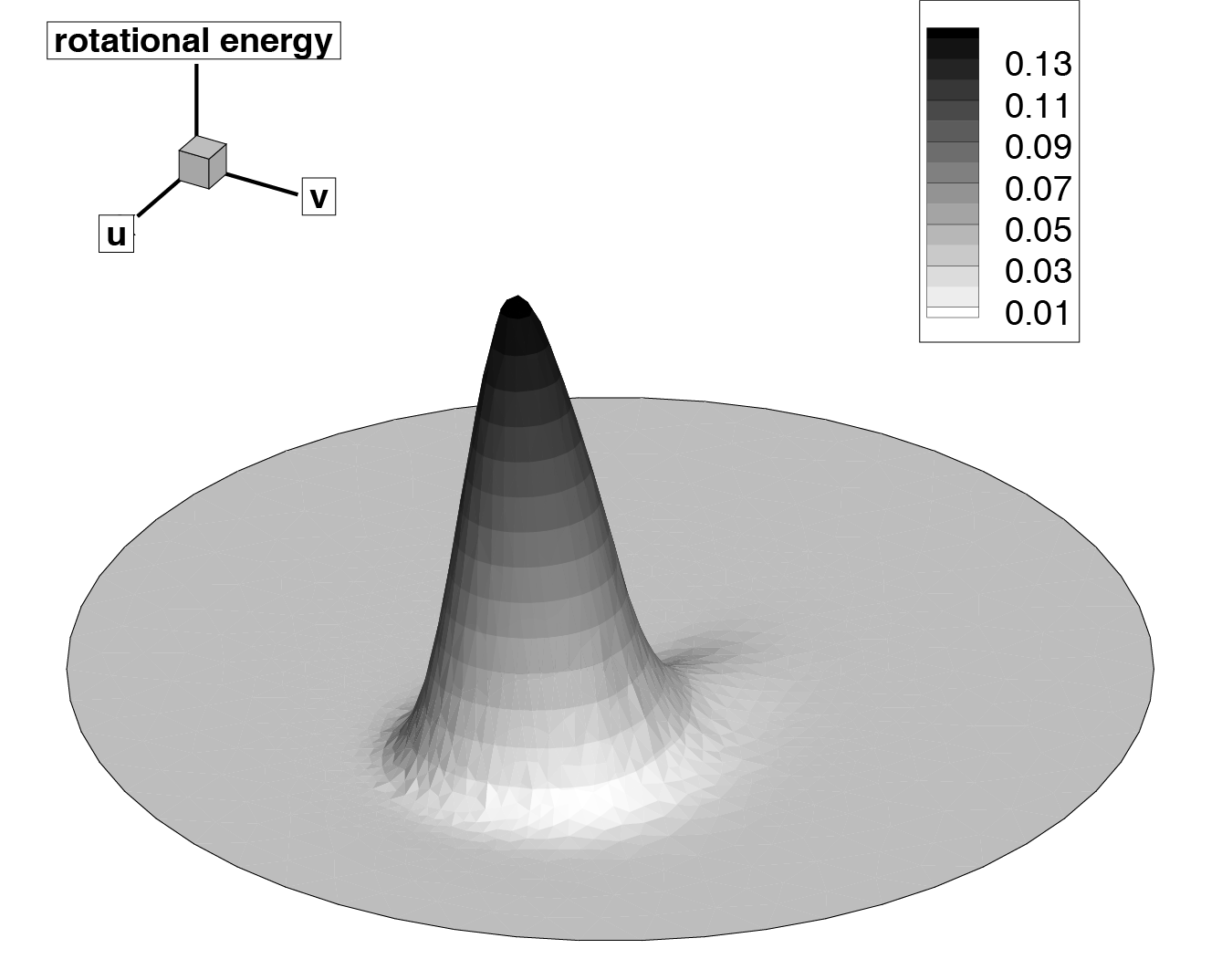}}
\caption{\label{fig:test3_x5y3}Distribution functions in the particle velocity space at $x=5.0\rm{mm}, y=3.0\rm{mm}$ for the hypersonic flow passing a flat plate.}
\end{figure}

A 3869-cell mesh for the physical space and a 2838-cell mesh for the velocity space are adopted, as shown in Fig.~\ref{fig:test3_mesh} and Fig.~\ref{fig:test3_vecmesh}. The computation is conducted on a computer with 24 cores' parallel execution (two-way Intel(R) Xeon(R) CPU E5-2678 v3 @ 2.50GHz). The residual criterion for convergence is set as $10^{-9}$. The calculation finished at the 136th step in 48 seconds, which is very efficient. The density, equilibrium temperature, translational temperature and rotational temperature contours are shown in Fig.~\ref{fig:test3_contour}. It is observed that the flow passing through the upper surface of the plate first experiences a compression near the leading edge and the maximum density comes to near $1.7\rho _\infty $, and then the flow expands with a decrease in density to around $0.7\rho _\infty $, forming a thick nonequilibrium layer above the plate. On the lower surface of the plate, the oblique shock wave merges with the boundary layer and the maximum density which is near $5.5\rho _\infty $ occurs on the slope surface of the leading edge due to the strong compression of the flow. The translational temperature reaches the maximum value 680K near the apex of the slope while the maximum rotational temperature comes later above the middle of the slope and has a value of around 430K. This is due to the energy transfer process from translational degrees of freedom to rotational degrees of freedom. The temperature profiles above the upper surface of the plate at two vertical cross sections $x=5\rm{mm},20\rm{mm}$ are shown in Fig.~\ref{fig:test3_Tprofile}. The rotational temperature profiles match quite well with the experimental results \cite{Tsuboi2005Experimental}. The thickness of the thermal nonequilibrium layer is around 10mm at $x=5\rm{mm}$ while 16mm at $x=20\rm{mm}$. The distribution functions of particles at two positions along the vertical line x=5mm are plotted in Fig.~\ref{fig:test3_x5y1} and Fig.~\ref{fig:test3_x5y3}. At the lower position the distribution functions demonstrate a large deviation from the Maxwell distribution due to the strong nonequilibrium effect, while at the higher position the distributions approach the Maxwell distribution due to the relaxation process.

\section{Conclusions}\label{sec:conclusions}

In this paper, a conservative implicit scheme for steady state solutions of diatomic gas flow is proposed. In the present method, the translational and rotational degrees of freedom are considered and the Rykov model equation is solved in a finite volume framework where the equation is discretized in time, physical space and particle velocity space. To get a multiscale numerical flux and release the cell size from the constraint of the cell Kn number, a difference scheme of the model equation is used to project the initial data inside the cell to the interface with a physical local time step $h_{ij}$. To get fast convergence rate in all flow regimes, the implicit macroscopic equation is solved along with the implicit microscopic system to evaluate a predicted equilibrium state $\tilde g_{i,k}^{ * ,n + 1}$. All of the implicit discrete equations are solved by SGS iterations. To make the scheme conservative, an integral error compensation is implemented when calculating the macroscopic variables from the discretized distribution function, which can reduce the accuracy requirement of the discretization for the particle velocity space thus the discrete velocity space can be more flexible.

In the numerical tests, the efficiency and accuracy of the method are first verified by the cases of cavity flows in several flow regimes. The results of the present method agree well with the results of UGKS and the present method is 1--2 orders of magnitude faster than the explicit UGKS in all flow regimes. The physical local time step technique is validated in a case with large cell size difference and it shows that without physical local time step the scheme will lose its multiscale property in the nonuniform mesh. The integral error compensation is also proved to be effective and it is indicated that the scheme cannot converge if without the compensation procedure when using an unstructured mesh in velocity space. Moreover, the test cases of shock structure and hypersonic flow passing a flat plate are performed, in which the present method shows good accuracy comparing with the results of DSMC and experiment.

In conclusion, the present method is efficient and accurate for computing steady solutions of diatomic gas flow in all flow regimes, with flexible discretization of the velocity space.

\clearpage


\bibliographystyle{yuan_implicit}
\bibliography{yuan_implicit}

\begin{thebibliography}{10}

\bibitem{Bird1994Molecular}
G.~A. Bird.
\newblock \emph{Molecular gas dynamics and the direct simulation of gas flows}.
\newblock Clarendon Press, 1994.

\bibitem{Goldstein1989Investigations}
D.~Goldstein, B.~Sturtevant, and J.~E. Broadwell.
\newblock Investigations of the motion of discrete-velocity gases.
\newblock \emph{Progress in Astronautics and Aeronautics}, 1989.
\newblock 117:100--117.

\bibitem{Yang1995Rarefied}
J.~Y. Yang and J.~C. Huang.
\newblock Rarefied flow computations using nonlinear model Boltzmann equations.
\newblock \emph{Journal of Computational Physics}, 1995.
\newblock 120(2):323--339.

\bibitem{Mieussens2000Discretev}
L.~Mieussens.
\newblock Discrete-velocity models and numerical schemes for the Boltzmann-BGK
  equation in plane and axisymmetric geometries.
\newblock \emph{Journal of Computational Physics}, 2000.
\newblock 162(2):429--466.

\bibitem{li2004Study}
Z.-H. Li and H.-X. Zhang.
\newblock Study on gas kinetic unified algorithm for flows from rarefied
  transition to continuum.
\newblock \emph{Journal of Computational Physics}, 2004.
\newblock 193(2):708--738.

\bibitem{Titarev2007Conservative}
V.~A. Titarev.
\newblock Conservative numerical methods for model kinetic equations.
\newblock \emph{Computers \& Fluids}, 2007.
\newblock 36(9):1446--1459.

\bibitem{Mieussens2000DISCRETE}
L.~Mieussens.
\newblock Discrete velocity model and implicit scheme for the BGK equation of
  rarefied gas dynamics.
\newblock \emph{Mathematical Models and Methods in Applied Sciences}, 2000.
\newblock 10(08):1121--1149.

\bibitem{Chen2017A}
S.~Chen, C.~Zhang, L.~Zhu, and Z.~Guo.
\newblock A unified implicit scheme for kinetic model equations. Part I. Memory
  reduction technique.
\newblock \emph{Science Bulletin}, 2017.
\newblock 62(2):119--129.

\bibitem{Xu2010A}
K.~Xu and J.~C. Huang.
\newblock A unified gas-kinetic scheme for continuum and rarefied flows.
\newblock \emph{Journal of Computational Physics}, 2010.
\newblock 229(20):7747--7764.

\bibitem{guo2013discrete}
Z.~Guo, K.~Xu, and R.~Wang.
\newblock Discrete unified gas kinetic scheme for all Knudsen number flows:
  Low-speed isothermal case.
\newblock \emph{Physical Review E}, 2013.
\newblock 88(3):033305.

\bibitem{guo2015discrete}
Z.~Guo, R.~Wang, and K.~Xu.
\newblock Discrete unified gas kinetic scheme for all Knudsen number flows. II.
  Thermal compressible case.
\newblock \emph{Physical Review E}, 2015.
\newblock 91(3):033313.

\bibitem{Mao2015STUDY}
M.~Mao, D.~Jiang, L.~Jin, and X.~Deng.
\newblock Study on implicit implementation of the unified gas kinetic scheme.
\newblock \emph{Chinese Journal of Theoretical and Applied Mechanics}, 2015.
\newblock 47(5):822--829.

\bibitem{Zhu2016Implicit}
Y.~Zhu, C.~Zhong, and K.~Xu.
\newblock Implicit unified gas-kinetic scheme for steady state solutions in all
  flow regimes.
\newblock \emph{Journal of Computational Physics}, 2016.
\newblock 315:16--38.

\bibitem{Zhu2017Unified}
Y.~Zhu, C.~Zhong, and K.~Xu.
\newblock Unified gas-kinetic scheme with multigrid convergence for rarefied
  flow study.
\newblock \emph{Physics of Fluids}, 2017.
\newblock 29(9):096102.

\bibitem{Yang2018An}
L.~M. Yang, C.~Shu, W.~M. Yang, and J.~Wu.
\newblock An implicit scheme with memory reduction technique for steady state
  solutions of DVBE in all flow regimes.
\newblock \emph{Physics of Fluids}, 2018.
\newblock 30(4):040901.

\bibitem{Huang2011Aconservative}
J.~C. Huang.
\newblock A conservative discrete ordinate method for model Boltzmann
  equations.
\newblock \emph{Computers \& Fluids}, 2011.
\newblock 45(1):261--267.

\bibitem{Huang2015Aconservative}
J.~C. Huang, T.~Y. Hsieh, and J.~Y. Yang.
\newblock A conservative discrete ordinate method for solving semiclassical
  Boltzmann-BGK equation with Maxwell type wall boundary condition.
\newblock \emph{Journal of Computational Physics}, 2015.
\newblock 290:112--131.

\bibitem{Jiang2015STUDY}
D.~Jiang, M.~Mao, L.~Jin, and X.~Deng.
\newblock Study on the numerical error introduced by dissatisfying the
  conservation constraint in UGKS and its effects.
\newblock \emph{Chinese Journal of Theoretical and Applied Mechanics}, 2015.
\newblock 47(1):163--168.

\bibitem{liu2014unified}
S.~Liu, P.~Yu, K.~Xu, and C.~Zhong.
\newblock Unified gas-kinetic scheme for diatomic molecular simulations in all
  flow regimes.
\newblock \emph{Journal of Computational Physics}, 2014.
\newblock 259:96--113.

\bibitem{Rykov1975A}
V.~A. Rykov.
\newblock A model kinetic equation for a gas with rotational degrees of
  freedom.
\newblock \emph{Fluid Dynamics}, 1975.
\newblock 10(6):959--966.

\bibitem{zhang2015Aunified}
H.~Zhang.
\newblock \emph{A unified gas-kinetic scheme based on a vibrational model (in
  Chinese)}.
\newblock Master's thesis, Northwestern Polytechnical University, 2015.

\bibitem{Wang2017Unified}
Z.~Wang, H.~Yan, Q.~Li, and K.~Xu.
\newblock Unified gas-kinetic scheme for diatomic molecular flow with
  translational, rotational, and vibrational modes.
\newblock \emph{Journal of Computational Physics}, 2017.
\newblock 350:237--259.

\bibitem{Xu2015Direct}
K.~Xu.
\newblock \emph{Direct modeling for computational fluid dynamics: construction
  and application of unified gas-kinetic schemes}.
\newblock World Scientifc, 2015.

\bibitem{Parker1959Rotational}
J.~G. Parker.
\newblock Rotational and vibrational relaxation in diatomic gases.
\newblock \emph{Physics of Fluids}, 1959.
\newblock 2(4):449--462.

\bibitem{Lordi1970Rotational}
J.~A. Lordi and R.~E. Mates.
\newblock Rotational relaxation in nonpolar diatomic gases.
\newblock \emph{Physics of Fluids}, 1970.
\newblock 13(2):291--308.

\bibitem{Koura1998Statistical}
K.~Koura.
\newblock Statistical inelastic cross-section model for the Monte Carlo
  simulation of molecules with discrete internal energy.
\newblock \emph{Physics of Fluids A: Fluid Dynamics}, 1992.
\newblock 4(8):1782--1788.

\bibitem{Ivanov1998Computational}
M.~S. Ivanov and S.~F. Gimelshein.
\newblock Computational hypersonic rarefied flows.
\newblock \emph{Annual Review of Fluid Mechanics}, 1998.
\newblock 30(1):469--505.

\bibitem{bhatnagar1954model}
P.~L. Bhatnagar, E.~P. Gross, and M.~Krook.
\newblock A model for collision processes in gases. I. Small amplitude
  processes in charged and neutral one-component systems.
\newblock \emph{Physical Review}, 1954.
\newblock 94(3):511.

\bibitem{Rogers1995Comparison}
S.~E. Rogers.
\newblock Comparison of implicit schemes for the incompressible Navier-Stokes
  equations.
\newblock \emph{AIAA Journal}, 1995.
\newblock 33(11):2066--2072.

\bibitem{Yuan2002Comparison}
L.~Yuan.
\newblock Comparison of implicit multigrid schemes for three-dimensional
  incompressible flows.
\newblock \emph{Journal of Computational Physics}, 2002.
\newblock 177(1):134--155.

\bibitem{xu2001gas}
K.~Xu.
\newblock A gas-kinetic BGK scheme for the Navier-Stokes equations and its
  connection with artificial dissipation and Godunov method.
\newblock \emph{Journal of Computational Physics}, 2001.
\newblock 171(1):289--335.

\bibitem{Li2005Application}
Q.~Li, S.~Fu, and K.~Xu.
\newblock Application of gas-kinetic scheme with kinetic boundary conditions in
  hypersonic flow.
\newblock \emph{AIAA Journal}, 2005.
\newblock 43(10):2170--2176.

\bibitem{alsmeyer1976density}
H.~Alsmeyer.
\newblock Density profiles in argon and nitrogen shock waves measured by the
  absorption of an electron beam.
\newblock \emph{Journal of Fluid Mechanics}, 1976.
\newblock 74(3):497--513.

\bibitem{Tsuboi2005Experimental}
N.~Tsuboi and Y.~Matsumoto.
\newblock Experimental and numerical study of hypersonic rarefied gas flow over
  flat plates.
\newblock \emph{AIAA Journal}, 2005.
\newblock 43(6):1243--1255.

\bibitem{Shan2006Kinetic}
X.~Shan, X.~Yuan, and H.~Chen.
\newblock Kinetic theory representation of hydrodynamics: a way beyond the
  Navier-Stokes equation.
\newblock \emph{Journal of Fluid Mechanics}, 2006.
\newblock 550:413--441.

\end{thebibliography}

\clearpage

\clearpage

\clearpage
\renewcommand{\multirowsetup}{\centering}

\begin{table}
\centering
\caption{\label{tab:caveff}Comparison of the efficiency between the explicit UGKS and the present method for cavity flow simulation in all flow regimes.}
\begin{tabular}{p{50pt} p{40pt}<{\raggedleft} p{40pt}<{\raggedleft} p{40pt}<{\raggedleft} p{40pt}<{\raggedleft} p{40pt}<{\raggedleft} p{30pt}<{\raggedleft}}
\hline

\hline
\multicolumn{1}{c}{\multirow{2}{*}{Case}} & \multirow{2}{40pt}{Velocity space} & \multicolumn{2}{c}{Explicit UGKS} & \multicolumn{2}{c}{Present} & \multicolumn{1}{c}{\multirow{2}{*}{Speedup}}\\\cline{3-4}\cline{5-6}
 ~&~& \multicolumn{1}{c}{Steps} & \multicolumn{1}{c}{Time (s)} & \multicolumn{1}{c}{Steps} & \multicolumn{1}{c}{Time (s)} &~ \\
\hline
Re=1000 & 12 & 608946 & 19820 & 973 & 99 & 200.2 \\
Kn=0.075 & 792 & 6580 & 10610 & 148 & 278 & 38.2 \\
Kn=1 & 6286 & 4581 & 61703 & 182 & 3441 & 17.9 \\
Kn=10 & 6286 & 35313 & 478653 & 181 & 3451 & 138.7 \\
\hline

\hline
\end{tabular}
\end{table}

\end{document}